\documentclass[twocolumn]{aastex62}

\usepackage{graphicx}

\graphicspath{{./}{figures/}}

\shorttitle{Ionization Mechanisms in Quasar Outflows}
\shortauthors{Hinkle, Veilleux, \& Rupke}

\defcitealias{2017ApJ..850...40R}{R17}
\defcitealias{1987ApJS..63...295V}{VO87}
\defcitealias{1981PASP...93....5B}{BPT}

\begin{document}

\title{Ionization Mechanisms in Quasar Outflows}

\author{Jason T. Hinkle}
\affiliation{Department of Physics, University of Maryland, College Park, MD 20742, USA}
\affiliation{Department of Astronomy, University of Maryland, College Park, MD 20742, USA}

\author{Sylvain Veilleux}
\altaffiliation{veilleux@astro.umd.edu}
\affiliation{Department of Astronomy, University of Maryland, College Park, MD 20742, USA}
\affiliation{Joint Space-Science Institute, University of Maryland, College Park, MD 20742, USA}
\affiliation{Institute of Astronomy and Kavli Institute for Cosmology Cambridge, University of Cambridge, Cambridge, CB3 0HA, United Kingdom}

\author{David S.~N. Rupke}
\affiliation{Department of Physics, Rhodes College, Memphis TN, 38112, USA}

\begin{abstract}
\noindent
The various ionization mechanisms at play in active galactic nuclei (AGN) and quasars have been well studied, but relatively little has been done to separately investigate the contributions of these ionization mechanisms within the host galaxy and outflowing components. Using Gemini integral field spectroscopy (IFS) data, we study the ionization properties of these two components in four nearby ($z \la$ 0.2) radio-quiet Type 1 quasars. Emission line ratios and widths are employed to identify the dominant ionization mechanisms for the host and outflow components in each object. We find that photoionization by the AGN often dominates the ionization of both gaseous components in these systems. In three cases, the outflowing gas is more highly ionized than the gas in the host, indicating that it is more strongly exposed to the ionizing radiation field of the AGN. In two objects, a positive correlation between the line widths and line ratios in the outflowing gas component indicates that shocks with velocities of order 100 $-$ 500 km~s$^{-1}$ may also be contributing to the ionization and heating of the outflowing gas component. The line ratios in the outflowing gas of one of these two objects also suggest a significant contribution from photoionization by hot, young stars in the portion of the outflow that is closest to star-forming regions in the host galaxy component. The data thus favor photoionization by hot stars in the host galaxy rather than stars formed in the outflow itself. 

\end{abstract}


\section{Introduction} \label{sec:intro}
Quasar feedback may play a fundamental role in determining the evolution of galaxies and their environments (e.g., Veilleux et al.\ \citeyear{2005ARA&A..43..769V}; Fabian \citeyear{2012ARA&A..50..455F}; Harrison \citeyear{2018NatAs...2..198H}). The intense ionizing radiation field from quasars may severely affect the thermodynamic properties of the gaseous component of the galaxy hosts and the surrounding circumgalactic medium (e.g., Bajtlik et al.\ \citeyear{1988ApJ...327..570B}; Curran \& Whiting \citeyear{2012ApJ...759..117C}; Liu et al.\ \citeyear{2013MNRAS.430.2327L}). The radiative and mechanical energy injected by quasars in the cores of their galaxy hosts also drive massive multi-phase outflows that often reach galactic scales and beyond (e.g., Rupke et al.\ \citeyear{2017ApJ..850...40R}, hereafter \citetalias{2017ApJ..850...40R}; Harrison et al.\ \citeyear{2018NatAs...2..198H}).

There is still some debate over how common large-scale outflows are in quasars. Greene et al.\ (\citeyear{2011ApJ...732....9G}), Rupke et al.\ (\citeyear{2011ApJ...729L..27R}, \citeyear{2013ApJ...768...75R}, \citeyear{2017ApJ..850...40R}), Liu et al.\ (\citeyear{2013MNRAS.430.2327L}, \citeyear{2013MNRAS.436.2576L}, \citeyear{2014MNRAS.442.1303L}), Harrison et al.\ (\citeyear{2014MNRAS..441...3306H}), McElroy et al.\ (\citeyear{2015MNRAS.446.2186M}), and Sun et al.\ (\citeyear{2017ApJ...835..222S}) among others argue that such outflows are common. Others, including Husemann et al.\ (\citeyear{2013A&A...549A..43H}), Karouzos et al.\ (\citeyear{2016ApJ...819..148K}), Bae et al.\ (\citeyear{2017ApJ...837...91B}), Fischer et al.\ (\citeyear{2018ApJ...856..102F}), and Rose et al.\ (\citeyear{2018MNRAS.474..128R}), suggest that large-scale outflows are not prevalent or very weak. Discrepancies between these surveys are likely due in part to sample selection. Regardless, the overall impact of quasar-driven outflows on galaxy evolution is still a subject of intense research. There is a large amount of theoretical work suggesting that these outflows have a negative effect on the star formation activity in galaxies, stripping quasar hosts of some of the gas that would have otherwise been used to create new stars (e.g., Di Matteo et al.\ \citeyear{2005Natur.433..604D}; Zubovas \& King \citeyear{2012ApJL...745L..34Z}; Hopkins et al.\ \citeyear{2016MNRAS.458..816H}; Pontzen et al.\ \citeyear{2017MNRAS.465..547P}). Observational evidence for such a relationship between quasar feedback and star formation rate has been largely indirect (e.g., Carniani et al.\ \citeyear{2016A&A...591A..28C}, Wylezalek \& Zakamska \citeyear{2016MNRAS..461...3724W}). In contrast, it has been recently reported that in some galaxies, star formation occurs within the outflows themselves (e.g., Maiolino et al.\ \citeyear{2017Nature..544...202M}; Gallagher et al.\ \citeyear{2019MNRAS.tmp..556G}).

A complicating factor in these studies is the existence of (circum-)nuclear starbursts in many quasar hosts (e.g., Mullaney et al.\ \citeyear{2012ApJ...753L..30M}; Leslie et al.\ \citeyear{2016MNRAS.455L..82L}; Aird et al.\ \citeyear{2018arXiv181004683A}). The conditions to power quasars -- large concentrations of gas at the center of galaxies -- are also well suited to trigger new star formation. Stellar winds and supernovae from these starbursts often work in unison with quasars to drive the most powerful outflows (e.g., Veilleux et al.\ \citeyear{2013ApJ...776...27V}; Cicone et al.\ \citeyear{2014A&A...562A..21C}; Gonz\'{a}lez-Alfonso et al.\ \citeyear{2017ApJ...836...11G}; Fluetsch et al.\ \citeyear{2019MNRAS.483.4586F}). These starbursts may also contribute to the heating and ionization of the outflowing material. This will be reflected in their characteristic emission line spectra. It is therefore important to study {\em separately} the emission-line properties of the outflowing material and the host galaxies to gain a full understanding of the roles of the quasars and starbursts in these outflows. 

Line ratio diagnostic diagrams, introduced by Baldwin et al.\ (\citeyear{1981PASP...93....5B}; hereafter \citetalias{1981PASP...93....5B}) and revised by Veilleux \& Osterbrock (\citeyear{1987ApJS..63...295V}; hereafter \citetalias{1987ApJS..63...295V}), are typically used to study the line-emitting material in galaxies. These diagrams serve multiple purposes. First, they help reveal the presence of active galactic nuclei (AGN) or starbursts at the centers of galaxies. Second, they help quantify the importance of the various ionization processes at play in these objects: photoionization by an AGN, photoionization by the hot, newly formed OB stars of a starburst, or shock heating and ionization (e.g., Kauffmann et al.\ \citeyear{2003MNRAS..346...1055K}; Kewley et al.\ \citeyear{2006MNRAS..372...961K}; Rich et al.\ \citeyear{2011ApJ..734...87R}, \citeyear{2014ApJ...781L..12R}, \citeyear{2015ApJS..221...28R}; Davies et al.\ \citeyear{2016MNRAS.462.1616D}). Lastly, these line ratio diagnostic diagrams may be used to constrain key physical parameters of the line-emitting gas (e.g., density, metallicity), when the data are compared with the predictions of theoretical models (e.g., Dopita \& Sutherland \citeyear{1995ApJ...455..468D}; Groves et al.\ \citeyear{2004ApJS..153...75G}; Dopita et al.\ \citeyear{2006ApJS..167..177D}; Allen et al.\ \citeyear{2008ApJS..178...20A}). 

Multiple ionization mechanisms are often at work in nearby AGN (e.g., Davies et al.\ \citeyear{2014MNRAS.444.3961D}, \citeyear{2016MNRAS.462.1616D}). Contributions from both stellar photoionization and AGN photoionization produce a ``mixing sequence'' in the line ratio diagrams. In these cases, the derived AGN contribution often decreases with distance from the nucleus (e.g., Davies et al.\ \citeyear{2014MNRAS.444.3961D}). Shocks may also contribute to the ionization of this gas, further complicating the analysis (e.g., D'Agostino et al. \citeyear{2018MNRAS.479.4907D}, \citeyear{2019MNRAS.tmpL..33D} and references therein). 

This is where the strategy to combine the information derived from these line ratio diagrams with the kinematic information becomes very useful (e.g., Veilleux et al.\ \citeyear{1995ApJS...98..171V}; Allen et al.\ \citeyear{1999ApJ...511..686A}; Monreal-Ibero et al.\ \citeyear{2006ApJ...637..138M}; Sharp \& Bland-Hawthorn \citeyear{2010ApJ...711..818S}; Rich et al.\ \citeyear{2011ApJ..734...87R}, \citeyear{2014ApJ...781L..12R}, \citeyear{2015ApJS..221...28R}, and references therein). This strategy is currently going through a revival thanks to the plethora of new high-quality imaging spectroscopic data obtained with the new generation of integral-field spectrographs such as WiFeS (Dopita et al.\ \citeyear{2007Ap&SS.310..255D}), SAMI (Croom et al.\ \citeyear{2012MNRAS.421..872C}), MUSE (Bacon et al.\ \citeyear{2010SPIE.7735E..08B}), and MaNGA (Bundy et al.\ \citeyear{2015ApJ...798....7B}). 

This strategy is particularly helpful for objects with line ratios residing in the portion of these diagnostic diagrams populated by the Low-Ionization Nuclear Emission-line Regions (LINERs\footnote{In this paper, we use the term LINER loosely to also include cases where the line emission is outside the nuclear regions}; Heckman \citeyear{1980A&A....87..152H}), where multiple different physical processes may be active. These include ionization by shocks, AGN photoionization with a low ionization parameter (e.g., Groves et al.\ \citeyear{2004ApJS..153...75G}; Kewley et al.\ \citeyear{2006MNRAS..372...961K}), or ionization by an older stellar population (e.g. Sarzi et al.\ \citeyear{2005ApJ...628..169S}). Shocks are favored when a correlation between line ratios and kinematics is observed. 

In this paper, we make use of Gemini integral field spectroscopy (IFS) data of nearby ($z <$ 0.2) radio-quiet Type 1 quasars presented in \citetalias{2017ApJ..850...40R}. Using methods described in \citetalias{2017ApJ..850...40R}, we extract the outflow from the host galaxy and study their properties separately. By constraining the ionization mechanisms in the host galaxy and outflowing components separately, we gain valuable insights into the environment of the quasar and its outflow, and how the two components interact.

The paper is organized as follows. Section 2 describes the sample of \citetalias{2017ApJ..850...40R} and explains how the present set of objects was chosen. Section 3 summarizes how the data were obtained and how they were reduced. In Section 4, we present the results for each of our four objects individually. Section 5 compares the data with theoretical models and discusses the results of these comparisons. Section 6 summarizes the main results from this analysis. 

\begin{deluxetable*}{cccccccc}
\tablecaption{Properties of the Galaxies in the Sample\label{tab:sample}}
\tablewidth{0pt}
\tablehead{\normalfont{Object} &\normalfont{$z$} & \normalfont{log($L_{bol}/L_{\odot})$} & \normalfont{AGN Fraction} & \normalfont{log($M_{BH}/M_{\odot})$} & \normalfont{$R_{obs}$} & \normalfont{Avg. $v_{98}$} & \normalfont{log[$(dM/dt)]$}}
\decimalcolnumbers
\startdata
F05189--2524 & 0.04288 & 12.22 & $0.71^{+0.21}_{-0.29}$ & 8.32 $\pm$ 0.5 & 2.8 & -926 & $0.40^{+0.07}_{-0.14}$\\
F13218+0552 & 0.2047 & 12.68 & $0.83^{+0.21}_{-0.17}$ & 8.55 $\pm$ 0.5 & 11.5 & -714 & $1.03^{+0.04}_{-0.09}$\\
F13342+3932 & 0.1797 & 12.49 & $0.69^{+0.23}_{-0.24}$ & 9.12 $\pm$ 0.5 & 7.9 & -473 & $1.37^{+0.03}_{-0.04}$\\
PG1613+658 & 0.12925 & 12.29 & $0.82^{+0.11}_{-0.09}$ &  $8.34^{+0.47}_{-0.52}$ & 7.5 & -457 & $0.04^{+0.16}_{-0.36}$\\
\enddata
\tablecomments{Data reproduced from \citetalias{2017ApJ..850...40R}. Column 2: Redshift measured from the data in \citetalias{2017ApJ..850...40R}. Column 3: Galaxy bolometric luminosity (Veilleux et al.\ \citeyear{2009ApJS..182..628V}). Column 4: Fraction of the bolometric luminosity due to an AGN (Veilleux et al.\ \citeyear{2009ApJS..182..628V}). Error bars encompass the full range of values derived from six mid-infrared measurements. Column 5: Black hole mass from reverberation mapping (Bentz \& Katz \citeyear{2015PASP..127...67B}) or {\em HST} photometric measurements (Veilleux et al.\ \citeyear{2009ApJ...701..587V}). Errors from reverberation mapping are the uncertainty in the virial coefficient $f$ summed in quadrature ($\delta f = 0.44$, Woo et al.\ \citeyear{2010ApJ...716..269W}). Errors for photometric measurements are from the scatter in the $M_{BH}-L_{H}$ relationship, from which the masses were derived (Marconi \& Hunt \citeyear{2003ApJ...589L..21M}). Column 6: Maximum observed wind radius. Column 7: Area-weighted average value of $v_{98}$ in component 2. Column 8: Logarithm of wind mass outflow rate. This value is computed using a time-averaged thin-shell model that is inversely dependent on the shell radius (\citetalias{2017ApJ..850...40R}, Rupke et al.\ \citeyear{2005ApJS..160..115R}, Shih \& Rupke \citeyear{2010ApJ...724.1430S}, Rupke \& Veilleux \citeyear{2013ApJ...768...75R}). This quantity depends on the electron density as $n_e^{-1}$, which is calculated for each spaxel based on the [\ion{S}{2}] $\lambda$6716/[\ion{S}{2}] $\lambda$6731 line ratio.} 
\end{deluxetable*}

\section{Sample}\label{sec:Original Sample}

The original sample of quasars in \citetalias{2017ApJ..850...40R} was selected from the Quasar and ULIRG  Evolution Study (QUEST; \citetalias{2017ApJ..850...40R} and references therein). This sample consists of ten nearby ($z <$ 0.2) radio-quiet Type 1 quasars. These objects were chosen based on observability, proximity, and diversity. In this paper, we select four of these quasars to analyze in greater detail. Our two selection criteria were (1) the detection of the seven emission lines used in our line ratio analysis and (2) the successful extraction of separate components from the IFS cubes corresponding to the host galaxy and outflow. The seven strong emission lines are: H$\alpha$, H$\beta$, [\ion{O}{3}] $\lambda$5007, [\ion{N}{2}] $\lambda$6583, [\ion{S}{2}] $\lambda\lambda$6716, 6731,  and [\ion{O}{1}] $\lambda$6300. 

In PG~1700+518, F21219$-$1757, and F07599+6508, the archival data did not cover the requisite emission lines. PG~1411+442 and I~Zw~1 have a neutral outflow, but no ionized outflow was present. For Mrk~231, data cubes for different wavelength ranges were taken at different times and could not be easily merged. Moreover, the ionized outflow in this object is insignificant in comparison with the neutral gas outflow. Table \ref{tab:sample} summarizes the properties of the four objects studied in this paper. 

\begin{deluxetable*}{cccccccc}
\tablecaption{Observation Details of Current Sample\label{tab:obs}}
\tablewidth{0pt}
\tablehead{\normalfont{Object} &\normalfont{PID} & \normalfont{Dates} & \normalfont{$t_{exp}$ ($\times$ 1800 s)} & \normalfont{PSF ('')} & \normalfont{Range (\AA)} & \normalfont{PA ($^{\circ}$)} & \normalfont{FOV}}
\decimalcolnumbers
\startdata
F05189--2524 & GS-2011B-Q-64 & 2011 Dec 03, 13, 31 & 6 & 0.6 & 4560--7430 & 0 & 5.2'' $\times$ 5.0''\\
F13218+0552 & GN-2012A-Q-15 & 2012 Apr 02, 08 & 8 & 0.8 & 5340--8230 & 315 & 3.9'' $\times$ 4.5''\\
F13342+3932 & GN-2012A-Q-15 & 2012 May 17 & 8 & 0.6 & 5340--8230 & 313 & 4.2'' $\times$
5.1''\\
PG1613+658 & GN-2012A-Q-15 & 2012 Apr 22, 24 & 6 & 0.9 & 5340--8220 & 135 & 5.1'' $\times$ 7.8''\\
\enddata
\tablecomments{Data reproduced from \citetalias{2017ApJ..850...40R}. All observations used the B600 grating, which has a spectral resolution of R = 1688. Column 2: Program ID. Column 3: UT dates of observations. Column 4: Number of exposures at 1800 seconds each. Column 5: FWHM of quasar PSF, except for F05189--2524, which is an estimate of the seeing from observing logs. Column 6: Wavelength range of observation for combined data cube. Column 7: East of north position angle of IFU. Column 8: size of combined FOV.} 
\end{deluxetable*}

Note that these selection criteria preferentially select quasars where both the host and outflow are bright and spatially well resolved. This favors nearby objects. Moreover, our current sample does not include any of the three objects in \citetalias{2017ApJ..850...40R} where archival data of slightly poorer quality were used for the analysis. Finally, our requirement to include all the strong emission lines potentially eliminates highly obscured objects for which weaker lines, i.e.\ those other than [\ion{O}{3}] $\lambda$5007 and H$\alpha$, have a low signal-to-noise (S/N) ratio. These selection criteria thus prevent broad conclusions from being drawn from this small sample of objects. 

\section{Emission Line Fitting}\label{Emission Line Fitting}

Each object was observed with the integral field unit (IFU) of the Gemini Multi-Object Spectrograph (GMOS; Allington-Smith et al.\ \citeyear{2002PASP..114..892A}, Hook et al.\ \citeyear{2004PASP..116..425H}) on the Gemini North and South telescopes. Table \ref{tab:obs} details the observations of the four objects studied in this paper. All observations used the one-slit mode, and were dithered to better sample the point spread function and increase the field of view (FOV). For the four objects in this paper, the FOV was centered on the quasar. The data were reduced using the Gemini IRAF and IFUDR GMOS packages as well as IFSRED (Rupke \citeyear{ascl:1409.004}). For accurate emission-line fitting, the scattered light was removed before flat field correction. In this paper, we use the continuum and line fits from \citetalias{2017ApJ..850...40R}.

The separation of light from the quasar and host galaxy is detailed in \citetalias{2017ApJ..850...40R} (and references therein). This process is vital since the light from the quasar overwhelms that of the host in most objects (see \S 2.3.1 in \citetalias{2017ApJ..850...40R} for more details). 

For each object, the spatially unresolved line emission from the narrow line region (NLR) was removed as part of the scaled quasar spectrum, while the line profiles from the spatially resolved emission  were modeled with two Gaussian velocity components at each spaxel. These profiles were convolved with the spectral resolution before fitting. For each spaxel, all emission lines were fixed to the same velocity in each component and kept a component only if it exceeded a 3-$\sigma$ threshold in at least one strong line. A 2-$\sigma$ cut was applied to every emission line.

The two velocity components were sorted into two maps by velocity dispersion. The first component (c1) had a smaller velocity dispersion than the second component (c2), with select spaxels being reassigned by hand. If present, outflows resided in c2. Because this method does not assign c1 or c2 {\em a priori}, but rather is based on what makes the best galaxy-rotation model, there is higher confidence that c2 contains an outflow. In the case of I Zw 1, there is a c2 component that has very small velocities and velocity gradients inconsistent with an outflow.

For every spaxel, and each velocity component of each galaxy, the line flux, velocity dispersion $\sigma$, and velocity fields as a function of distance from the galaxy nucleus were calculated. The velocity fields were derived using the 50-, 84-, and 98-percentile of the velocity distributions. These values, called $v_{50}$, $v_{84}$, and $v_{98}$, are calculated in blueshifted (redshifted) spaxels by integrating from the red (blue) side of the velocity distribution until reaching 50, 84, or 98\% of the Gaussian distribution, respectively. For a Gaussian distribution $v_{84} = v_{50} + \sigma$ and $v_{98} = v_{50} + 2\sigma$. We use $v_{98}$ as a representative maximum velocity in the outflow. Each of these velocities as well as the velocity dispersion were fixed between the various emission lines within each component.

From the fluxes of the strong emission lines, logarithmic line ratios are calculated. These line ratios were not corrected for reddening because they produced noisy flux maps. For [\ion{N}{2}]/H$\alpha$, artificial limits were placed at $-$1.0 and 0.6, with values outside of this range being pegged to the appropriate limit. This was done because of the potential degeneracy between [\ion{N}{2}] and H$\alpha$, which is especially problematic with broad-line components. The upper limit is the apparent maximum seen in Kewley et al.\ (\citeyear{2006MNRAS..372...961K}) and Rich et al.\ (\citeyear{2014ApJ...781L..12R}). The lower limit is appropriate for ultra-luminous infrared galaxies (ULIRGs), where the metallicity is not low enough to produce line ratios below this value. As is common, the [\ion{S}{2}]/H$\alpha$ ratios presented here use the sum of the fluxes from [\ion{S}{2}] $\lambda$6716 and $\lambda$6731.

\section{Results}\label{sec:results}

\begin{figure*}[ht!]
\gridline{\includegraphics[height=0.3\linewidth,trim=3.0cm 0.25cm 4.25cm 1cm,clip]{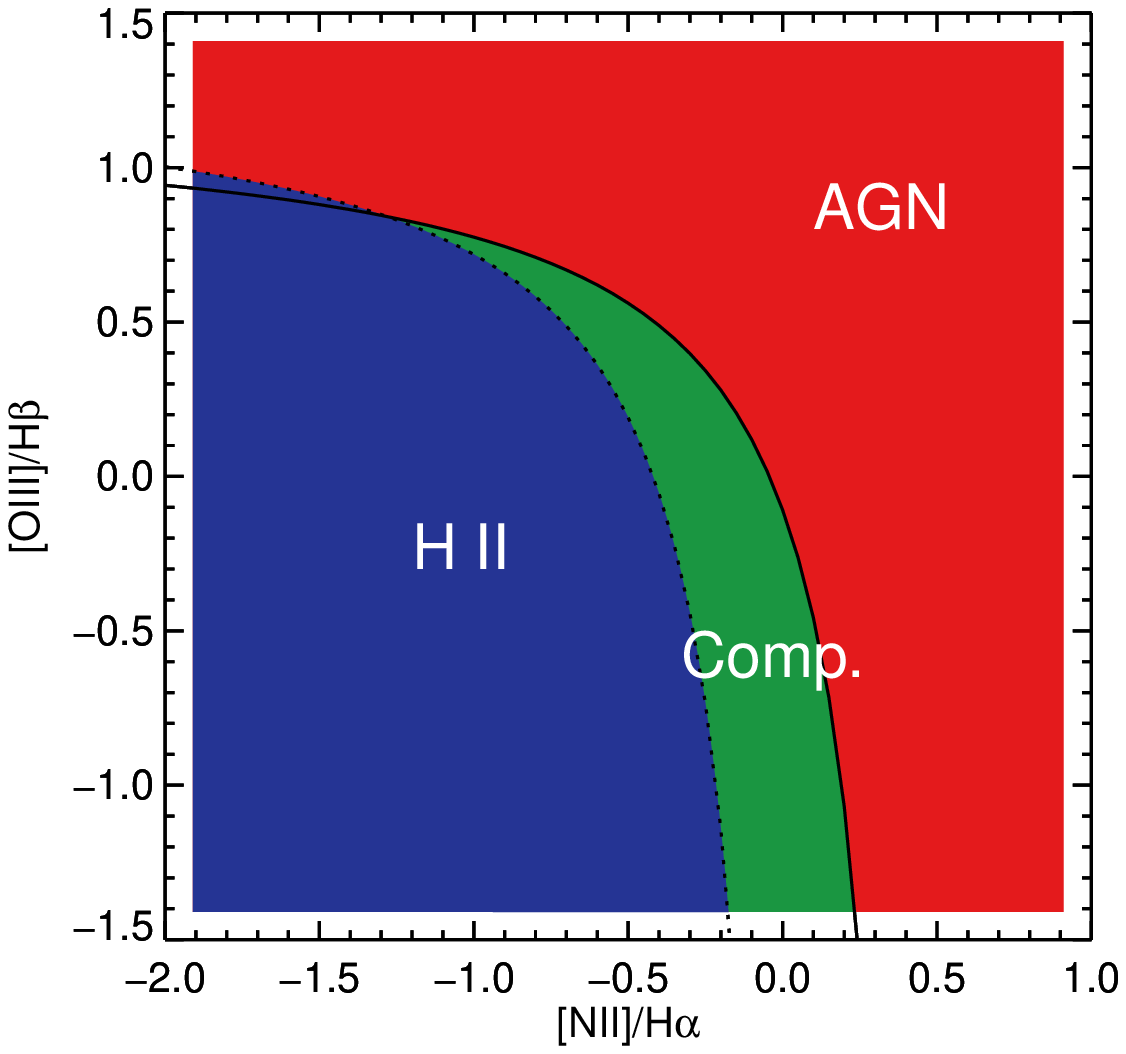}
		  \includegraphics[height=0.3\linewidth,trim=3.0cm 0.25cm 4.25cm 1cm,clip]{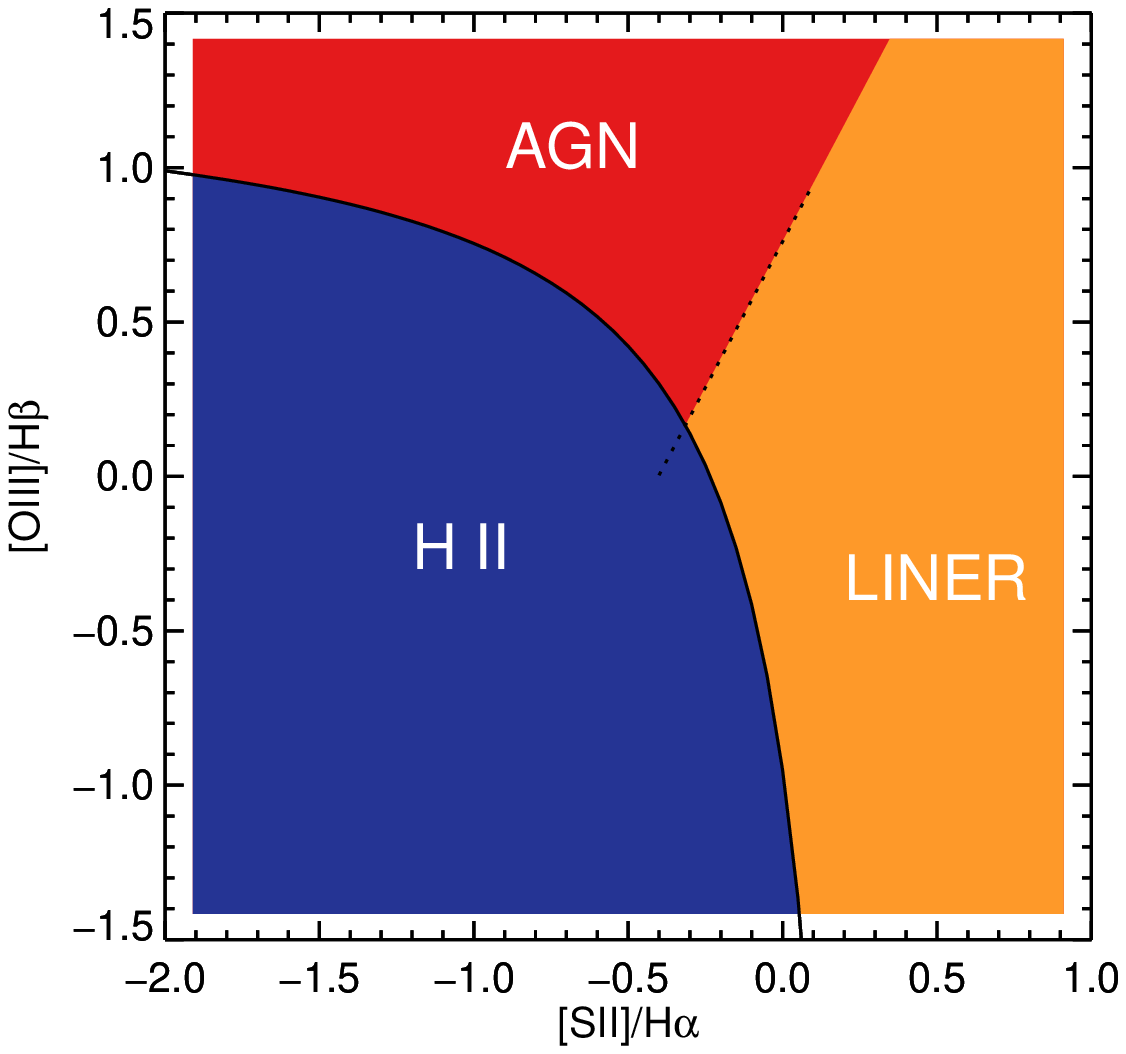}
          \includegraphics[height=0.3\linewidth,trim=3.0cm 0.25cm 4.25cm 1cm,clip]{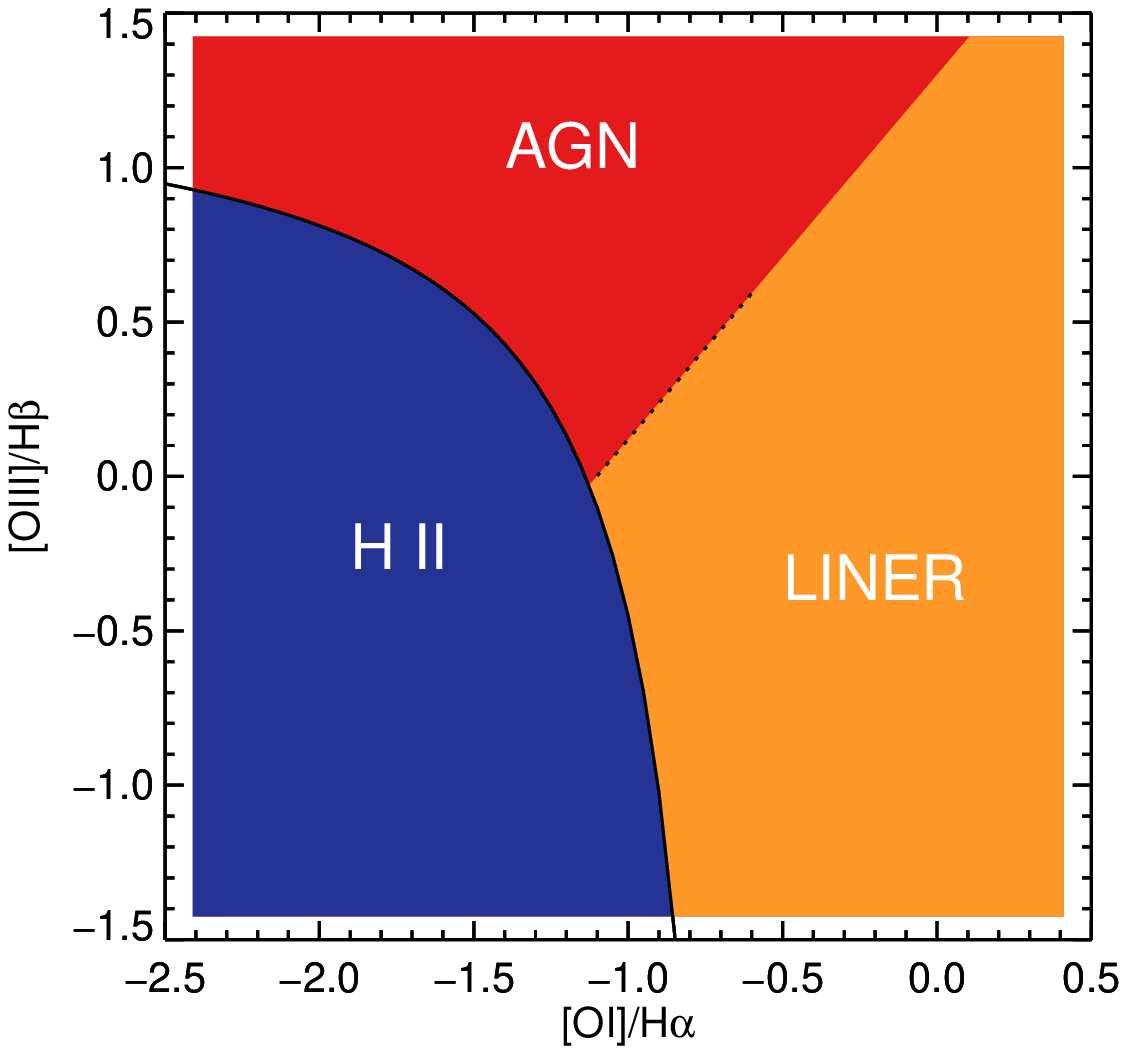}}
\caption{Colors and classifications used later in Figures \ref{fig:f05189map}, \ref{fig:f13218map}, \ref{fig:f13342map}, and \ref{fig:pg1613map}. The three panels and classification lines are the same as in Figure \ref{fig:f05189sdss}\label{fig:mapkey}}
\end{figure*}

\begin{figure*}[ht!]
\gridline{\includegraphics[height=0.3\linewidth,trim=3.5cm 0.25cm 4.25cm 1cm,clip]{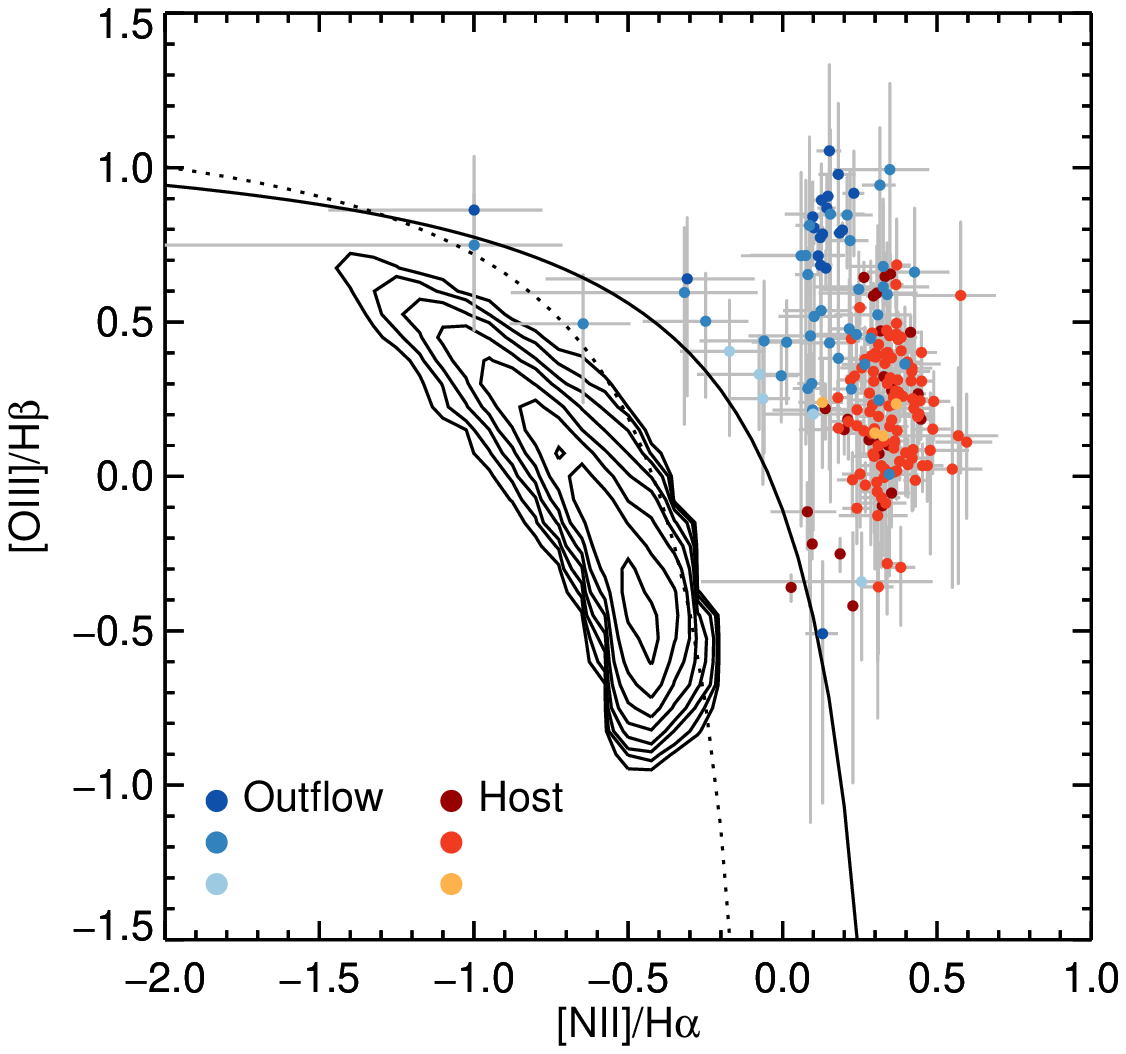}
		  \includegraphics[height=0.3\linewidth,trim=3.5cm 0.25cm 4.25cm 1cm,clip]{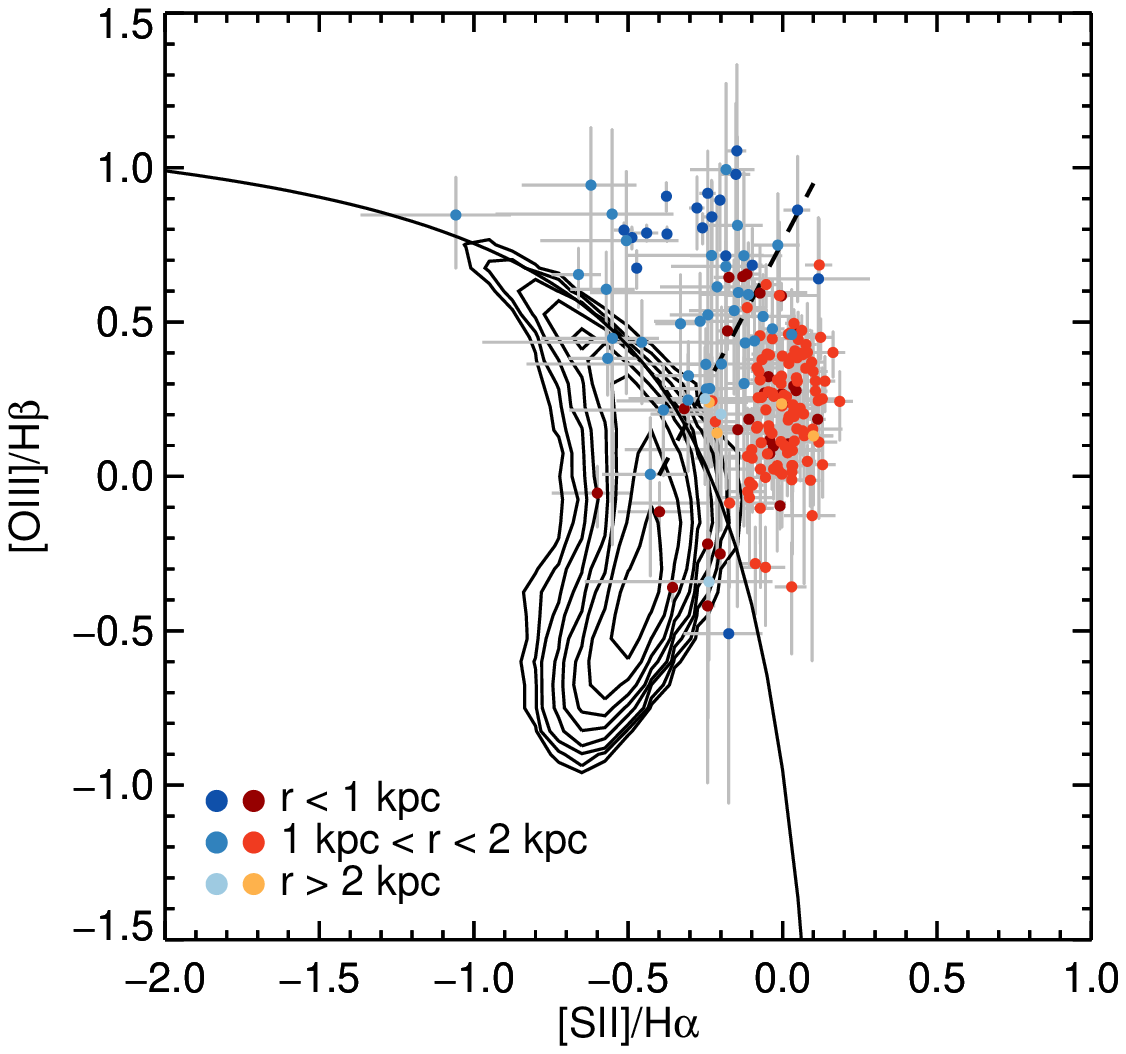}
          \includegraphics[height=0.3\linewidth,trim=3.5cm 0.25cm 4.25cm 1cm,clip]{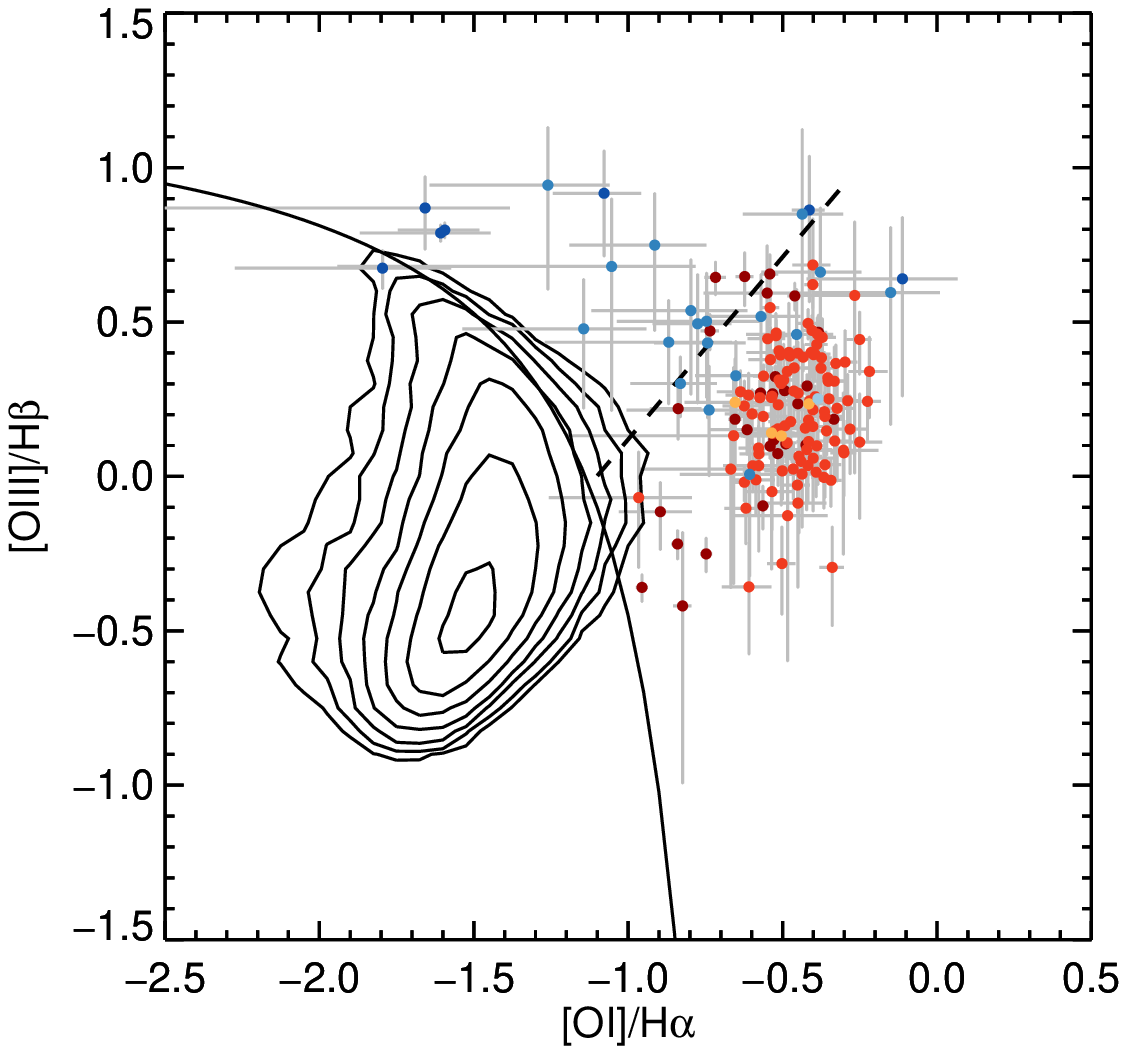}}
\caption{Line ratio diagrams for F05189$-$2524. The three panels are log~[\ion{O}{3}] $\lambda$5007/H$\beta$ vs.\ log~[\ion{N}{2}] $\lambda$6583/H$\alpha$ (Left), log~[\ion{S}{2}] $\lambda\lambda$6716,  6731/H$\alpha$ (Middle), and log~[\ion{O}{1}] $\lambda$6300/H$\alpha$ (Right).  The line ratios in the host galaxy (c1) are in red and the outflow (c2) line ratios are in blue. The symbols are color-coded by radius from the center of the galaxy. In all panels, the solid line is the theoretical line separating AGN (above right) and \ion{H}{2}-regions (below left) from Kewley et al.\ (\citeyear{2001ApJ..556...121K}). In the left panel, the dotted line is the empirical line from Kauffmann et al.\ (\citeyear{2003MNRAS..346...1055K}) showing the same separation. Objects between the dotted and solid lines are classified as composites. In the middle and right panels, the diagonal dashed line is the theoretical line separating Seyferts (above left) and LINERs (below right) from Kewley et al.\ \citeyear{2006MNRAS..372...961K}. Also shown in all three panels are density contours created from the SDSS spectra of starburst galaxies. Each contour is separated by a factor of two.\label{fig:f05189sdss}}
\end{figure*}

\begin{figure*}[htp!]
\centering
\includegraphics[width=0.75\linewidth,trim=1.0cm 0.4cm 0.1cm 0.1cm,clip]{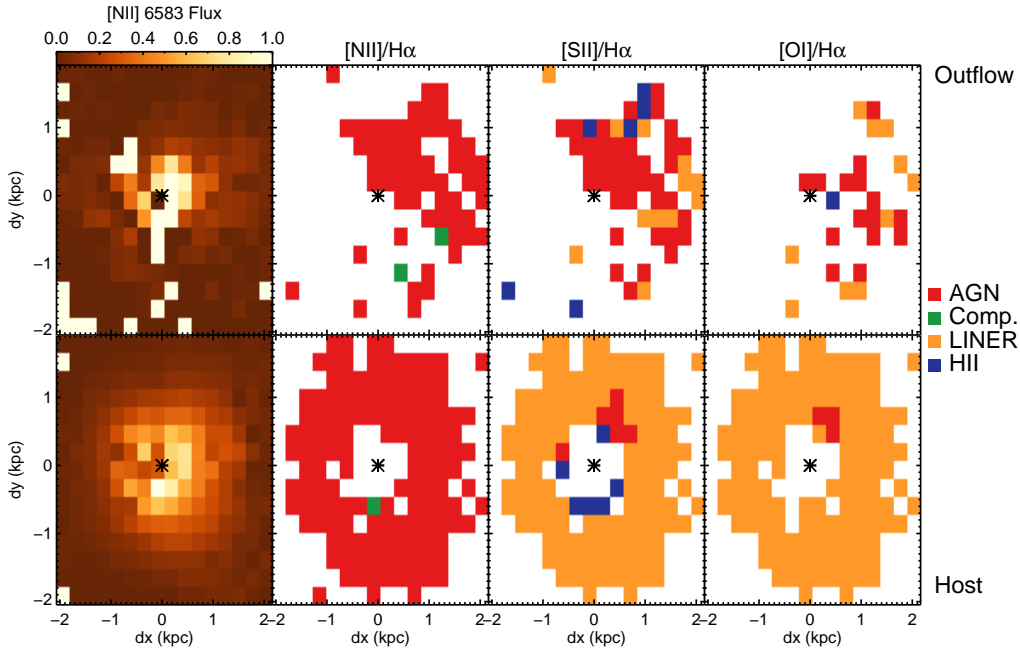}
\caption{[\ion{N}{2}] $\lambda$6583 flux maps (First Column) of the host galaxy and outflow of F05189$-$2524. The fluxes for both the host and outflow components are in units of 4.00 $\times$ 10$^{-15}$ erg s$^{-1}$ cm$^{-2}$ arcsec$^{-2}$. Maps of the dominant ionization processes in F05189$-$2524 based on (Second Column) the [\ion{O}{3}] $\lambda$5007/H$\beta$ versus [\ion{N}{2}] $\lambda$6583/H$\alpha$ diagram, (Third Column) the [\ion{O}{3}] $\lambda$5007/H$\beta$ versus [\ion{S}{2}]/H$\alpha$ diagram, and (Fourth Column) the [\ion{O}{3}]/H$\beta$ versus [\ion{O}{1}]/H$\alpha$ diagram. The top panels show the outflow component while the bottom panels show the host galaxy component. White means that there is no data for these particular spaxels. The other colors correspond to the spectral types shown in Figure \ref{fig:mapkey}. Red indicates AGN-like line ratios, blue indicates \ion{H}{2} region-like ratios, green indicates composite, and orange indicates LINER-like line ratios. The black asterisk at the center of each panel marks the nucleus of the galaxy.\label{fig:f05189map}}
\end{figure*}

\begin{figure}[ht!]
\centering
\bigskip
\includegraphics[width=1\linewidth,trim=1cm 0.3cm 12.5cm 1.0cm,clip]{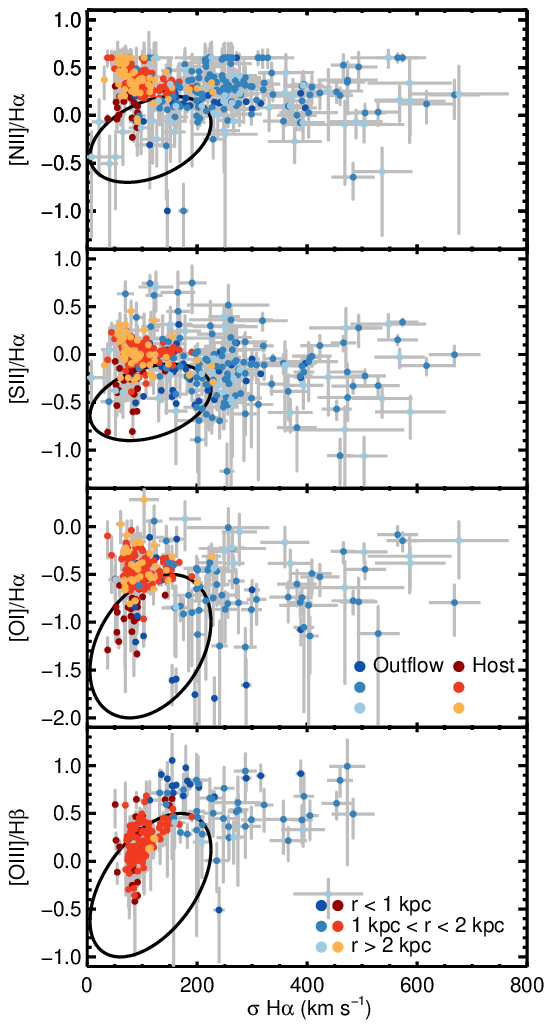}
\caption{Line ratios versus H$\alpha$ velocity dispersions for F05189$-$2524. The panels show (in order from top to bottom) log~[\ion{N}{2}] $\lambda$6583/H$\alpha$, [\ion{S}{2}] $\lambda\lambda$6716, 6731/H$\alpha$, log~[\ion{O}{1}] $\lambda$6300/H$\alpha$, log~[\ion{O}{3}] $\lambda$5007/H$\beta$ vs.\ $\sigma$(H$\alpha$). The black ellipses encompass $\sim$90\% of the points in the shock-dominated galaxies of Rich et al.\ \citeyear{2011ApJ..734...87R}, \citeyear{2014ApJ...781L..12R}, \citeyear{2015ApJS..221...28R}. The host galaxy (c1) data points are in red and the outflow (c2) data points are in blue. The symbols are color-coded by distance from the center of the galaxy, following Figure \ref{fig:f05189sdss}.\label{fig:f05189rich}}
\end{figure}

In this section, we present three sets of plots for each of the four objects in our sample to provide a first assessment of the ionization mechanism(s) at play in the host and outflow of each object. These data are compared with theoretical models in \S 5. 

The first figure for each object is comprised of the three two-dimensional line ratio diagrams of \citetalias{1987ApJS..63...295V}, i.e., log([\ion{O}{3}]/H$\beta$) vs.\ log([\ion{N}{2}]/H$\alpha$), log([\ion{O}{3}]/H$\beta$) vs.\ log([\ion{S}{2}]/H$\alpha$), and log([\ion{O}{3}]/H$\beta$) vs.\ log([\ion{O}{1}]/H$\alpha$), where the data on the host galaxy and outflow are presented simultaneously using different colors. In the following discussion, we refer to [\ion{N}{2}]/H$\alpha$, [\ion{S}{2}]/H$\alpha$, and [\ion{O}{1}]/H$\alpha$ as the low-ionization line ratios, while [\ion{O}{3}]/H$\beta$ is the high-ionization line ratio. In these diagrams, we also show number density contours of the star-forming galaxies in the line ratio space using spectra taken from the MPA-JHU (Max Planck Institute for Astrophysics and Johns Hopkins University) collaboration (Brinchmann et al.\ \citeyear{2004MNRAS.351.1151B}) from the Sloan Digital Sky Survey (SDSS) data release 8 (Gunn et al.\ \citeyear{2006AJ....131.2332G}, Aihara et al.\ \citeyear{2011ApJS..193...29A}, Eisenstein et al.\ \citeyear{2011AJ....142...72E}, Smee et al.\ \citeyear{2013AJ....146...32S}). From the entire SDSS DR8, we chose 105,506 star-forming galaxies with a S/N of the strong emission lines $\geq$ 3 and redshifts of $0.04 < z < 0.1$ (Kewley et al.\ \citeyear{2006MNRAS..372...961K}). These contours show the region of each line ratio diagram where ionization due to hot young stars is dominant. This information will help us explore the possibility of on-going star formation within the outflows and in the hosts. 

The second figure for each object is a map of the host galaxy and the outflow with appropriate classifications derived from each of the three two-dimensional line ratio diagrams. To create these maps, we used the boundaries from Kewley et al.\ (\citeyear{2001ApJ..556...121K}, \citeyear{2006MNRAS..372...961K}) and Kauffmann et al.\ (\citeyear{2003MNRAS..346...1055K}) to classify the line ratios measured at each individual spaxel into three classes. Each of the three line ratio diagrams has an AGN and an \ion{H}{2}-region classification. The diagram involving [\ion{N}{2}]/H$\alpha$ also has a composite classification that lies between AGN and \ion{H}{2}-region, while the diagrams involving [\ion{S}{2}]/H$\alpha$ and [\ion{O}{1}]/H$\alpha$ have a LINER classification characterized by strong low-ionization line ratios but weak [\ion{O}{3}]/H$\beta$ ratios. Figure \ref{fig:mapkey} shows the regions of line ratio space occupied by these classifications.

The third figure for each object compares four different line ratios with the H$\alpha$ velocity dispersions $\sigma$. Here we use data presented in Rich et al.\ (\citeyear{2011ApJ..734...87R}, \citeyear{2014ApJ...781L..12R}, \citeyear{2015ApJS..221...28R}) to draw ellipses encompassing $\sim$90\% of the Rich et al.\ shock-dominated galaxies in line ratio space. These ellipses illustrate the expected positive correlation between the line ratios and line widths if the line widths are a measure of shock velocity. In the figures of Rich et al., there are branches of points at high values of $\sigma$ that are likely associated with AGN. Therefore, when creating the ellipses for these plots, we focused on the large concentrations of data points with $\sigma$ $\la$ 300 km s$^{-1}$. 

\subsection{F05189--2524}\label{sec:f05189}

\begin{figure*}[ht!]
\gridline{\includegraphics[height=0.3\linewidth,trim=3.5cm 0.25cm 4.25cm 1cm,clip]{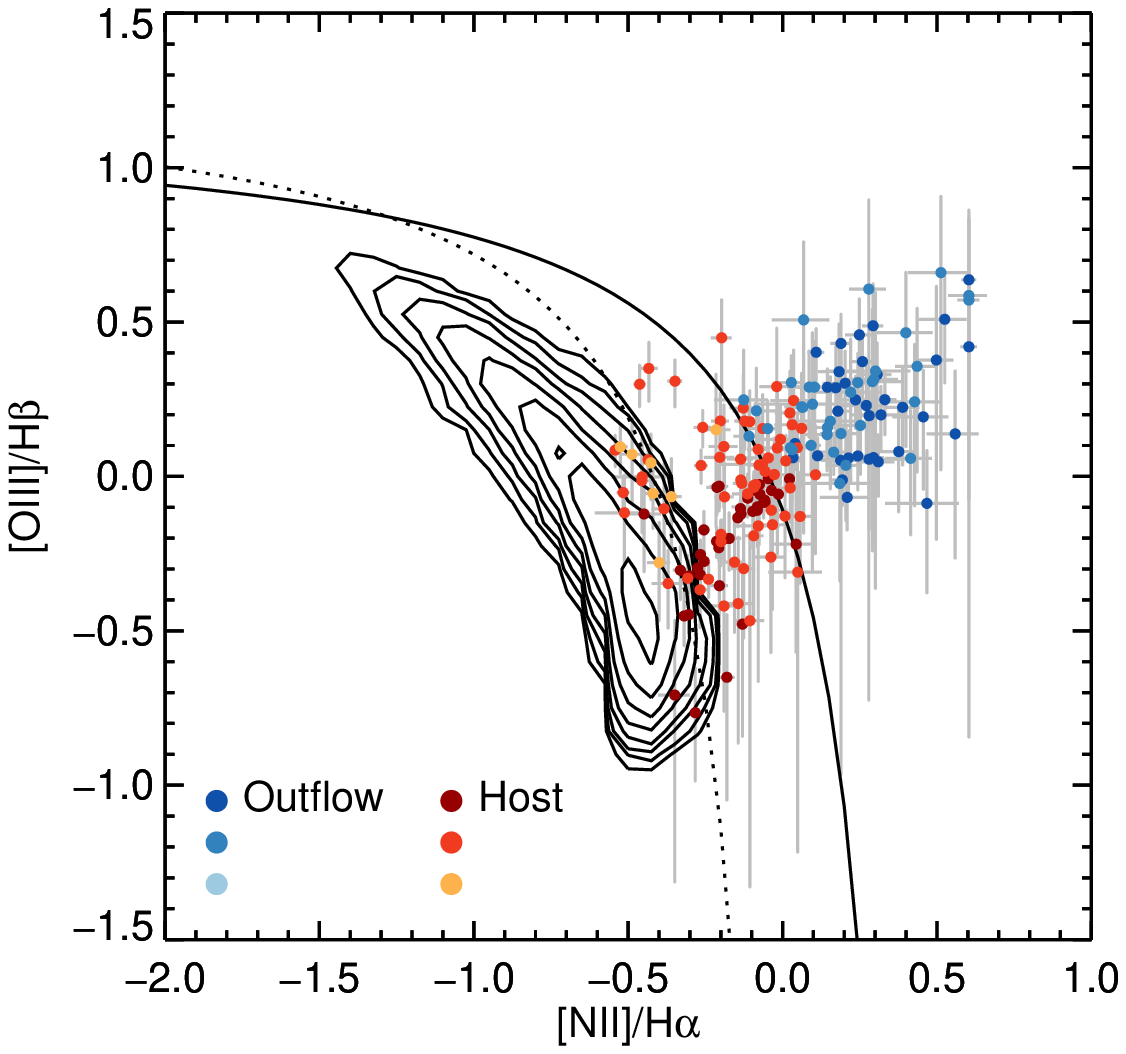}
		  \includegraphics[height=0.3\linewidth,trim=3.5cm 0.25cm 4.25cm 1cm,clip]{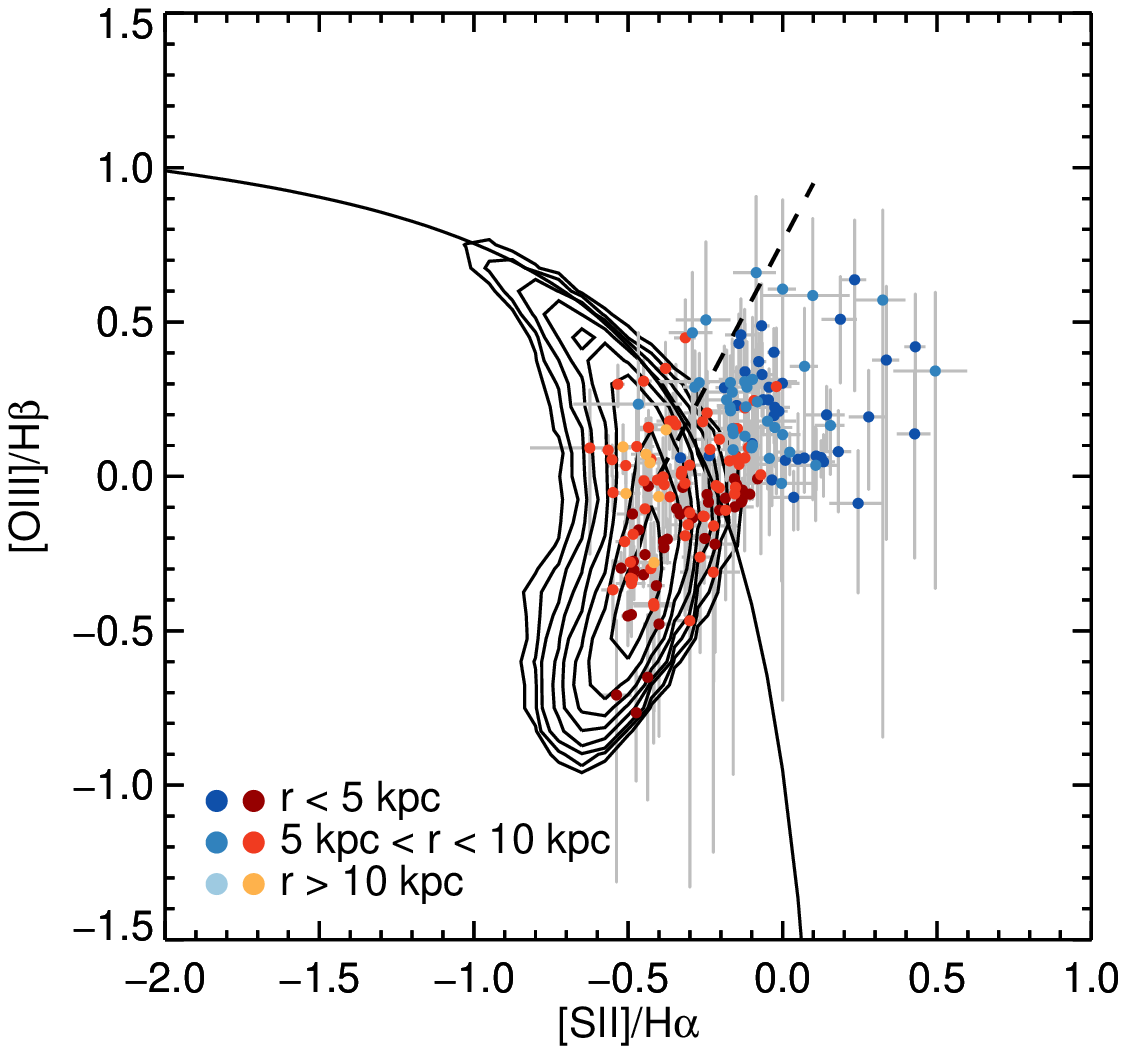}
          \includegraphics[height=0.3\linewidth,trim=3.5cm 0.25cm 4.25cm 1cm,clip]{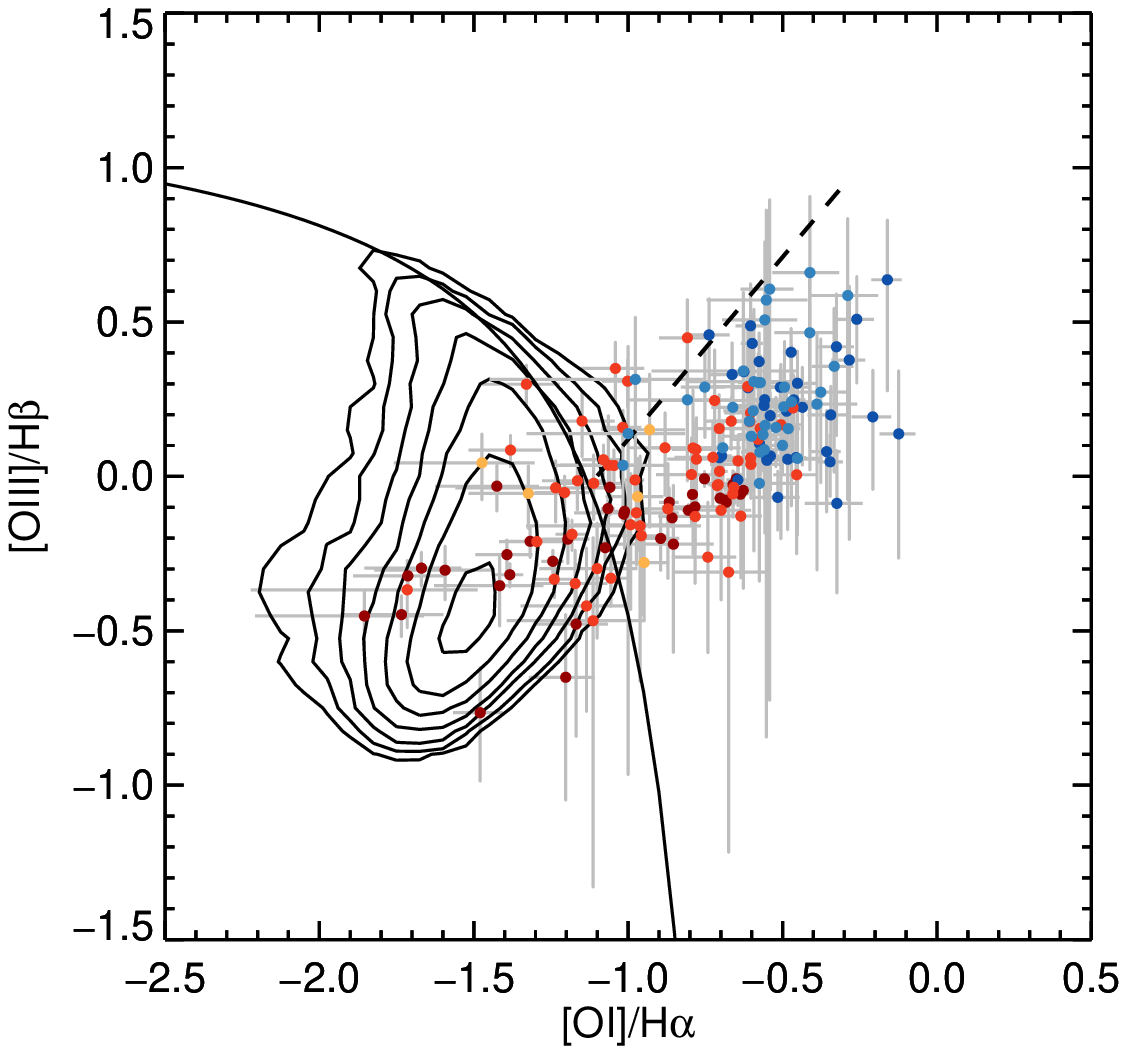}}
\caption{Same as Figure \ref{fig:f05189sdss} but for F13218+0552.\label{fig:f13218sdss}}
\end{figure*}

\begin{figure*}[htp!]
\centering
\includegraphics[width=0.75\linewidth,trim=0.75cm 0.4cm 0.1cm 0.1cm,clip]{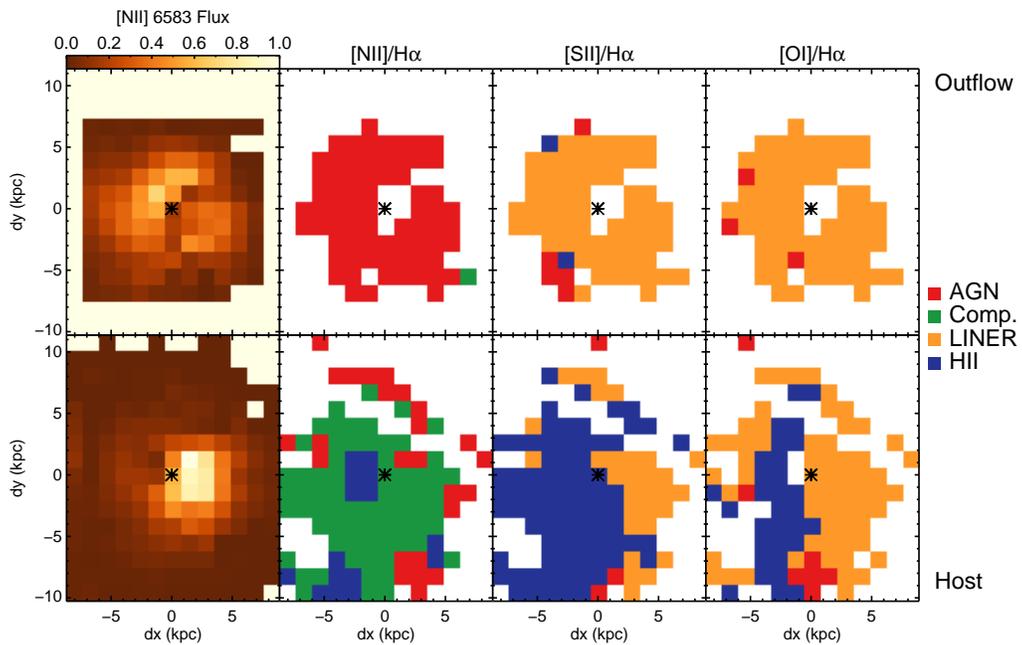}
\caption{Same as Figure \ref{fig:f05189map} but for F13218+0552 and here the fluxes for the host and outflow components are in units of 4.64 $\times$ 10$^{-15}$ and 3.32 $\times$ 10$^{-15}$ erg s$^{-1}$ cm$^{-2}$ arcsec$^{-2}$, respectively.}\label{fig:f13218map}
\end{figure*}

\begin{figure}[ht!]
\centering
\bigskip
\includegraphics[width=1\linewidth,trim=1cm 0.3cm 12.5cm 1.0cm,clip]{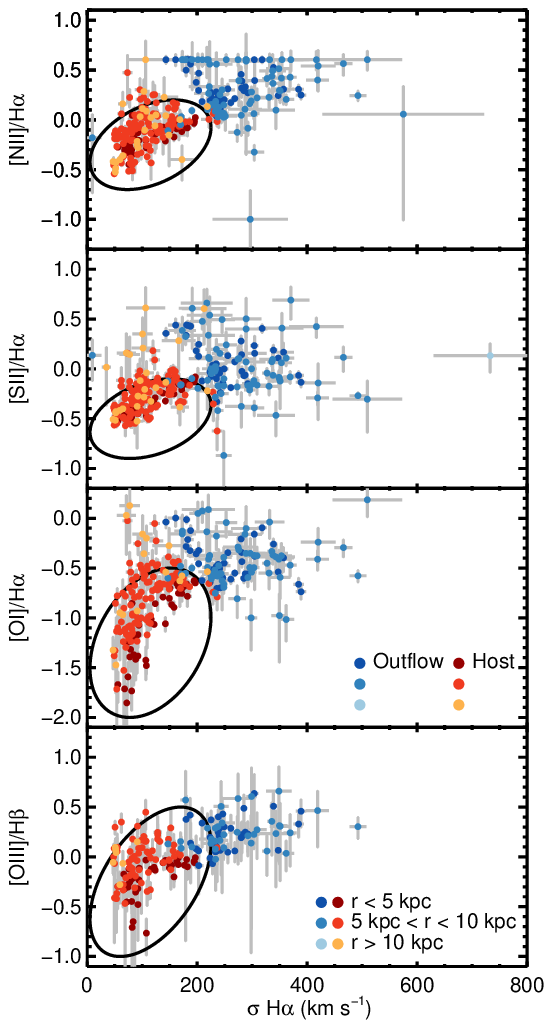}
\caption{Same as Figure \ref{fig:f05189rich} but for F13218+0552.\label{fig:f13218rich}}
\end{figure}

In Figure \ref{fig:f05189sdss}, we show the line ratio diagrams of the host galaxy (red filled circles) and outflow (blue filled circles) of F05189$-$2524. Besides a few data points in the [\ion{S}{2}]/H$\alpha$ diagram, we see very little overlap between the measured line ratios and the starburst galaxies from SDSS. These diagrams suggest that F05189$-$2524 possesses a LINER-like host galaxy and an outflow that is dominated by AGN photoionization. Using the color code shown in Figure \ref{fig:mapkey}, Figure \ref{fig:f05189map} confirms the different ionization conditions between the outflow and the host galaxy. While there is no obvious trend between line ratios and distance from the nucleus within the host component, the outflowing gas closest to the nucleus exhibits the highest values of [\ion{O}{3}]/H$\beta$. Note also that the ionized outflow in this object is largely one-sided, visible only north-west of the nucleus (\citetalias{2017ApJ..850...40R}), and aligned with the ionization cone traced by [\ion{O}{3}]/H$\beta$. Dust in the host galaxy likely obscures our view of the south-east outflow and ionization cone. 

In Figure \ref{fig:f05189rich}, we compare the observed line ratios with the line widths. No obvious trends are seen between these quantities, suggesting that shocks do not play a significant role in ionizing the material in this galaxy. In addition, there is no noticeable relationship between radial distance and velocity dispersion.

Of interest is the stark separation between the host and outflow components in the line ratio diagrams. These two components are almost perfectly separated by the ``mixing line'' (Kewley et al.\ \citeyear{2006MNRAS..372...961K}), which distinguishes Seyferts from LINERs. The lack of evidence for shock ionization in this galaxy from Figure \ref{fig:f05189rich} suggests that the LINER-like line ratios in the host are due to photoionization by a diluted AGN radiation field. We return to this point in \S 5.1.   

\subsection{F13218+0552}\label{sec:f13218}

\begin{figure*}[ht!]
\gridline{\includegraphics[height=0.3\linewidth,trim=3.5cm 0.25cm 4.25cm 1cm,clip]{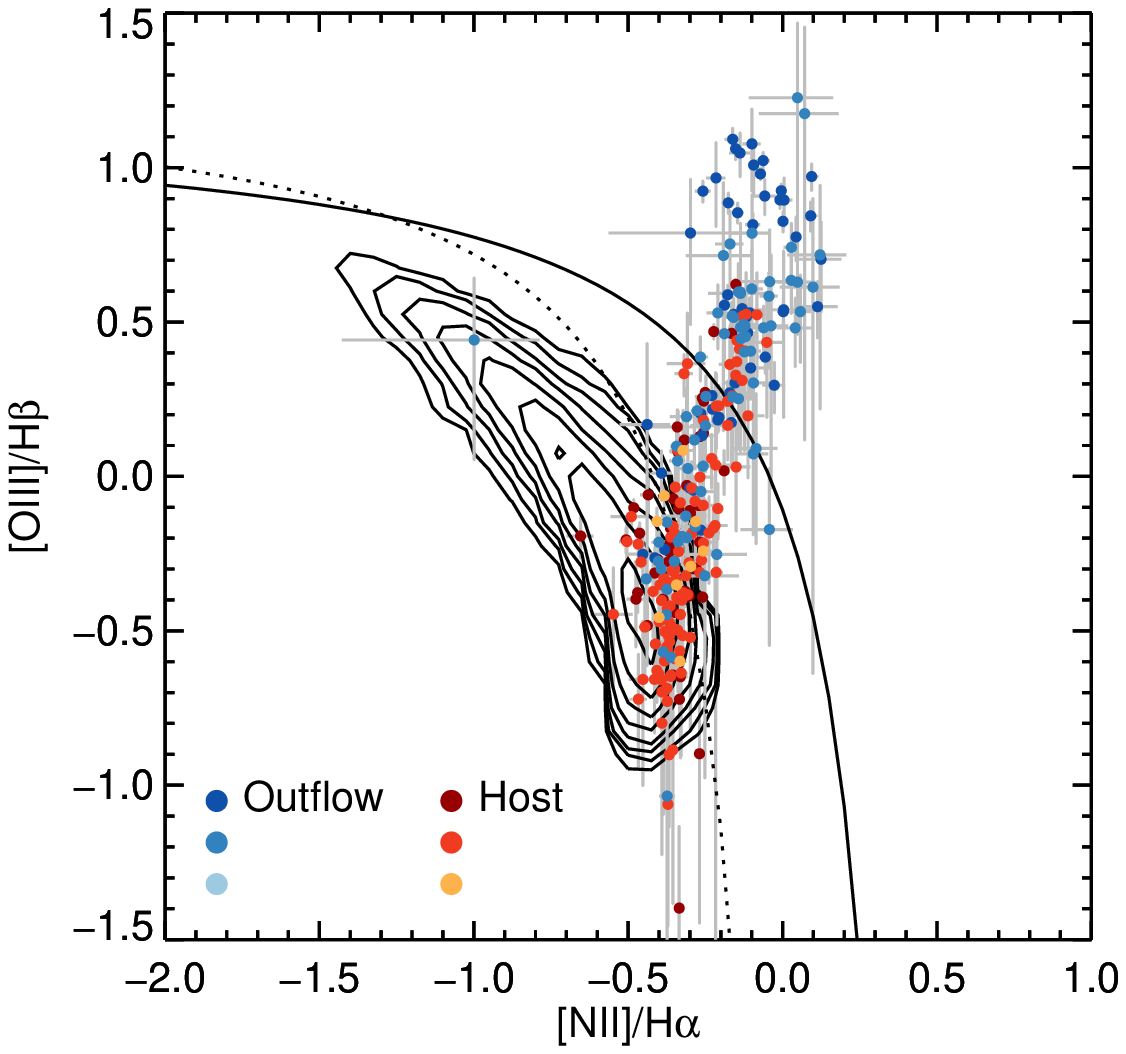}
		  \includegraphics[height=0.3\linewidth,trim=3.5cm 0.25cm 4.25cm 1cm,clip]{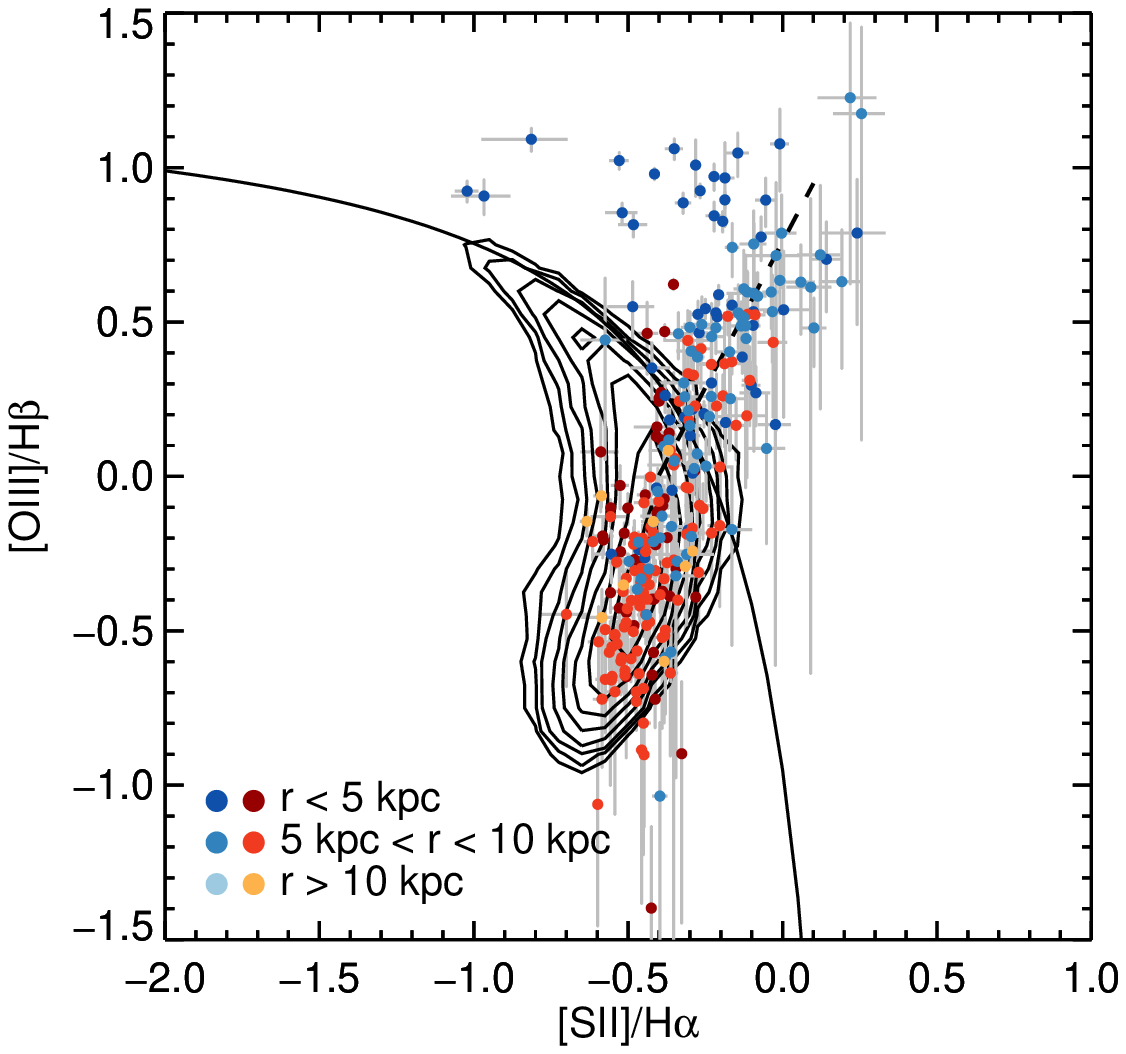}
          \includegraphics[height=0.3\linewidth,trim=3.5cm 0.25cm 4.25cm 1cm,clip]{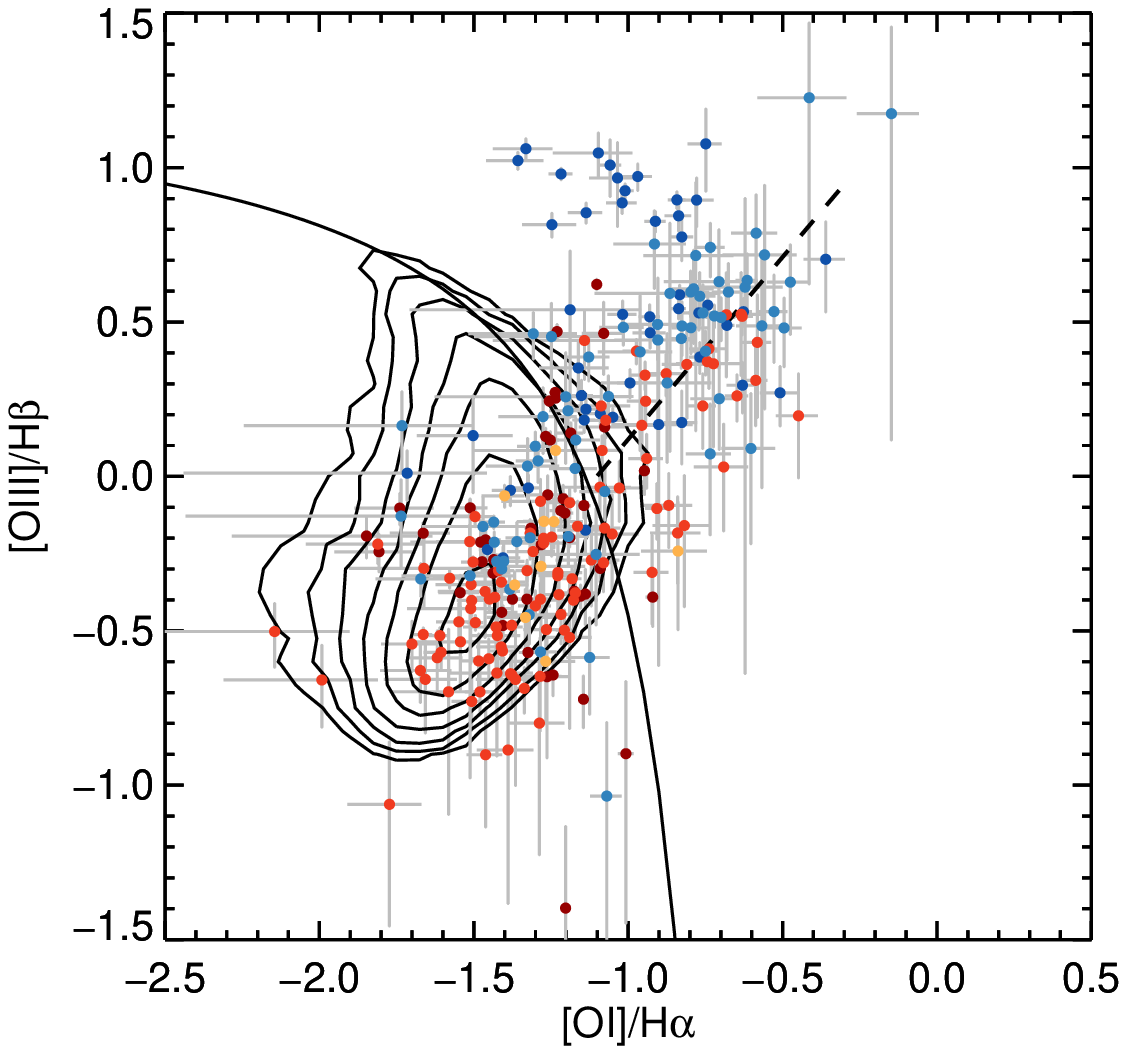}}
\caption{Same as Figure \ref{fig:f05189sdss} but for F13342+3932.\label{fig:f13342sdss}}
\end{figure*}

\begin{figure*}[htp!]
\centering
\includegraphics[width=0.75\linewidth,trim=0.75cm 0.4cm 0.1cm 0.1cm,clip]{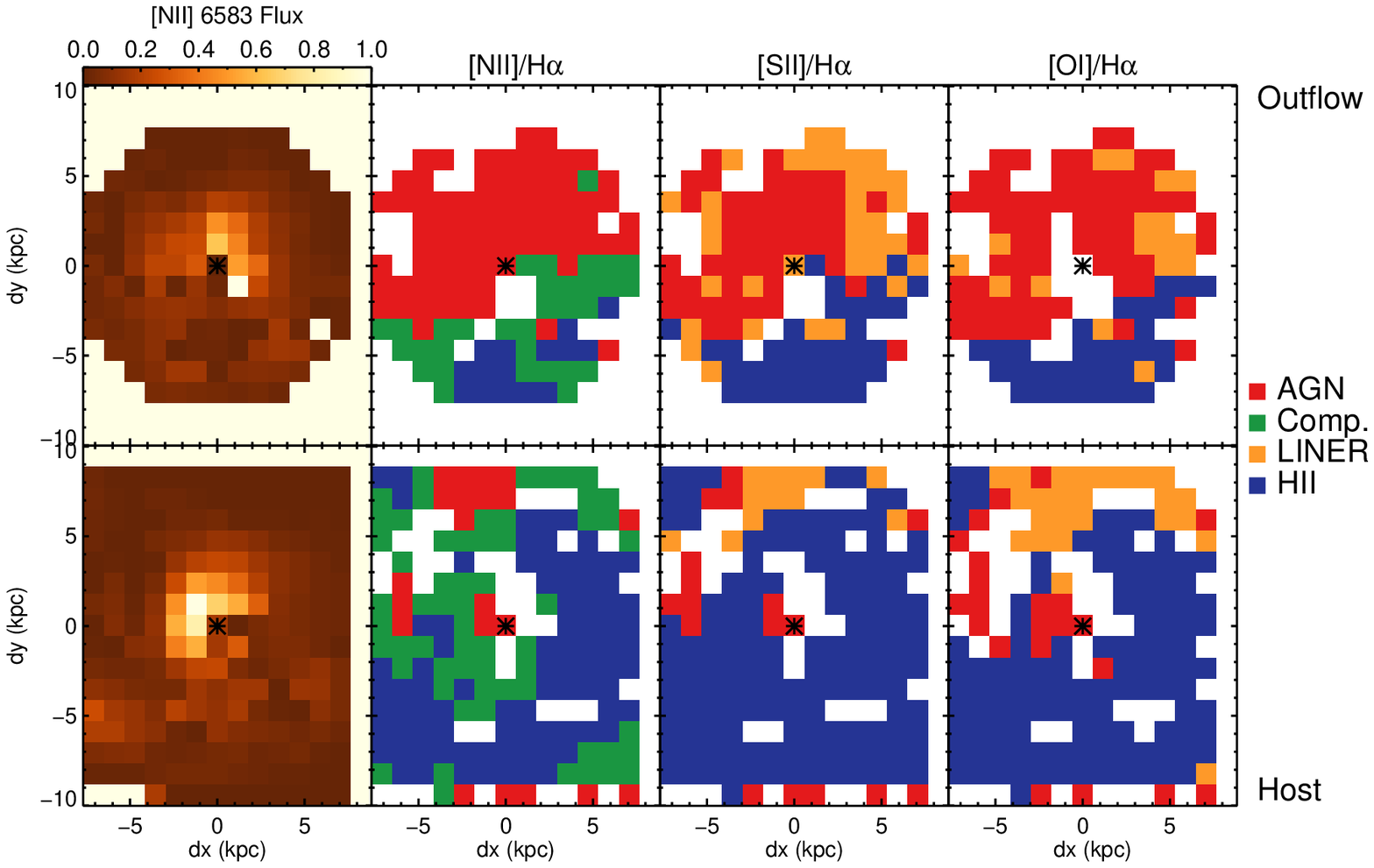}
\caption{Same as Figure \ref{fig:f05189map} but for F13342+3932 and here the fluxes of the host and outflow components are in units of 1.00 $\times$ 10$^{-14}$ and 6.50 $\times$ 10$^{-15}$ erg s$^{-1}$ cm$^{-2}$ arcsec$^{-2}$, respectively.\label{fig:f13342map}}
\end{figure*}

\begin{figure}[ht!]
\centering
\bigskip
\includegraphics[width=1\linewidth,trim=1cm 0.3cm 12.5cm 1.0cm,clip]{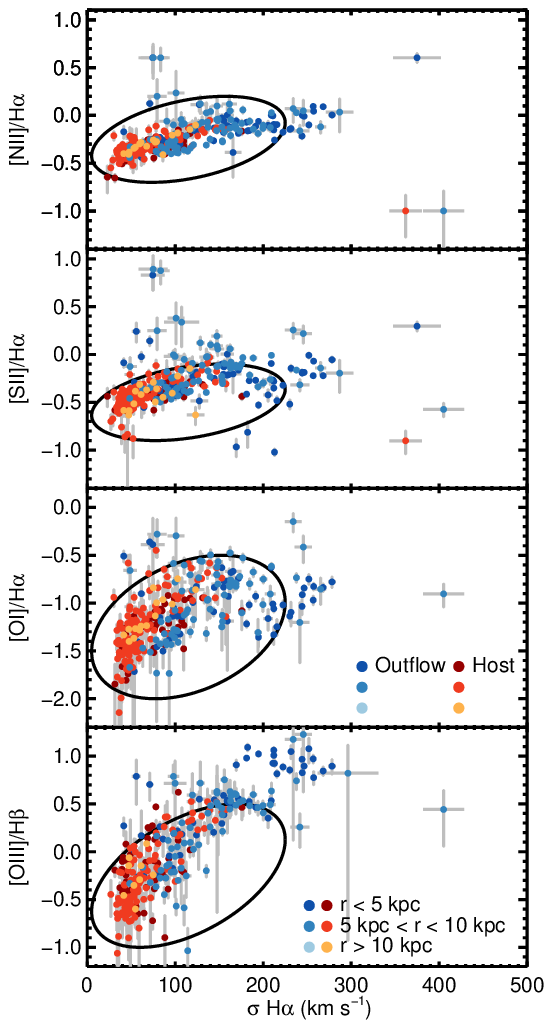}
\caption{Same as Figure \ref{fig:f05189rich} but for F13342+3932.\label{fig:f13342rich}}
\end{figure}

F13218+0552 is similar to F05189$-$2524 in the fact that there is a marked separation between the host and outflow components in the line ratio diagrams. However, rather than being split about the mixing line as in the case of F05189$-$2524, the two components in F13218+0552 are separated roughly by the theoretical starburst line (Kewley et al.\ \citeyear{2001ApJ..556...121K}). Figures \ref{fig:f13218sdss} and \ref{fig:f13218map} show that the line ratios of the host galaxy of F13218+0552 straddles the regions populated by \ion{H}{2} regions, composite galaxies, and LINERs. In contrast, most of the line ratios measured in the outflow lie squarely in the region populated by LINERs. In this object, the distance from the galaxy nucleus does not seem to affect the line ratios in the outflow. In the host, however, the line ratios tend to increase with increasing distance, especially for [\ion{O}{1}]/H$\alpha$ and [\ion{O}{3}]/H$\beta$. 

A straightforward application of the mixing sequence presented in Davies et al.\ (\citeyear{2014MNRAS.444.3961D}, \citeyear{2016MNRAS.462.1616D}) suggests a AGN contribution between 0\% and 30\% to the ionization of the gas in the host galaxy, while the outflowing gas has an AGN contribution between 25\% and 100\%. However, the values of [\ion{N}{2}]/H$\alpha$ are higher than those shown in Davies et al.\ (\citeyear{2014MNRAS.444.3961D}, \citeyear{2016MNRAS.462.1616D}), possibly indicating a contribution from shock ionization.

Indeed, Figure \ref{fig:f13218rich} shows significant correlations between velocity dispersion $\sigma$ and the line ratios in both the galaxy and outflow components. Many of the line ratios measured in the host galaxy fall within the loci of shock-dominated systems. The outflow generally exhibits larger line ratios and $\sigma$ that extend the general trend seen among shock-dominated systems to large velocities. Shock ionization may therefore play a role in the overall ionization of material in both the galaxy and outflow, but AGN photoionization most likely contribute to the large scatter seen in these plots. The lack of obvious correlation between line widths and radial distances suggests that shocks, if present, are distributed throughout the galaxy and outflow.

\subsection{F13342+3932}\label{sec:f13342}
\begin{figure*}[ht!]
\gridline{\includegraphics[height=0.3\linewidth,trim=3.5cm 0.25cm 4.25cm 1cm,clip]{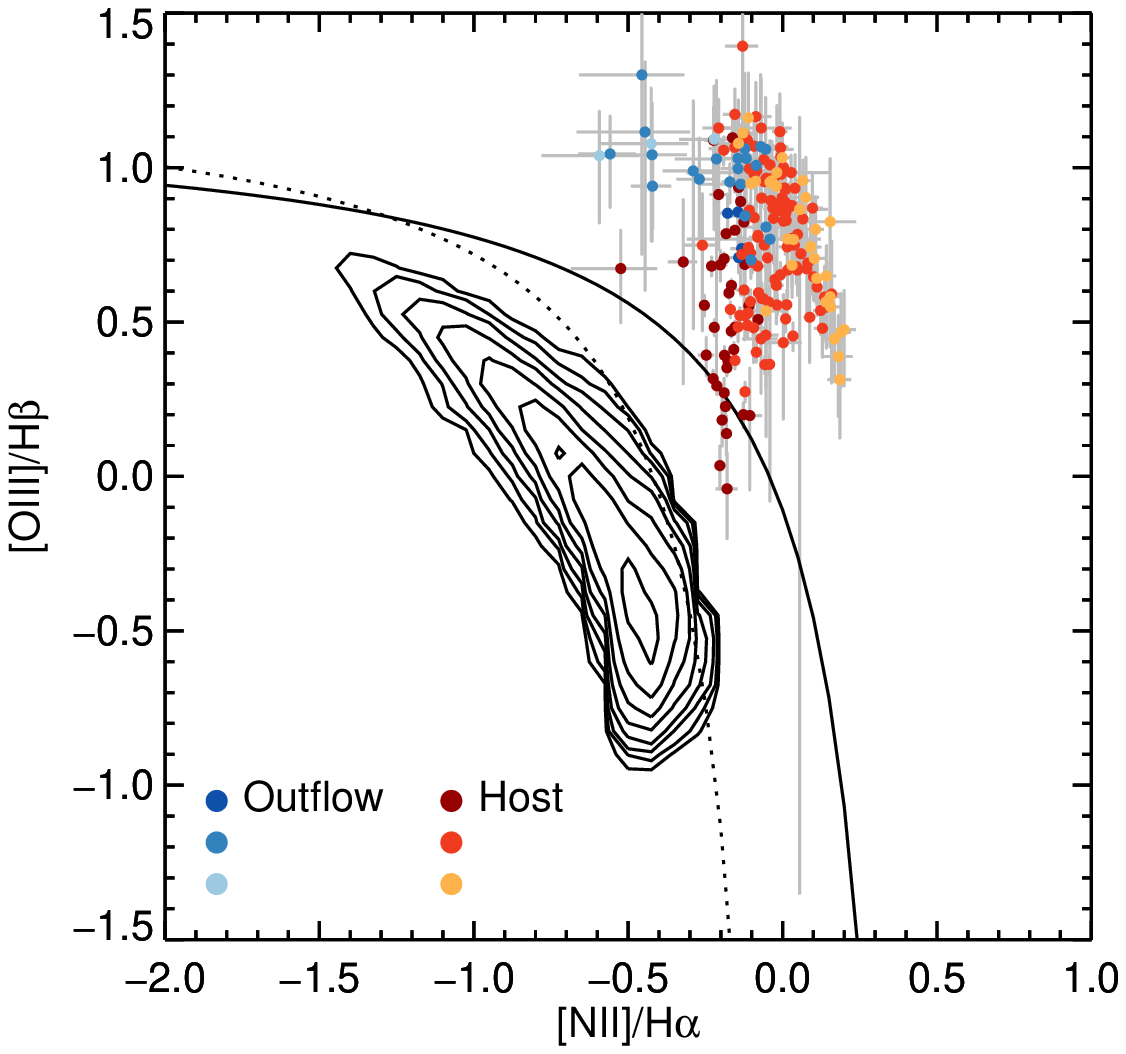}
		  \includegraphics[height=0.3\linewidth,trim=3.5cm 0.25cm 4.25cm 1cm,clip]{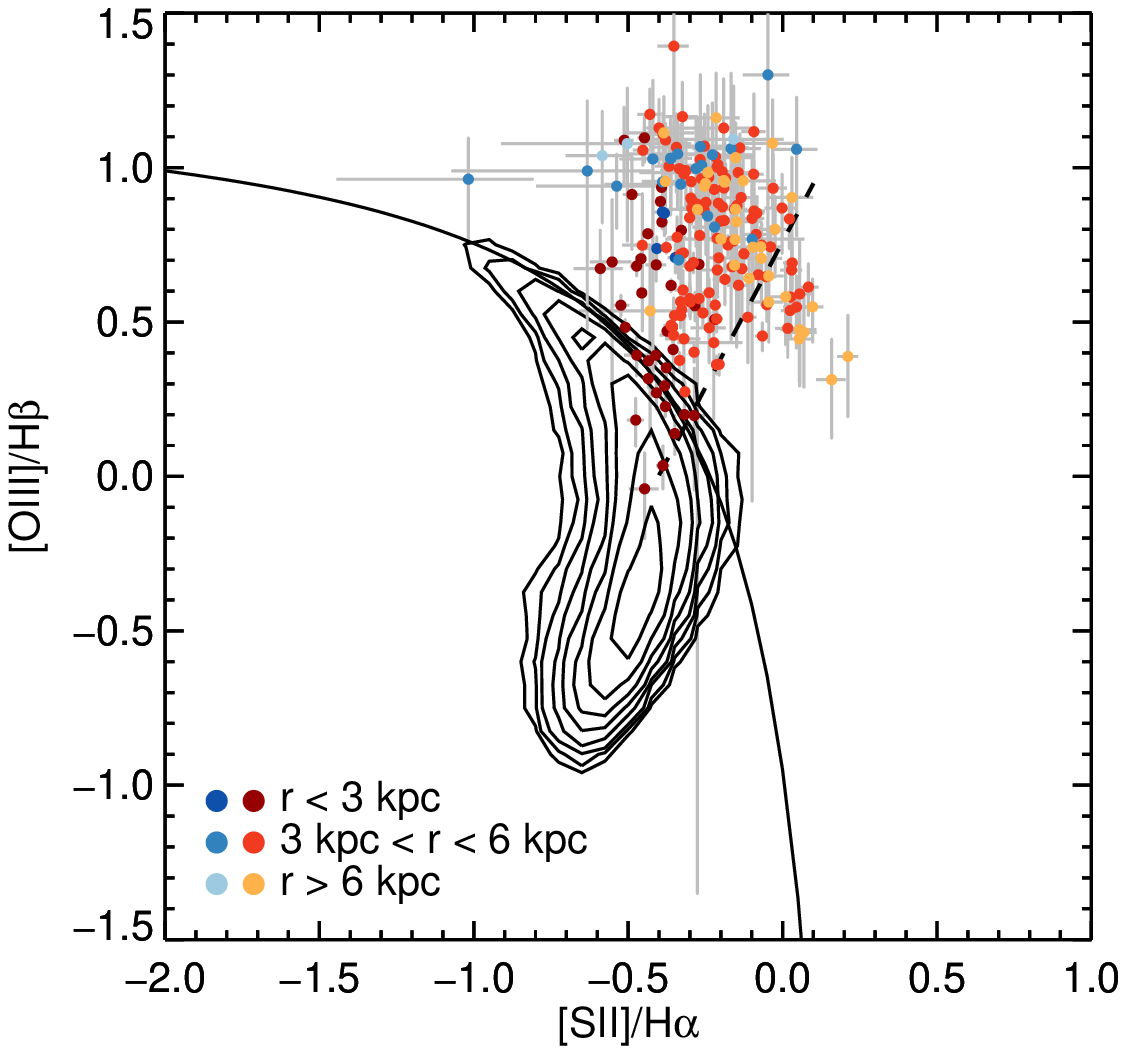}
          \includegraphics[height=0.3\linewidth,trim=3.5cm 0.25cm 4.25cm 1cm,clip]{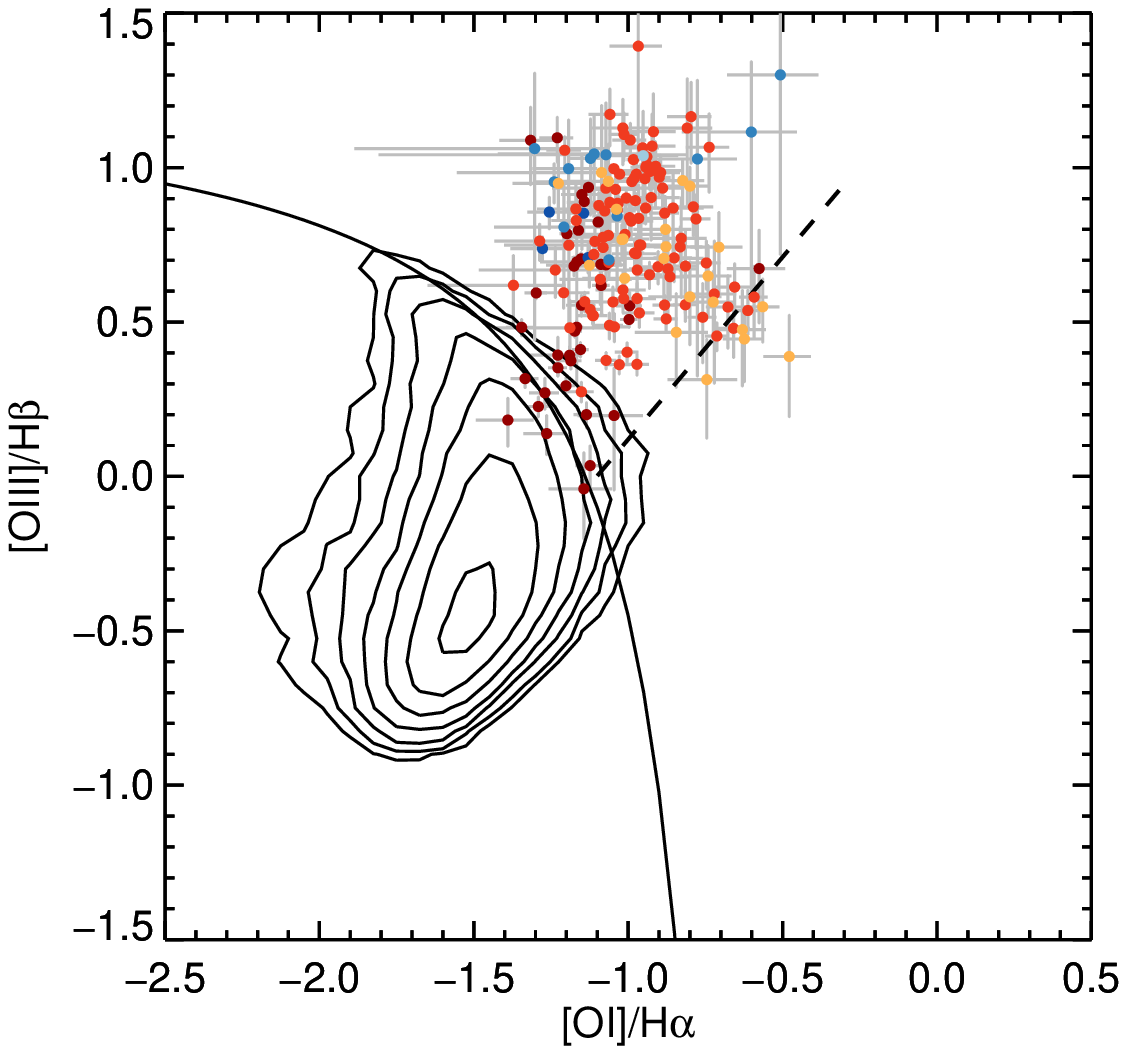}}
\caption{Same as Figure \ref{fig:f05189sdss} but for PG1613+658.\label{fig:pg1613sdss}}
\end{figure*}

\begin{figure*}[htp!]
\centering
\includegraphics[width=0.75\linewidth,trim=0.75cm 0.4cm 0.1cm 0.1cm,clip]{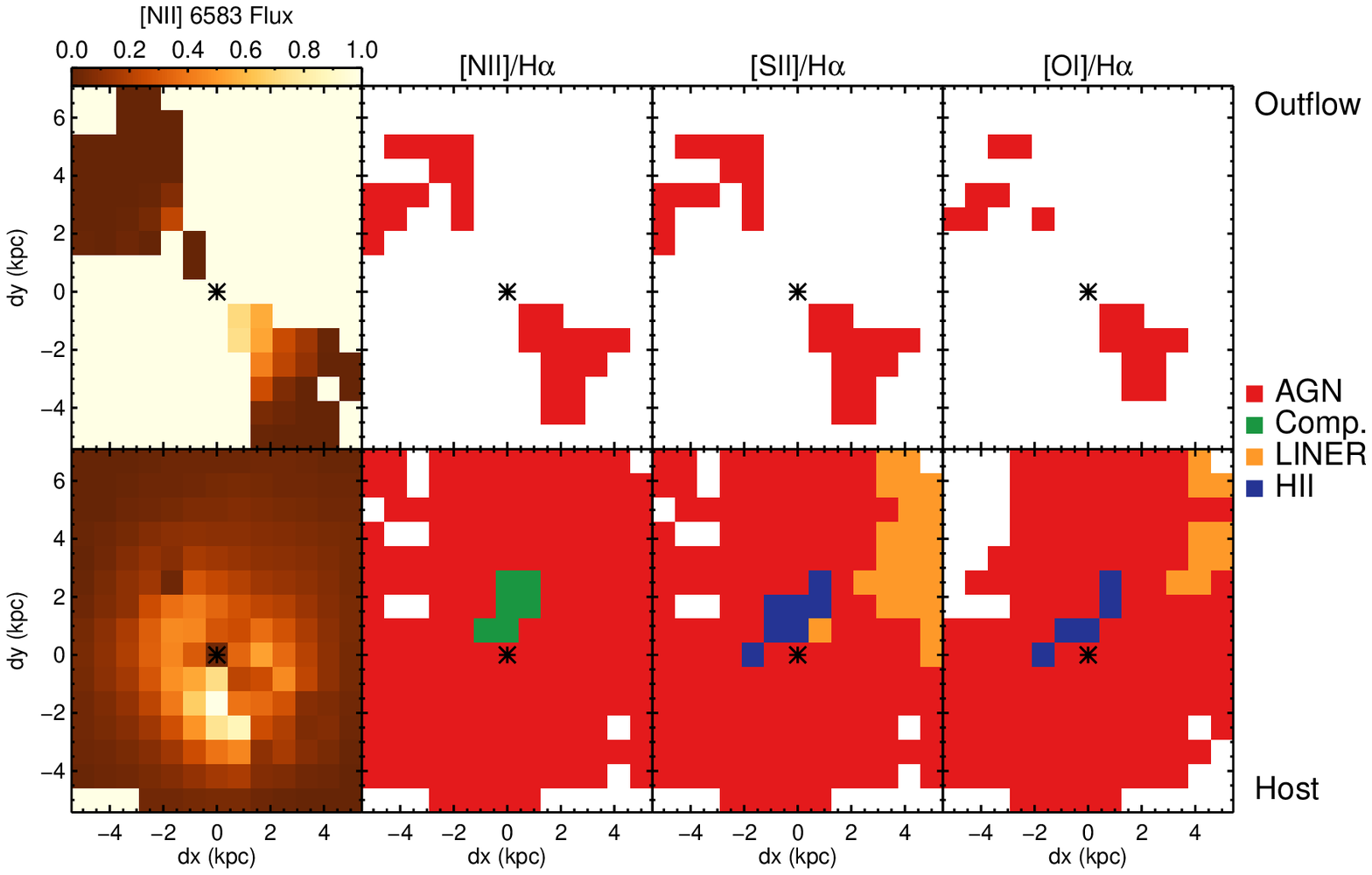}
\caption{Same as Figure \ref{fig:f05189map} but for PG1613+658 and here the fluxes of the host and outflow components are in units of 9.41 $\times$ 10$^{-15}$ and 7.02 $\times$ 10$^{-15}$ erg s$^{-1}$ cm$^{-2}$ arcsec$^{-2}$, respectively.\label{fig:pg1613map}}
\end{figure*}

\begin{figure}[ht!]
\centering
\bigskip
\includegraphics[width=1\linewidth,trim=1cm 0.3cm 12.5cm 1.0cm,clip]{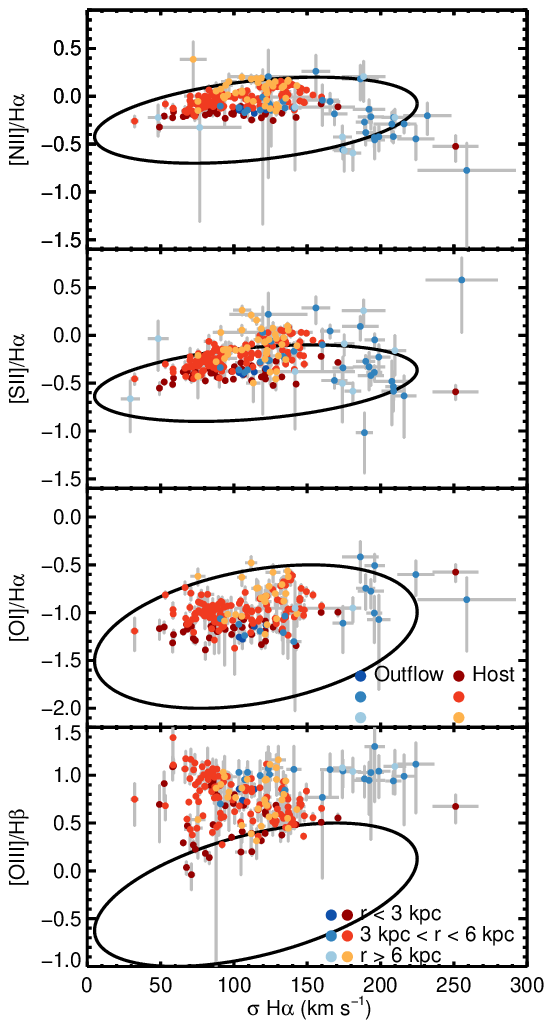}
\caption{Same as Figure \ref{fig:f05189rich} but for PG1613+658.\label{fig:pg1613rich}}
\end{figure}

Unlike F05189$-$2524 and F13218+0552, there is considerable, although not one-to-one, overlap between the host and outflow in the line ratio diagrams. Figure \ref{fig:f13342sdss} indicates that the host galaxy of F13342+3932 generally has \ion{H}{2} region-like line ratios, while the outflow has line ratios that are characteristic of AGN photoionization, LINER-like, and \ion{H}{2} region-like. Interestingly, some of the [\ion{O}{1}]/H$\alpha$ line ratios in the host are anomalously high, almost as high as those detected in the outflowing component.

Figure \ref{fig:f13342map} confirms that the gas in the host galaxy of this object is mostly \ion{H}{2} region-like, with small patches of LINER-like and AGN-photoionized material. In contrast, the outflow is predominantly photoionized by the AGN, but there is also a contiguous \ion{H}{2} region-like region south of the nucleus, and a transition zone between the \ion{H}{2} region-like and AGN-like data points that show either composite or LINER-like line ratios. It is interesting to note that the \ion{H}{2} region-like outflowing material coincides (as projected on the sky) with host material that is characterized by very similar line ratios and thus similar physical conditions (i.e gas metallicity, ionization level). This suggests that the same stars that are ionizing the gas in the host are also ionizing some of the outflowing material. We return to this point in \S 5.2. 

Figure \ref{fig:f13342rich} shows strong correlations between the velocity dispersion $\sigma$ and the line ratios of the host and outflow components. As in F13218+0552, most of the line ratios of the host fall within the ellipses of shock-dominated galaxies and follow the general slope of these ellipses, while the line ratios of the outflow component extend these relations to larger values of $\sigma$ but with considerably more scatter. The largest line widths in the outflows are found closest to the central AGN where the line ratios are also the largest. On the other hand, no obvious trend with distance from the nucleus is found in the host galaxy. These correlations suggest that shock ionization plays a significant role in the overall ionization of material in the galaxy and outflow, despite the fact that the line ratios in the host are mostly \ion{H}{2}-region like while those in the outflow are mostly AGN-like. 

There is likely mixing between OB star photoionization and shock ionization in the host component. As the line ratios increase and move away from those of pure stellar photoionization, the values of sigma also increase. The starburst-AGN mixing sequence of Davies et al.\ (\citeyear{2014MNRAS.444.3961D}, \citeyear{2016MNRAS.462.1616D}) suggests a contribution from AGN photoionization between 0\% and 50\% in the host galaxy and between 10\% and 100\% in the outflow. This very wide spread indicates that mixing between the various ionization mechanisms is highly significant.

\subsection{PG1613+658}\label{sec:pg1613}

Figures \ref{fig:pg1613sdss} and \ref{fig:pg1613map} show that both the host and outflow components of PG1613+658 display line ratios that are largely consistent with AGN photoionization. The only exception is the host galaxy material immediately north of the center, which exhibits composite or \ion{H}{2} region-like line ratios. This material coincides with a dusty structure that runs diagonally in this object along the south-east $-$ north-west direction (\citetalias{2017ApJ..850...40R}). The outflow is almost exactly perpendicular to this structure. It also coincides with a ionization bi-cone that is centered on the AGN. The outflowing gas is more highly ionized than the host gas, displaying stronger [\ion{O}{3}]/H$\beta$ and weaker [\ion{N}{2}]/H$\alpha$, [\ion{S}{2}]/H$\alpha$, and [\ion{O}{1}]/H$\alpha$. There is also a relationship between the line ratios in the host galaxy and distances from the nucleus, particularly in the [\ion{N}{2}]/H$\alpha$ diagram. As the distance from the center of the galaxy increases, the value of [\ion{N}{2}]/H$\alpha$, and to a lesser degree [\ion{S}{2}]/H$\alpha$ and [\ion{O}{1}]/H$\alpha$, also increases, becoming increasingly AGN-like.  As expected, the [\ion{O}{3}]/H$\beta$ line ratios in the host and outflow component are too high to fall within the ellipses of shock-dominated galaxies (Figure \ref{fig:pg1613rich}). Interestingly, this object has the lowest velocity dispersions of any in our sample, further indicating that shocks likely do not play a significant role in this object.

\begin{figure*}[htp!]
\gridline{\includegraphics[height=0.41\linewidth,trim=0.25cm 0.0cm 13.5cm 0.25cm,clip]{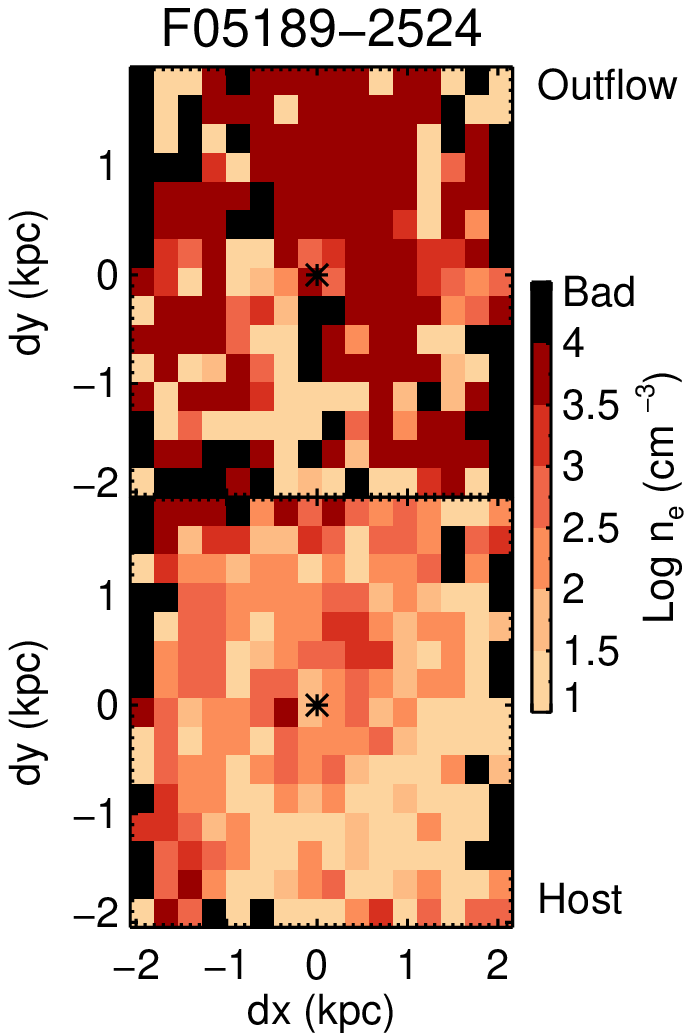}
		  \includegraphics[height=0.41\linewidth,trim=0.25cm 0.0cm 13.5cm 0.25cm,clip]{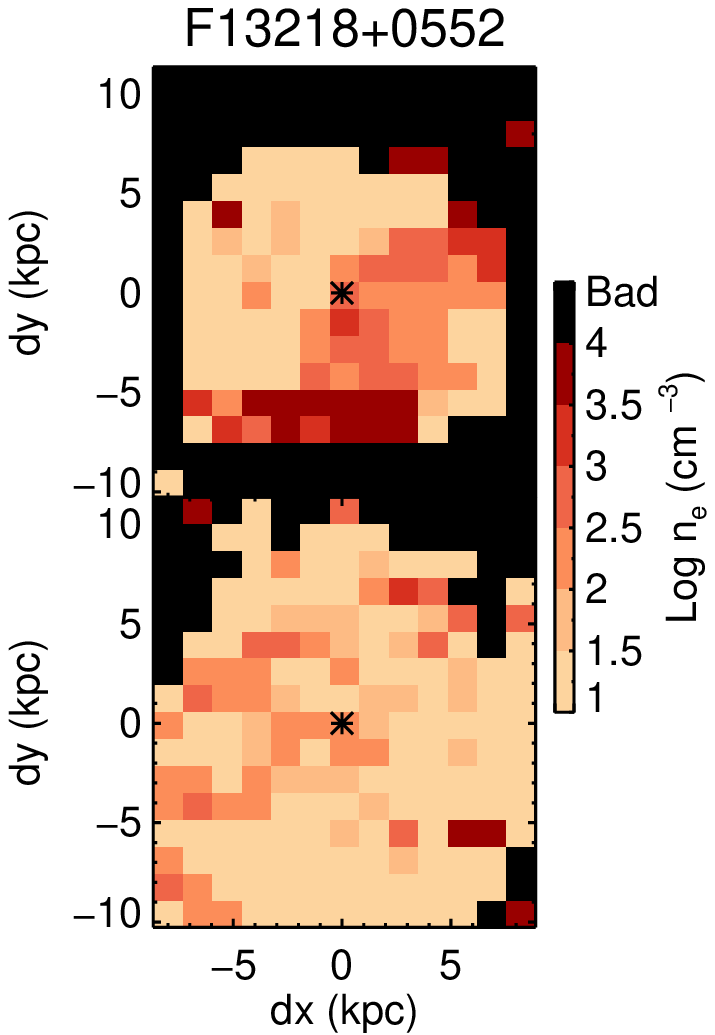}
          \includegraphics[height=0.41\linewidth,trim=0.25cm 0.0cm 13.5cm 0.25cm,clip]{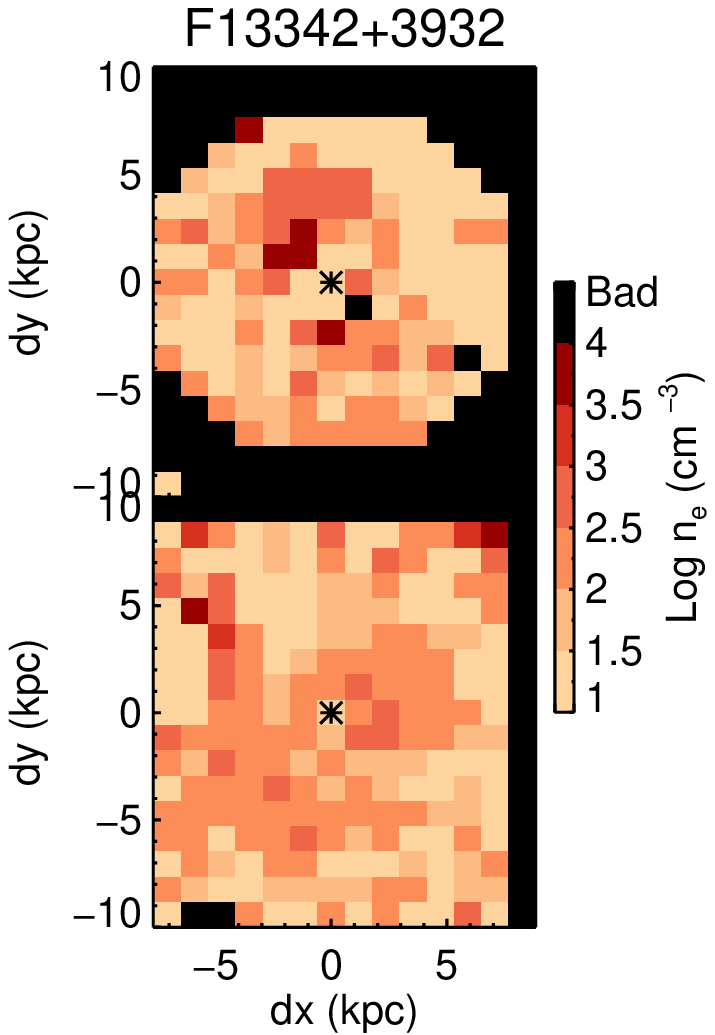}
          \includegraphics[height=0.41\linewidth,trim=0.25cm 0.0cm 10.0cm 0.25cm,clip]{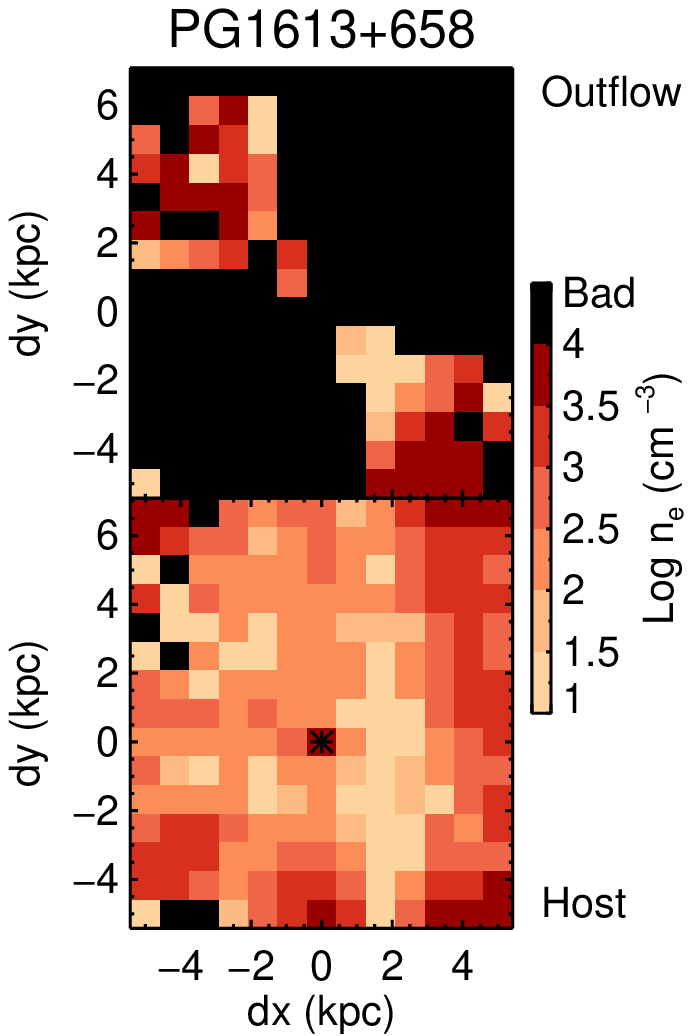}}
\caption{Maps of the electron density in the host galaxy and outflowing components of the four galaxies in our sample. The top panels show the outflow component while the bottom panels show the host galaxy component. Note the different spatial scales for each object. The logarithms of the densities are shown ranging from 10 cm$^{-3}$ (light) to 10$^4$ cm$^{-3}$ (dark). Black spaxels indicate either no data or that the spectra at these spaxels were not of high enough quality to calculate an electron density. The black asterisk at the center of each panel marks the nucleus of the galaxy. \label{fig:density}}
\end{figure*}

\begin{figure*}[htp!]
\gridline{\includegraphics[width=0.5\linewidth,trim=1cm 0cm 1cm 0.1cm,clip]{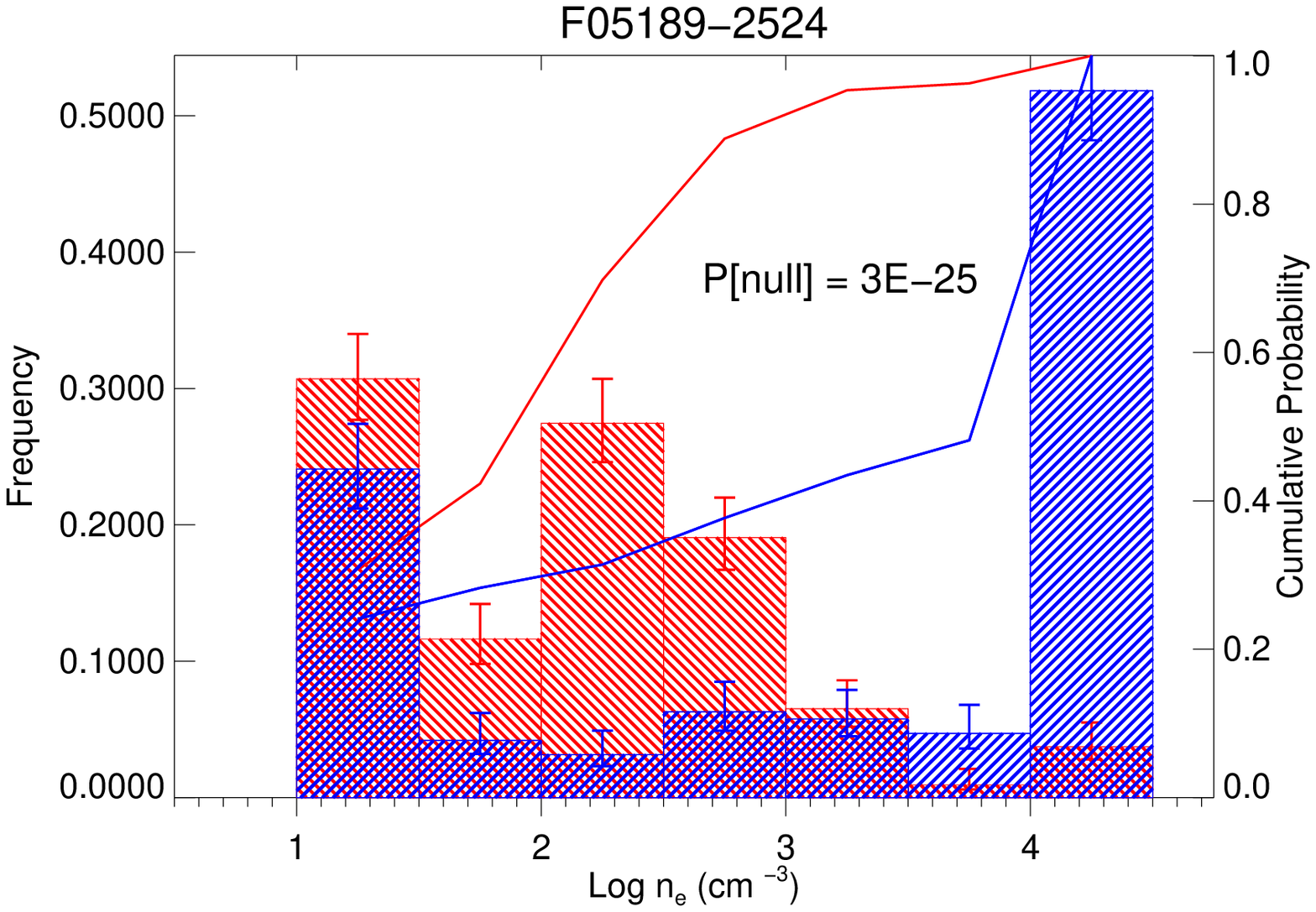}
		  \includegraphics[width=0.5\linewidth,trim=1cm 0cm 1cm 0.1cm,clip]{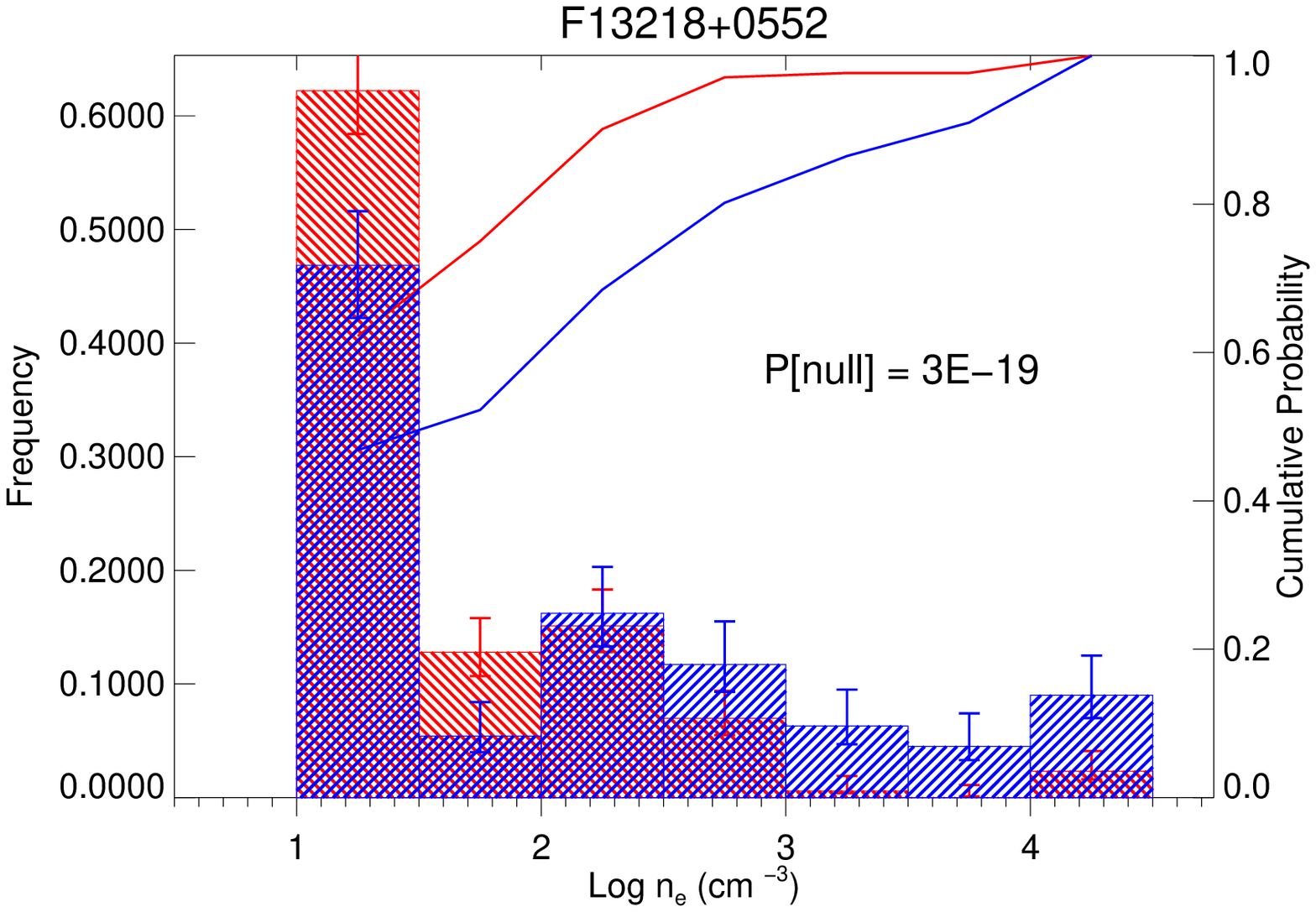}}
\gridline{\includegraphics[width=0.5\linewidth,trim=1cm 0cm 1cm 0.1cm,clip]{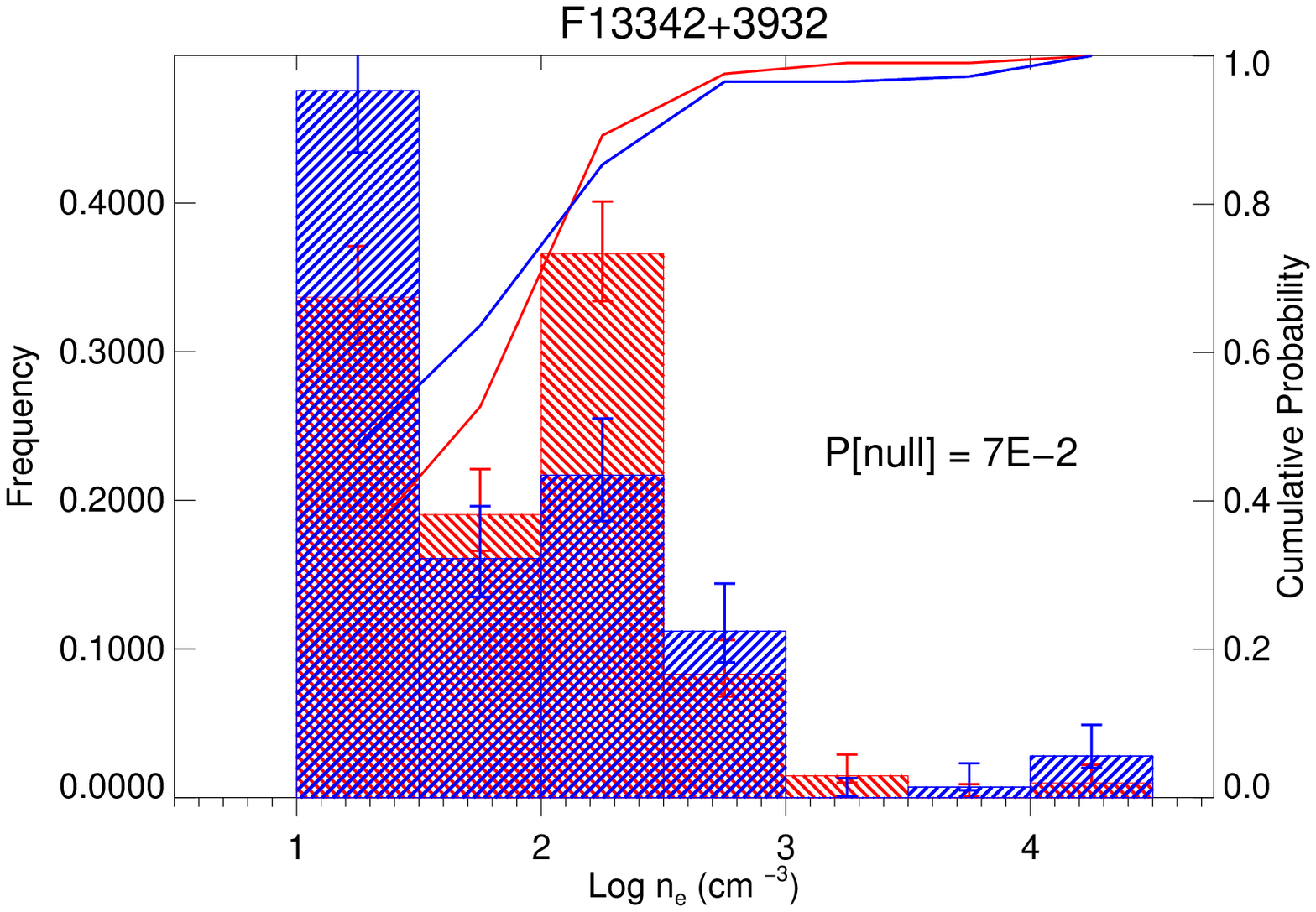}
          \includegraphics[width=0.5\linewidth,trim=1cm 0cm 1cm 0.1cm,clip]{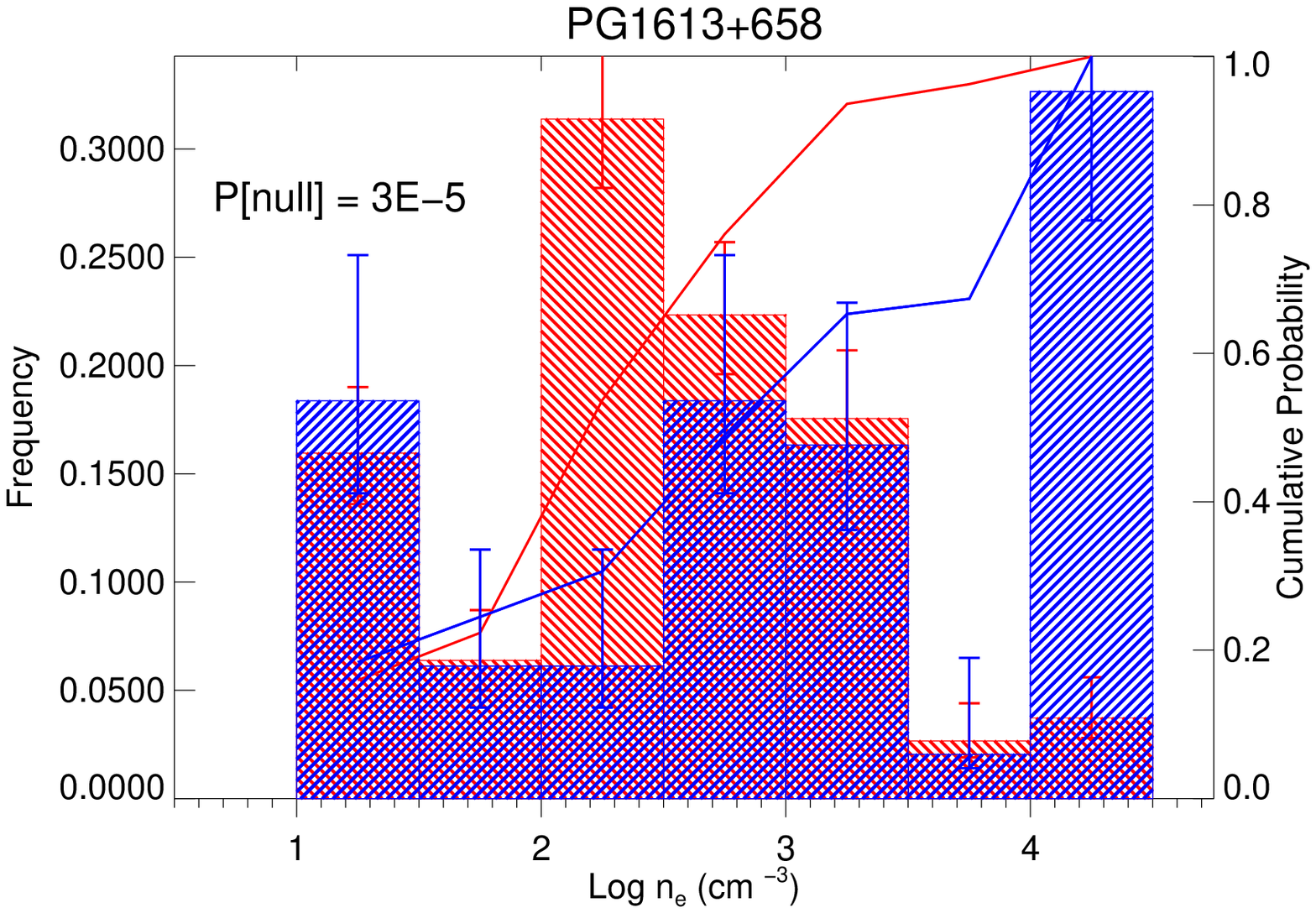}}
\caption{Histogram and cumulative distribution function of electron density in the host galaxy (red) and outflowing component (blue) of the four galaxies in our sample. The bins show the same ranges as in Figure \ref{fig:density} except for the upper limit of 10$^4$ cm$^{-3}$ being represented in a separate bar. Uncertainties are calculated using the method presented in Cameron (\citeyear{2011PASA...28..128C}). The probability of the null hypothesis that the host and outflow density are drawn from the same distribution, calculated using the Kolmogorov-Smirnov test, is shown. \label{fig:denhist}}
\end{figure*}

\section{Comparison With Theoretical Models}\label{theo}

\begin{figure*}[htp!]
\gridline{\includegraphics[height=0.3\linewidth,trim=4.0cm 0.25cm 4.5cm 1cm,clip]{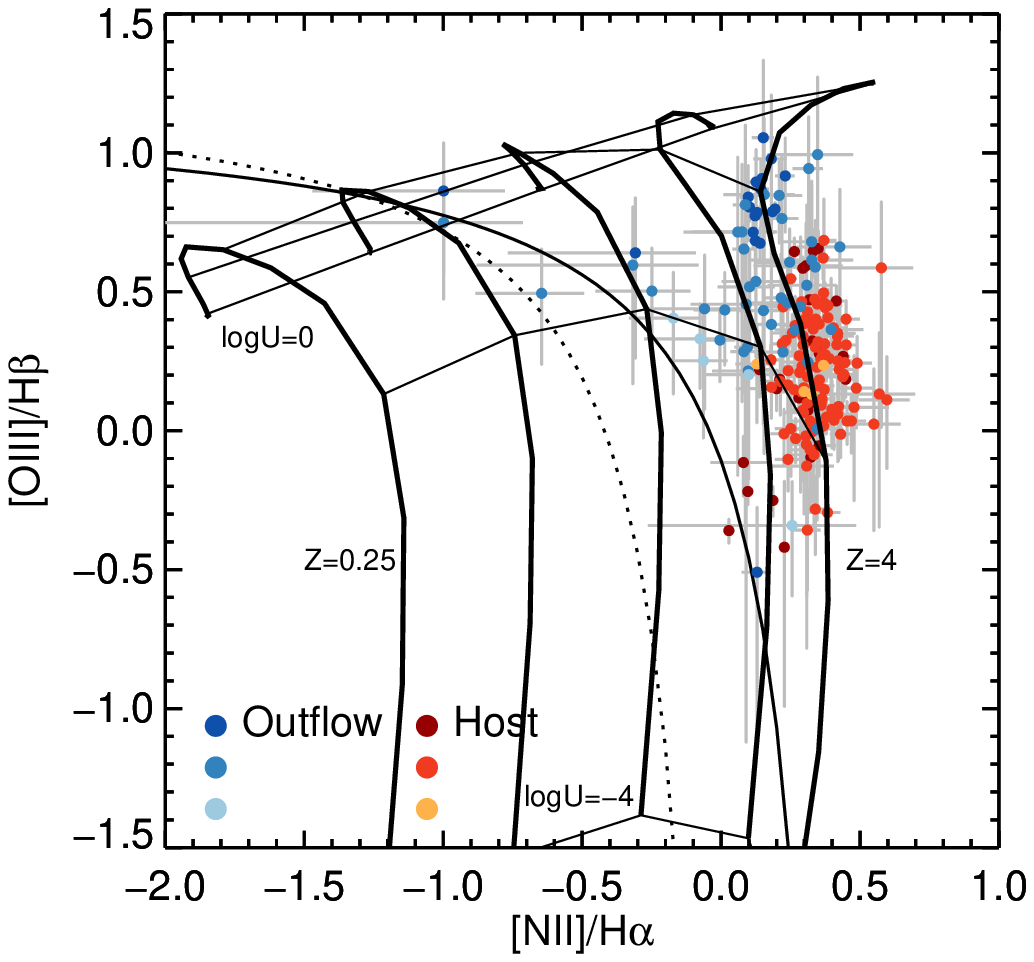}
		  \includegraphics[height=0.3\linewidth,trim=4.0cm 0.25cm 4.5cm 1cm,clip]{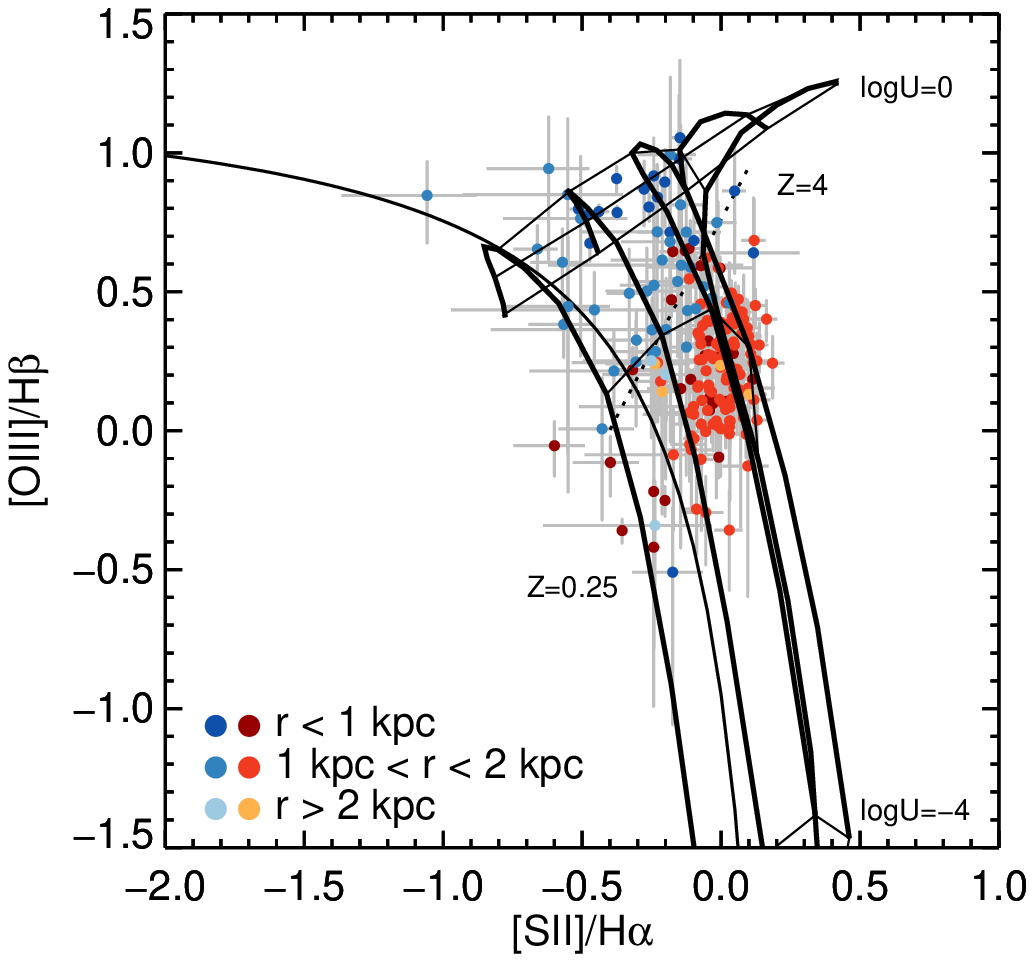}
          \includegraphics[height=0.3\linewidth,trim=4.0cm 0.25cm 4.5cm 1cm,clip]{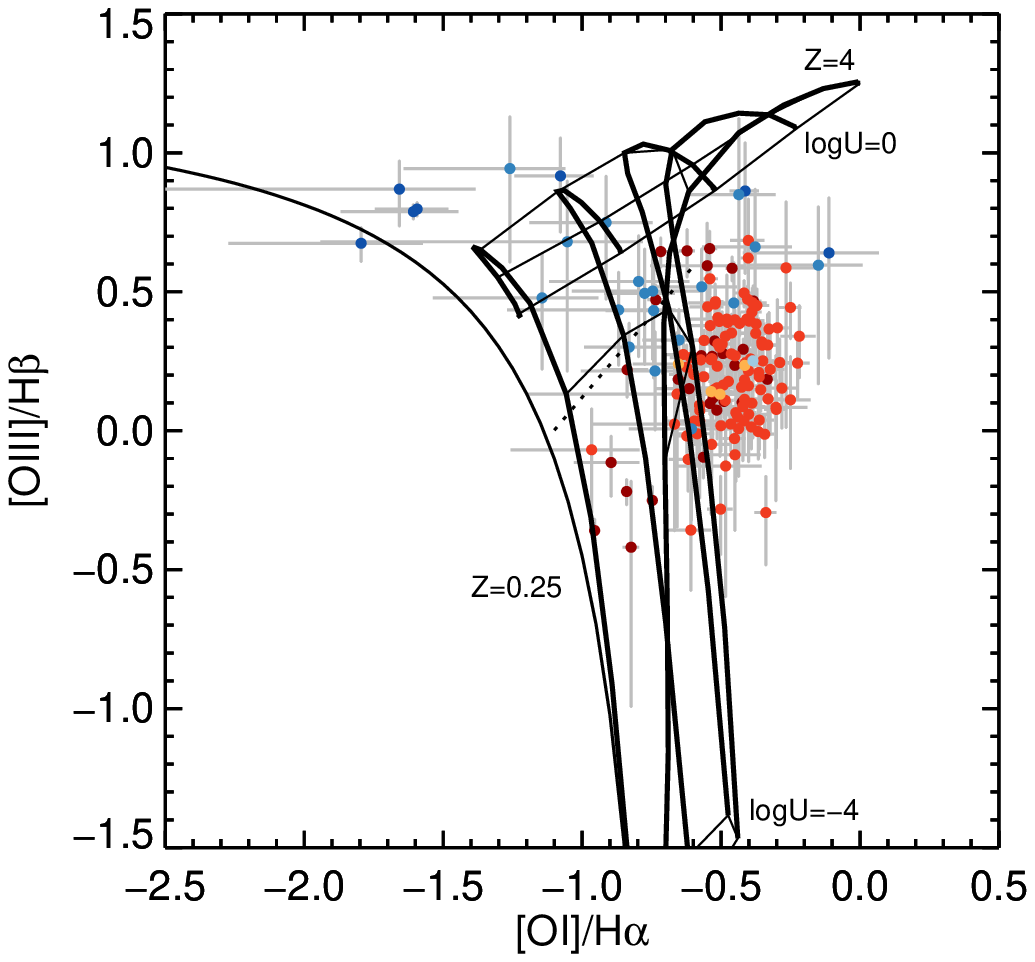}}
\caption{Comparisons between the line ratios observed in the host and outflow of F05189$-$2524 and the theoretical predictions from the dusty AGN models of Groves et al.\ \citeyear{2004ApJS..153....9G}. The color-coding of the data points is the same as in Figure \ref{fig:f05189sdss}. The thick lines show constant metallicity $Z$ from 0.25 to 4 times solar, spaced by factors of two. The thin lines show constant logarithm of dimensionless ionization parameter $U$ from $-$4 to 0 spaced by 1 dex. The neutral hydrogen column density is frozen to $n_H$=1000 cm$^{-3}$ and the power law index is frozen to $\alpha=-1.4$. \label{fig:f05189agn}}
\end{figure*}

\begin{figure*}[htp!]
\gridline{\includegraphics[height=0.3\linewidth,trim=4.0cm 0.25cm 4.5cm 1cm,clip]{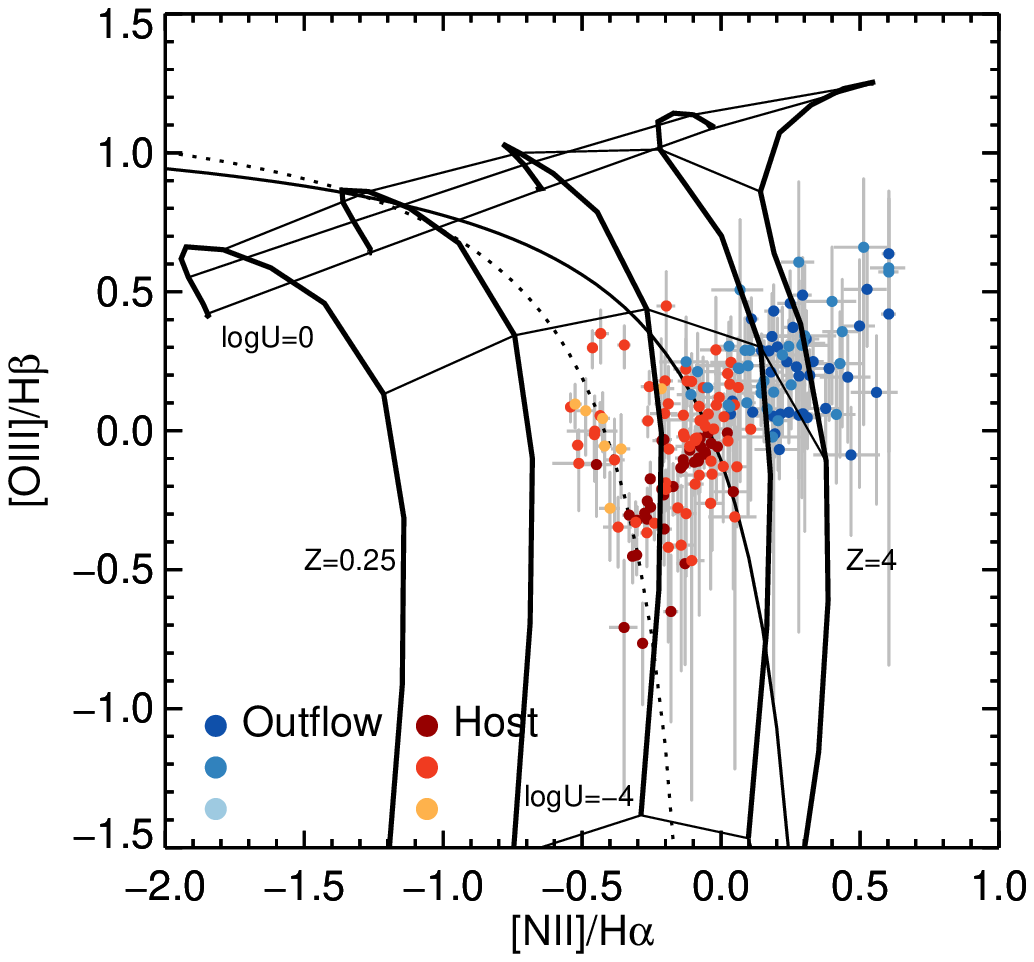}
		  \includegraphics[height=0.3\linewidth,trim=4.0cm 0.25cm 4.5cm 1cm,clip]{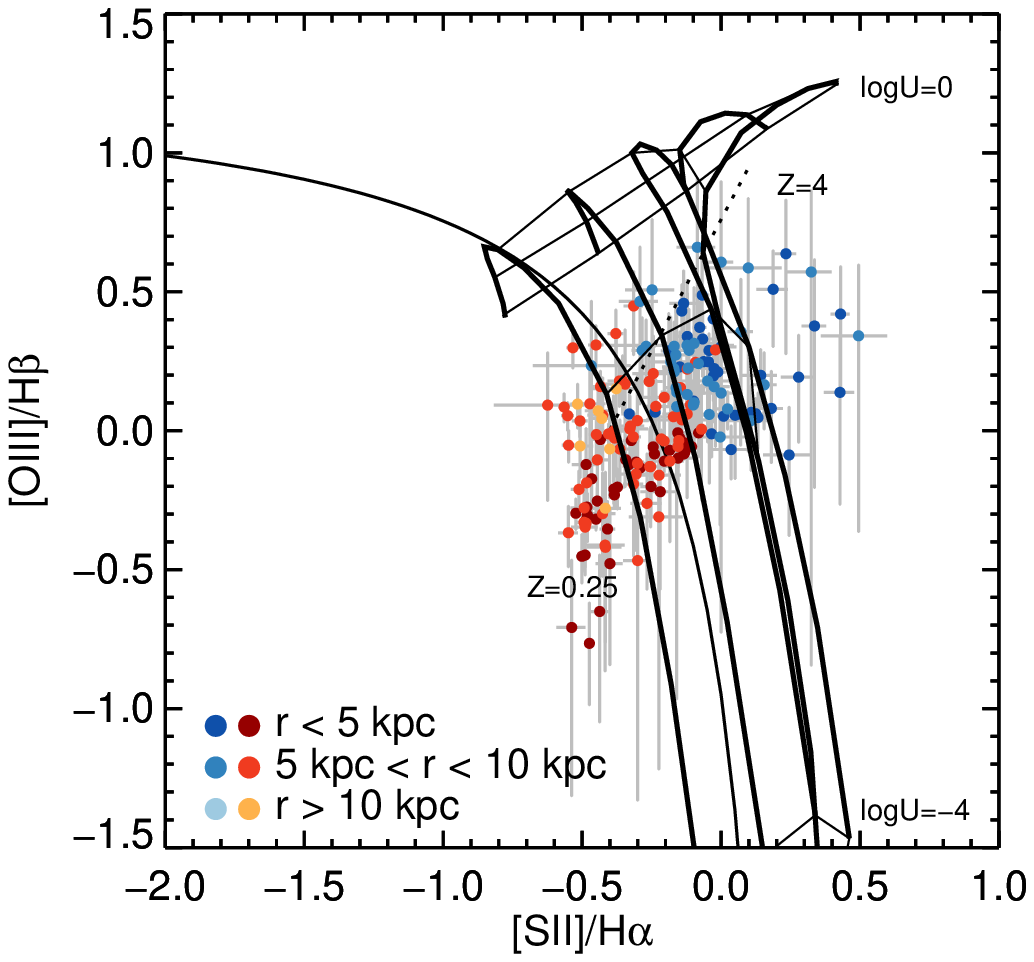}
          \includegraphics[height=0.3\linewidth,trim=4.0cm 0.25cm 4.5cm 1cm,clip]{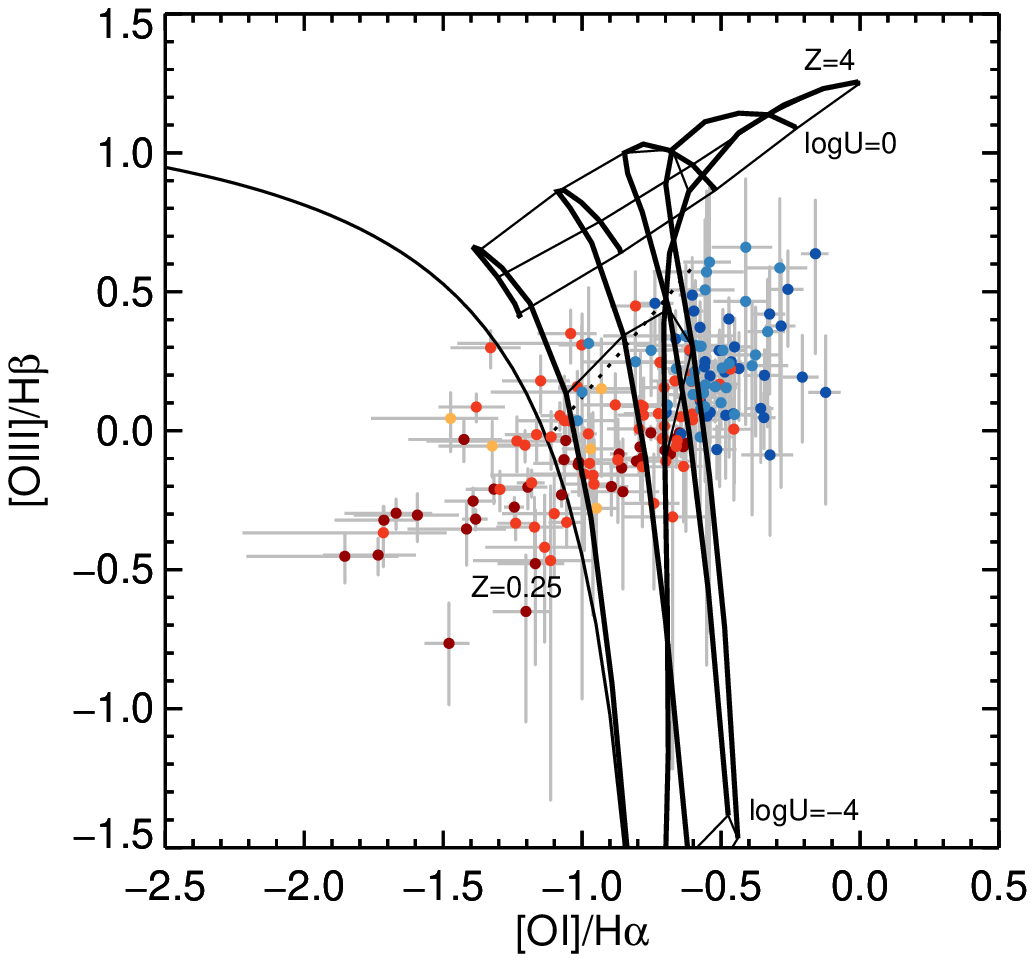}}
\caption{Same as Figure \ref{fig:f05189agn}, but for F13218+0552.\label{fig:f13218agn}}
\end{figure*}

\begin{figure*}[htp!]
\gridline{\includegraphics[height=0.3\linewidth,trim=4.0cm 0.25cm 4.5cm 1cm,clip]{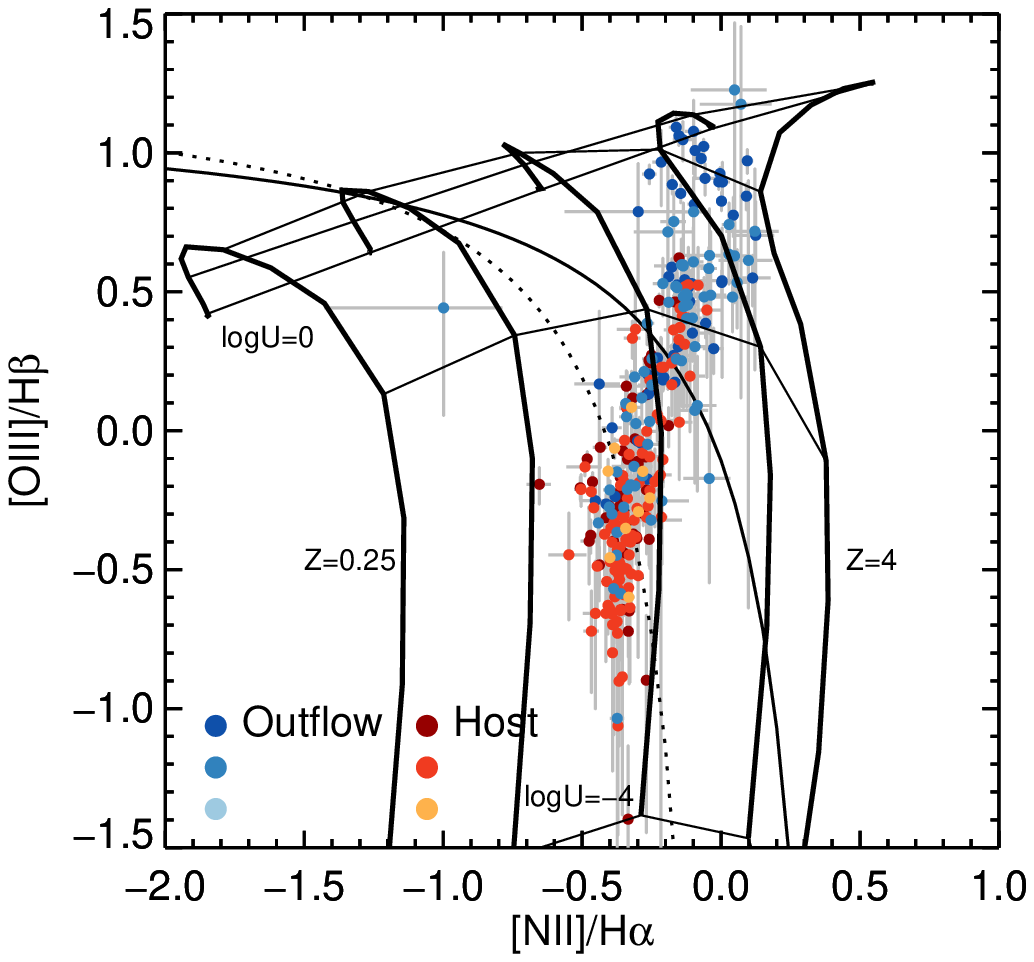}
		  \includegraphics[height=0.3\linewidth,trim=4.0cm 0.25cm 4.5cm 1cm,clip]{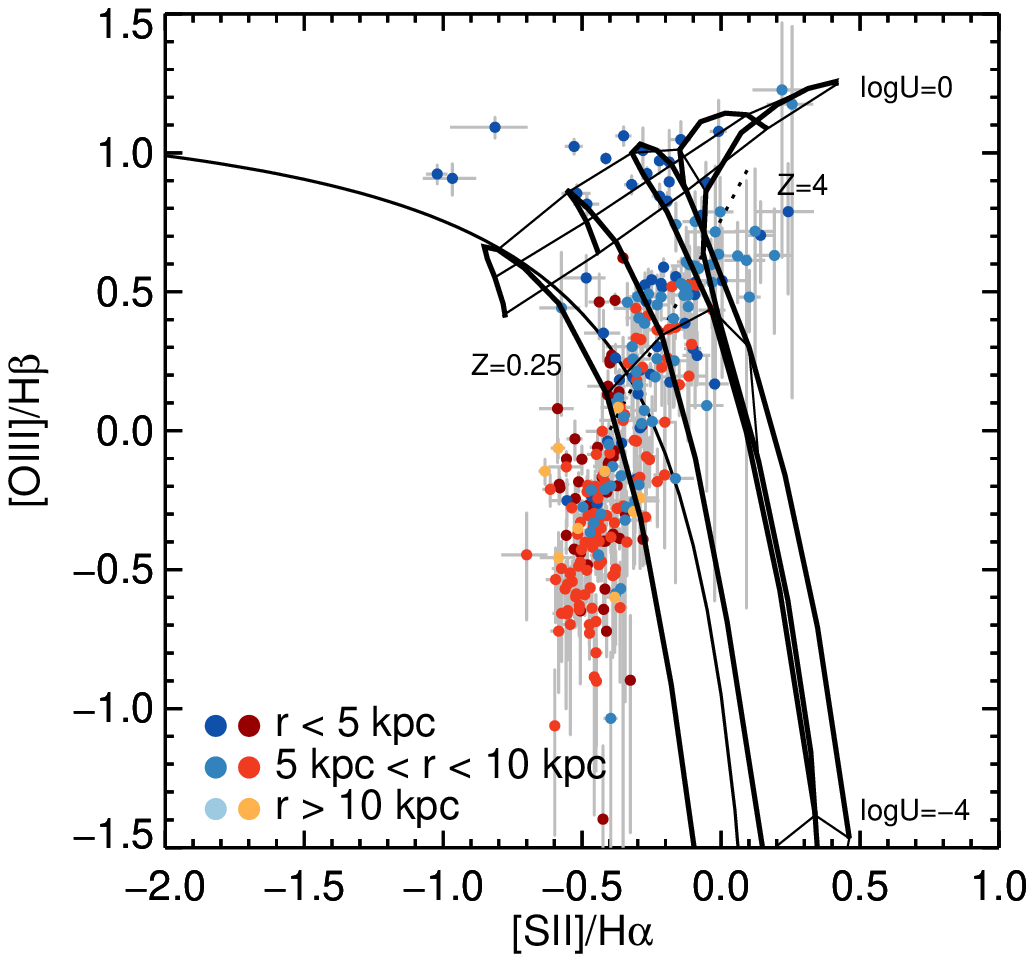}
          \includegraphics[height=0.3\linewidth,trim=4.0cm 0.25cm 4.5cm 1cm,clip]{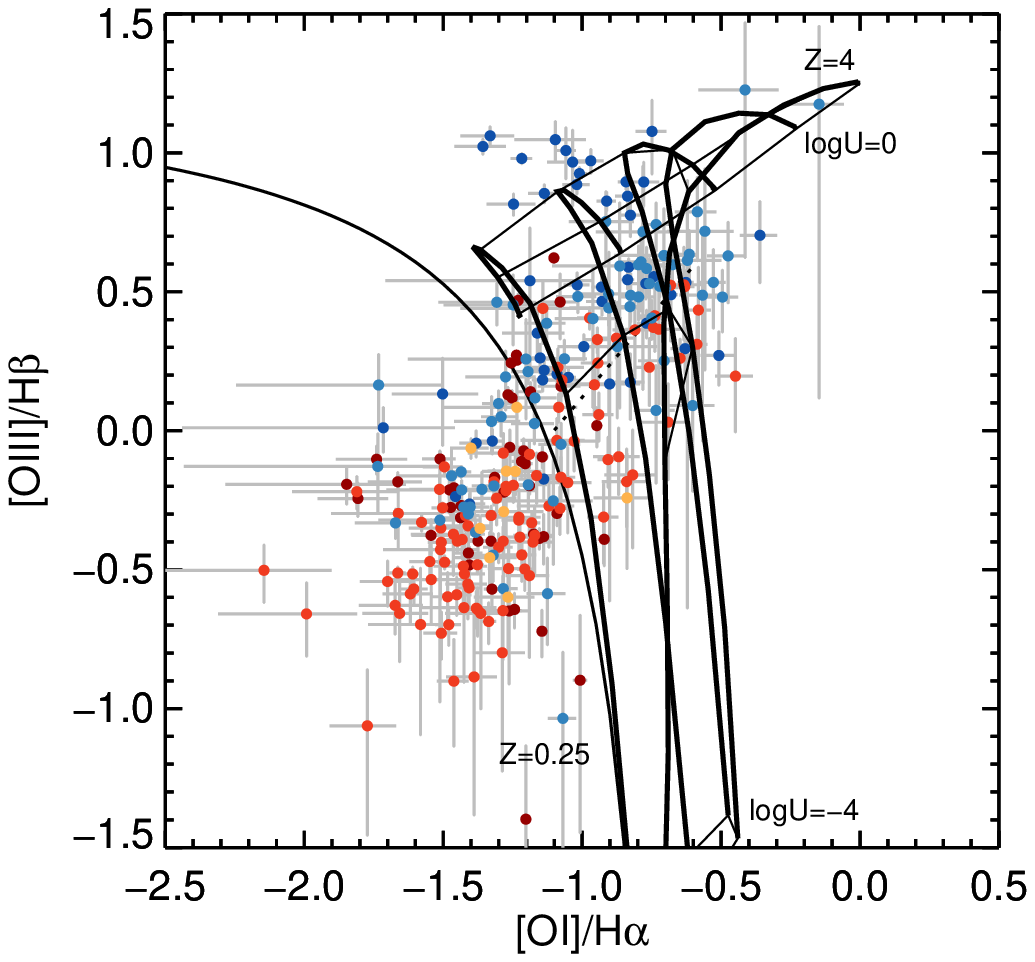}}
\caption{Same as Figure \ref{fig:f05189agn}, but for F13342+3932.\label{fig:f13342agn}}
\end{figure*}

\begin{figure*}[htp!]
\gridline{\includegraphics[height=0.3\linewidth,trim=4.0cm 0.25cm 4.5cm 1cm,clip]{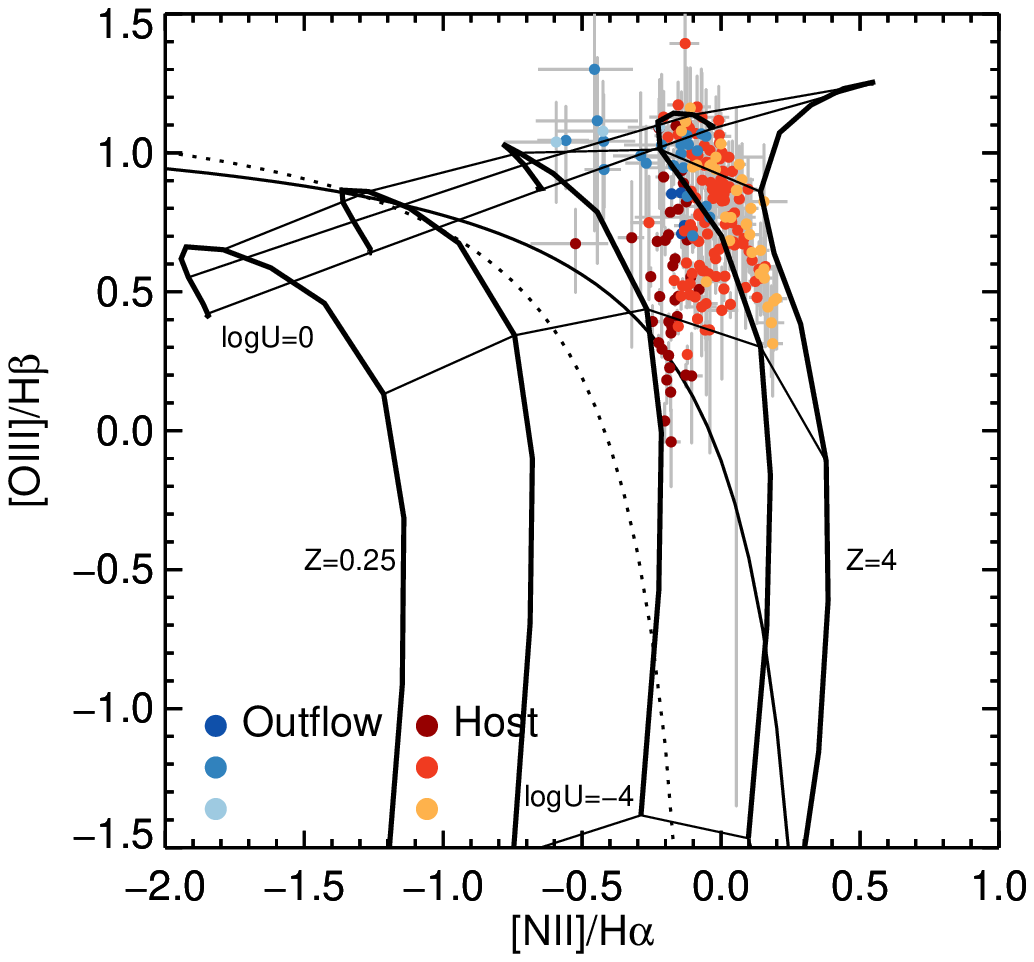}
		  \includegraphics[height=0.3\linewidth,trim=4.0cm 0.25cm 4.5cm 1cm,clip]{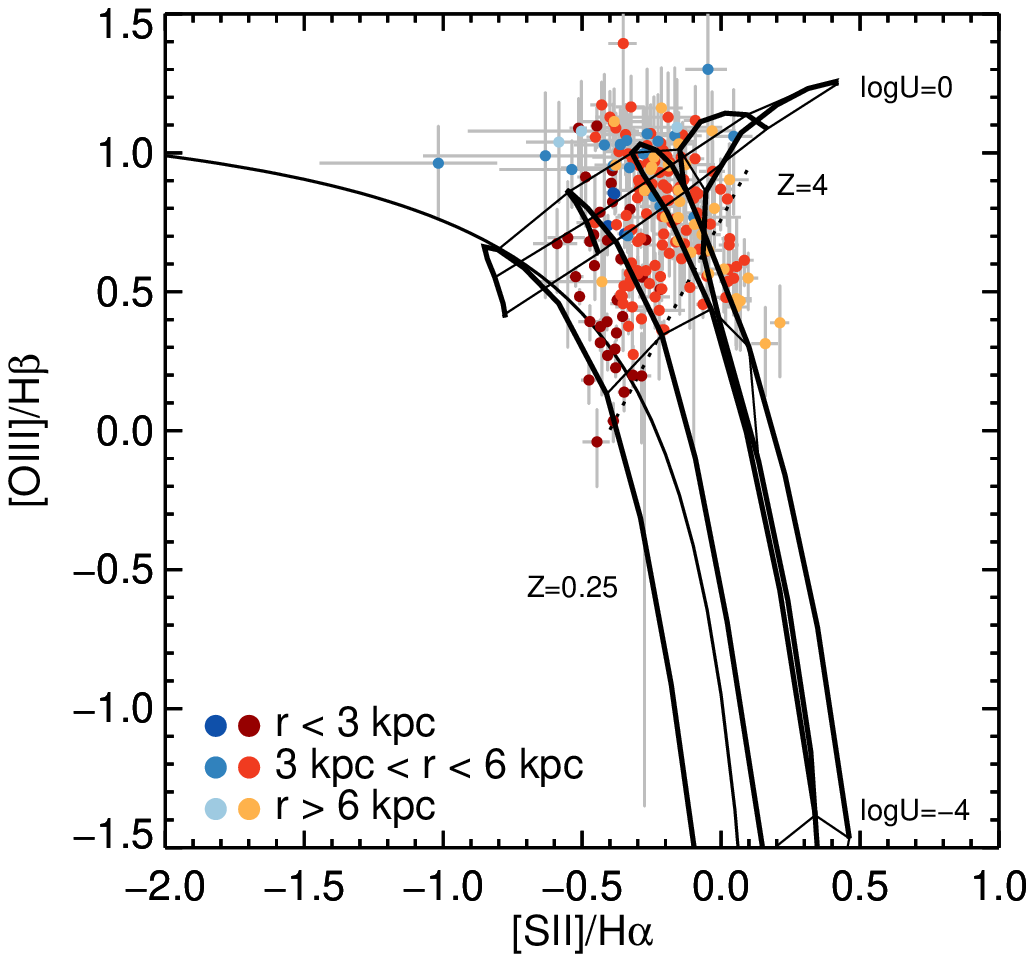}
          \includegraphics[height=0.3\linewidth,trim=4.0cm 0.25cm 4.5cm 1cm,clip]{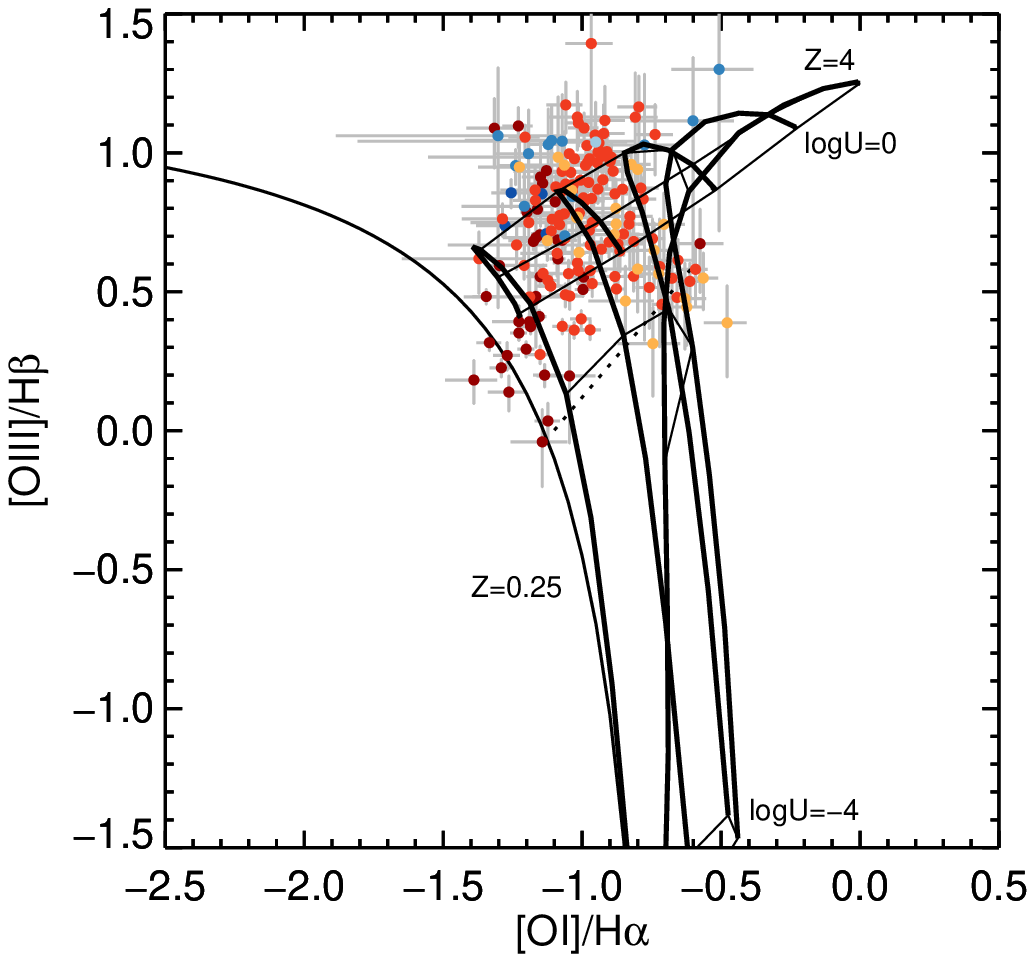}}
\caption{Same as Figure \ref{fig:f05189agn}, but for PG~1613+658.\label{fig:pg1613agn}}
\end{figure*}

\begin{figure*}[htp!]
\gridline{\includegraphics[height=0.3\linewidth,trim=4.0cm 0.25cm 4.5cm 1cm,clip]{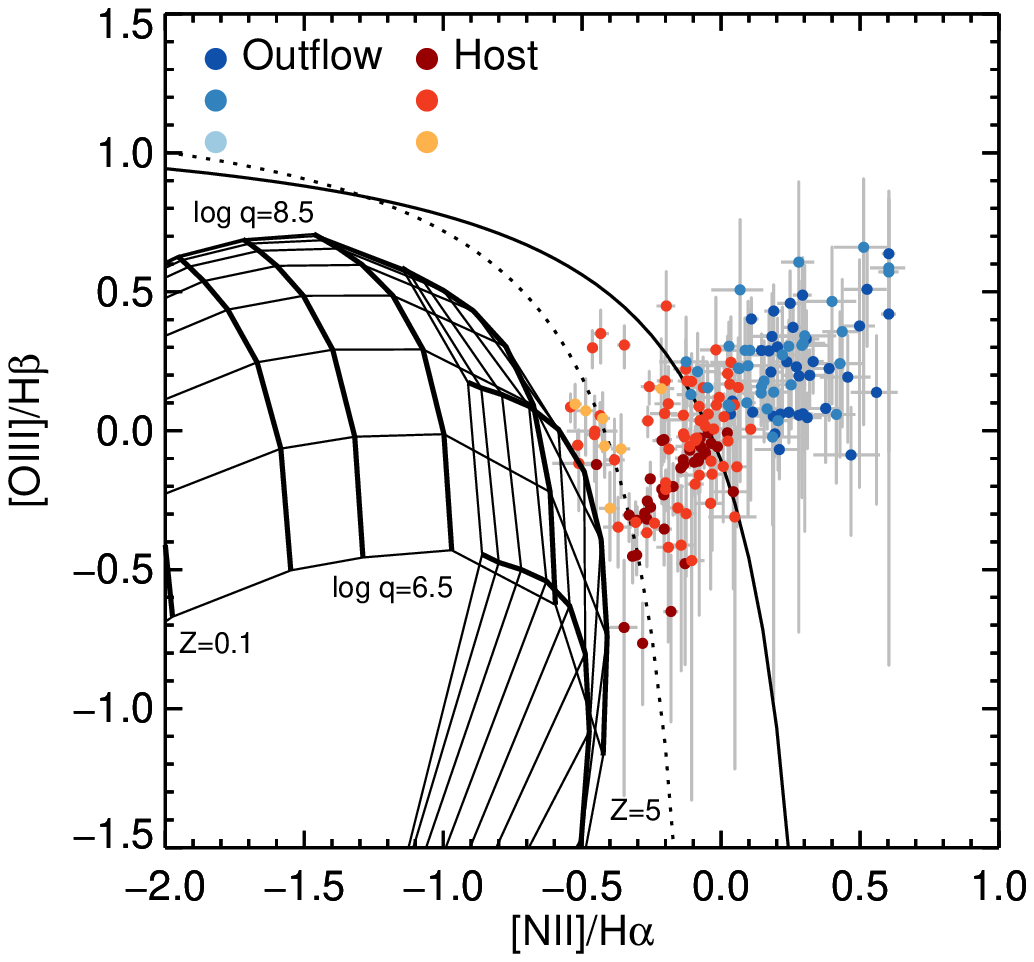}
		  \includegraphics[height=0.3\linewidth,trim=4.0cm 0.25cm 4.5cm 1cm,clip]{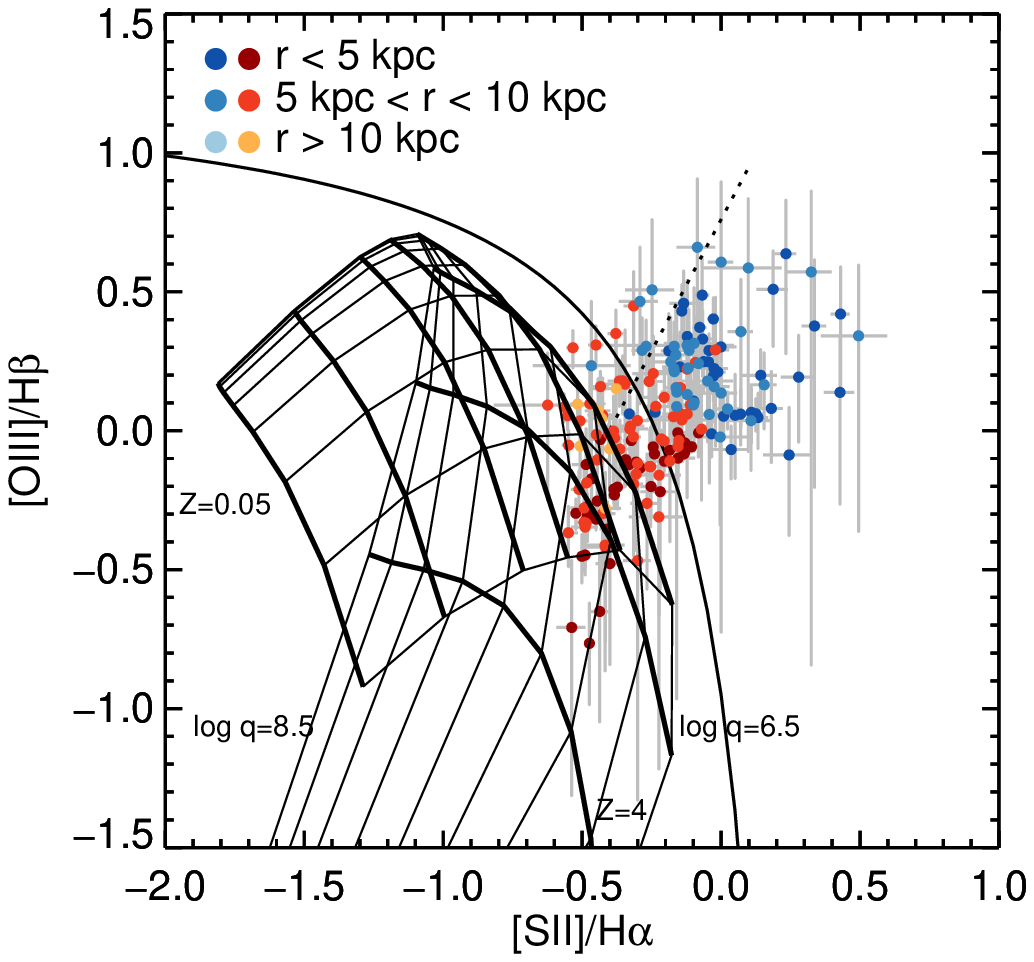}
          \includegraphics[height=0.3\linewidth,trim=4.0cm 0.25cm 4.5cm 1cm,clip]{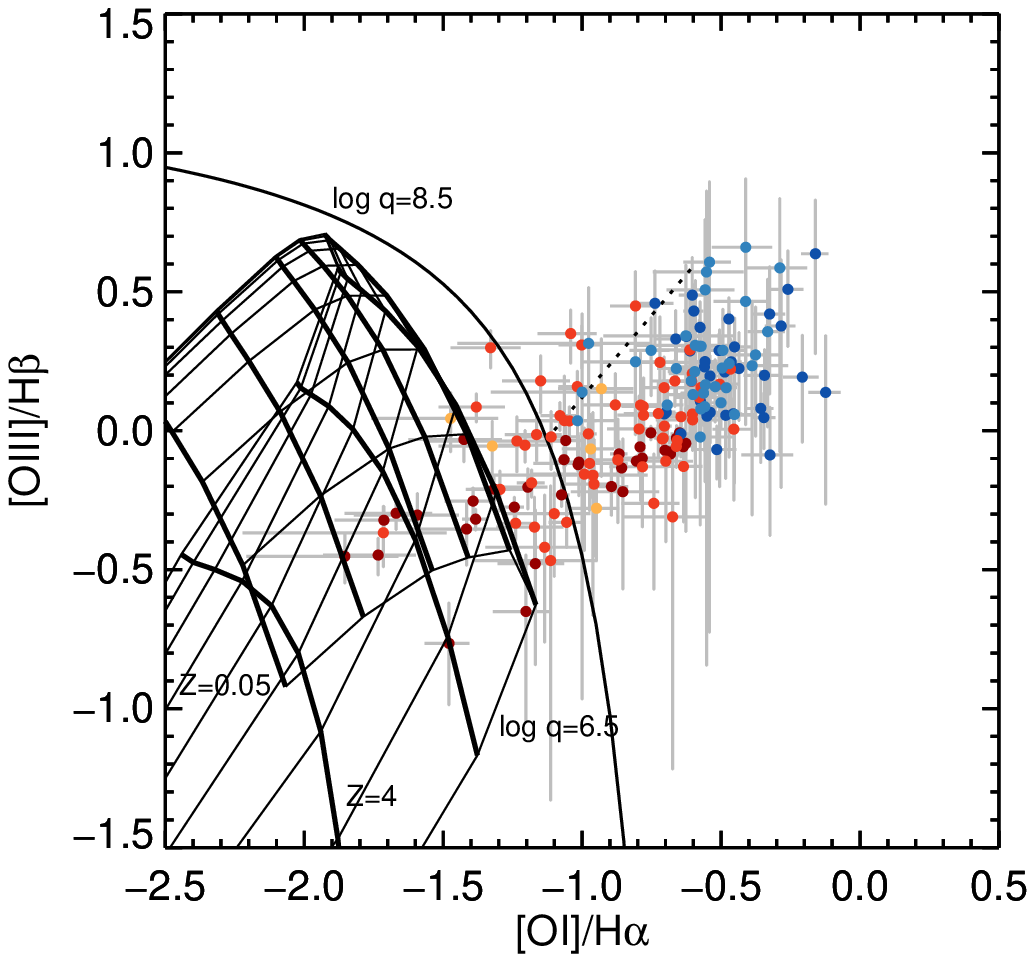}}
\caption{Comparisons between the line ratios observed in the host and outflow of F13218+0552 and the theoretical predictions from the starburst models of Dopita et al.\ (\citeyear{2013ApJS..208...10D}). The color-coding of the data points is the same as in Figure \ref{fig:f05189sdss}. The thick lines show constant metallicity $Z$ of 0.05, 0.1, 0.2, 0.3, 0.5, 1, 2, 3, and 5 times solar. The thin lines show constant logarithm of the ionization parameter $q = U c$ from 6.5 to 8.5 spaced by intervals of 0.25, where $q$ is in units of cm~s$^{-1}$. The electron density is frozen to $n_{e}$ = 10 cm$^{-3}$ with a Maxwell-Boltzmann distribution ($\kappa=\infty$). \label{fig:f13218hii}}
\end{figure*}

\begin{figure*}[htp!]
\gridline{\includegraphics[height=0.3\linewidth,trim=4.0cm 0.25cm 4.5cm 1cm,clip]{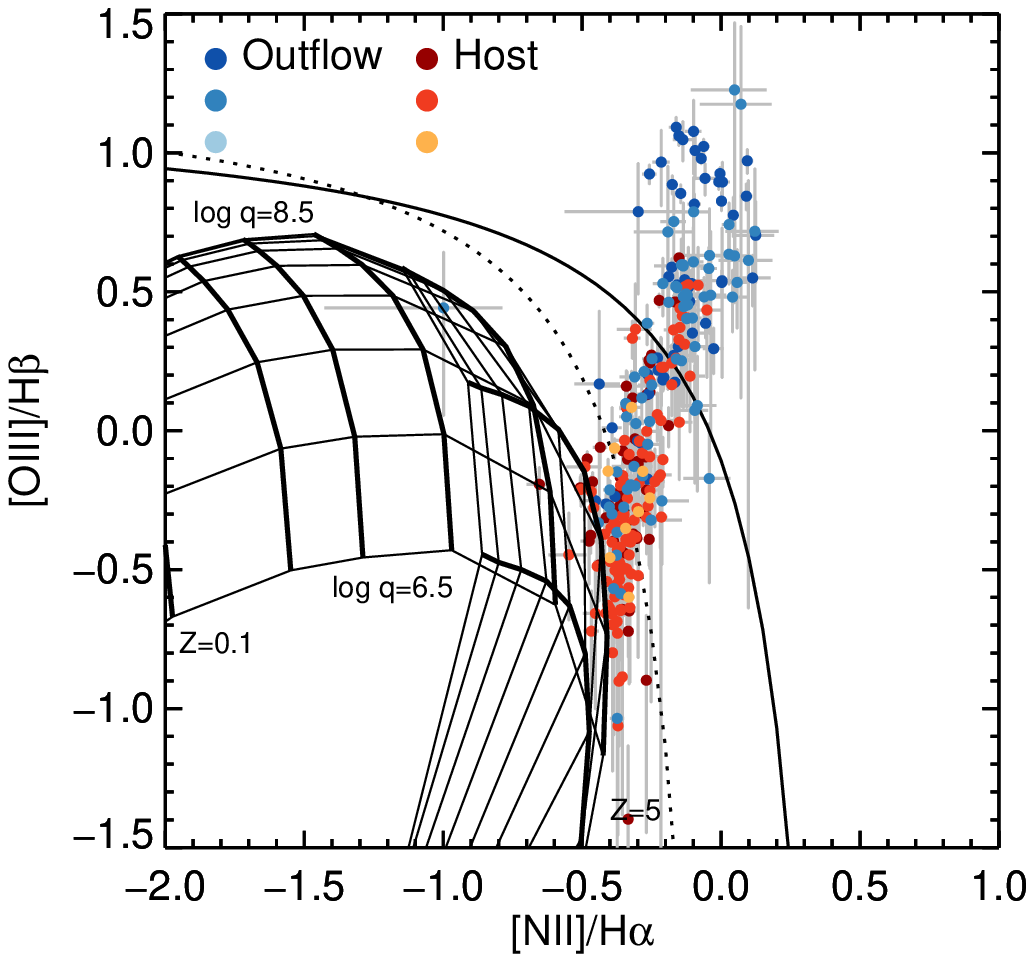}
		  \includegraphics[height=0.3\linewidth,trim=4.0cm 0.25cm 4.5cm 1cm,clip]{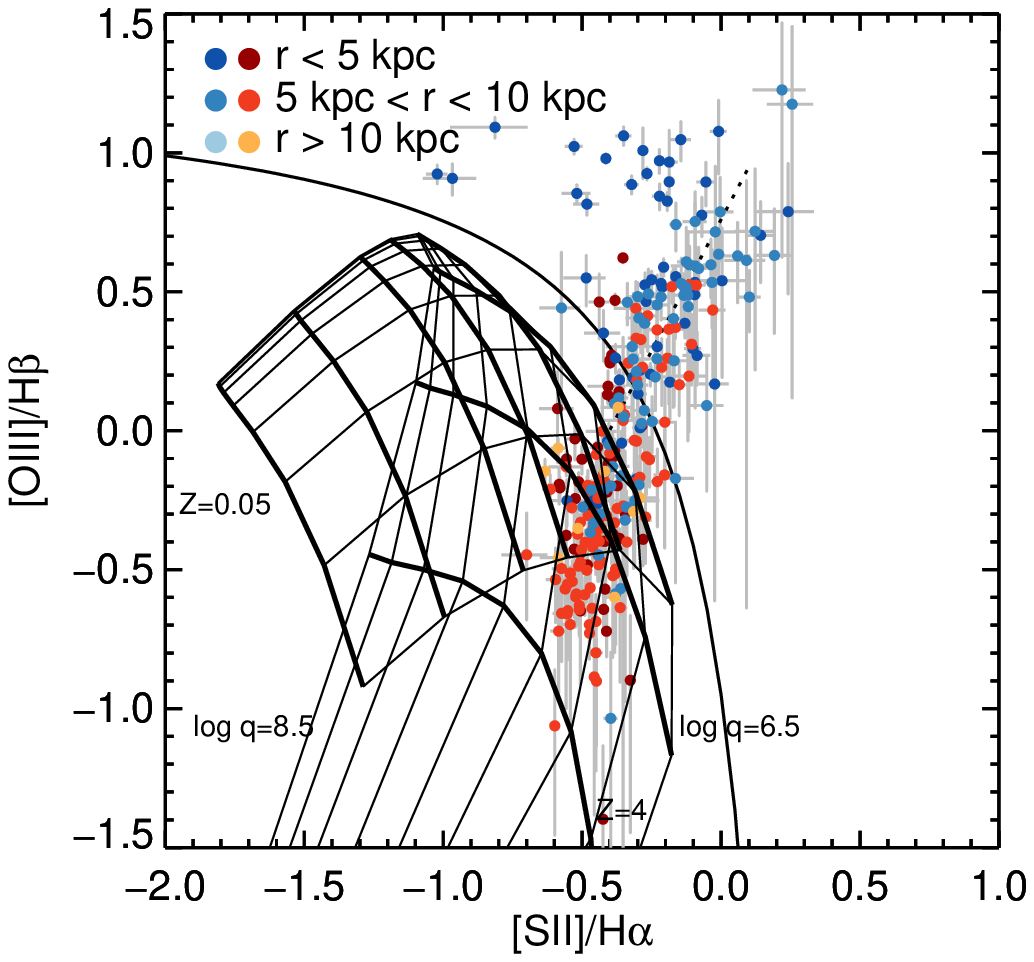}
          \includegraphics[height=0.3\linewidth,trim=4.0cm 0.25cm 4.5cm 1cm,clip]{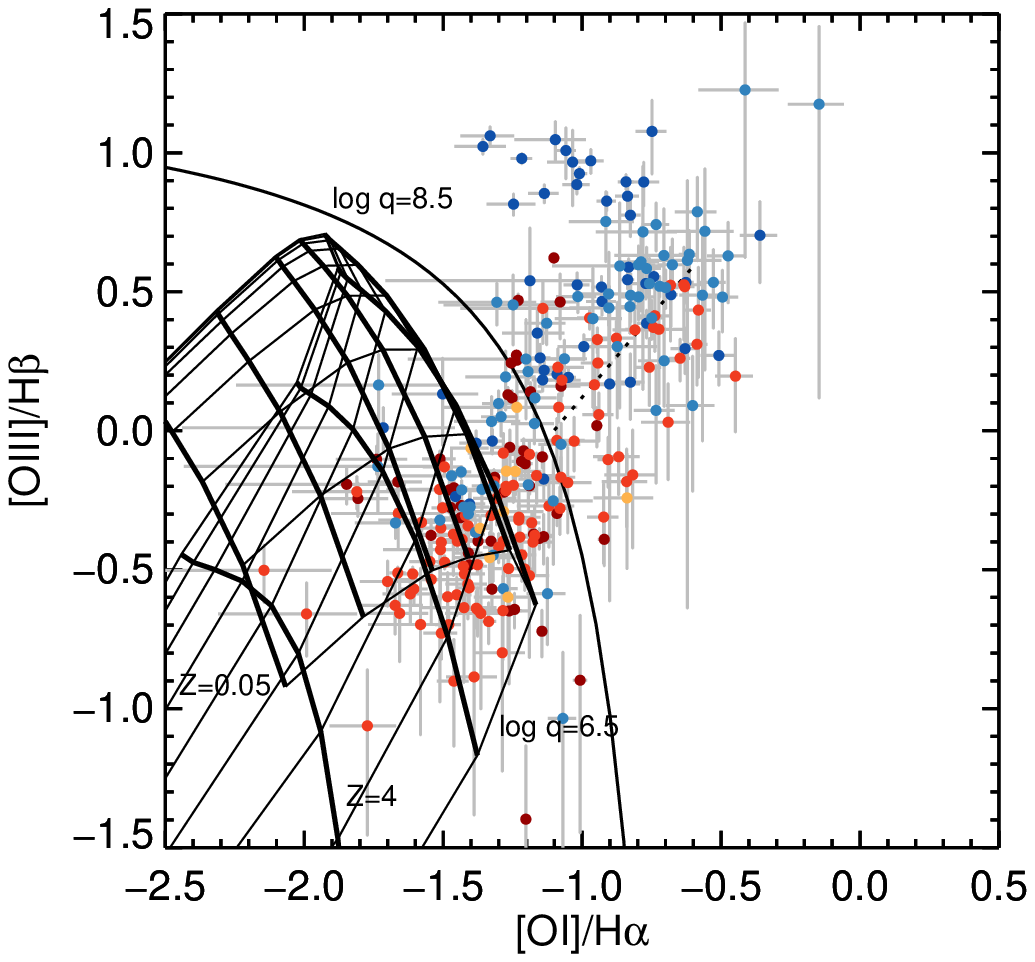}}
\caption{Same as Figure \ref{fig:f13218hii}, but for F13342+3932.\label{fig:f13342hii}}
\end{figure*}


\begin{figure*}[htp!]
\gridline{\includegraphics[height=0.3\linewidth,trim=3.5cm 0.25cm 4.25cm 1cm,clip]{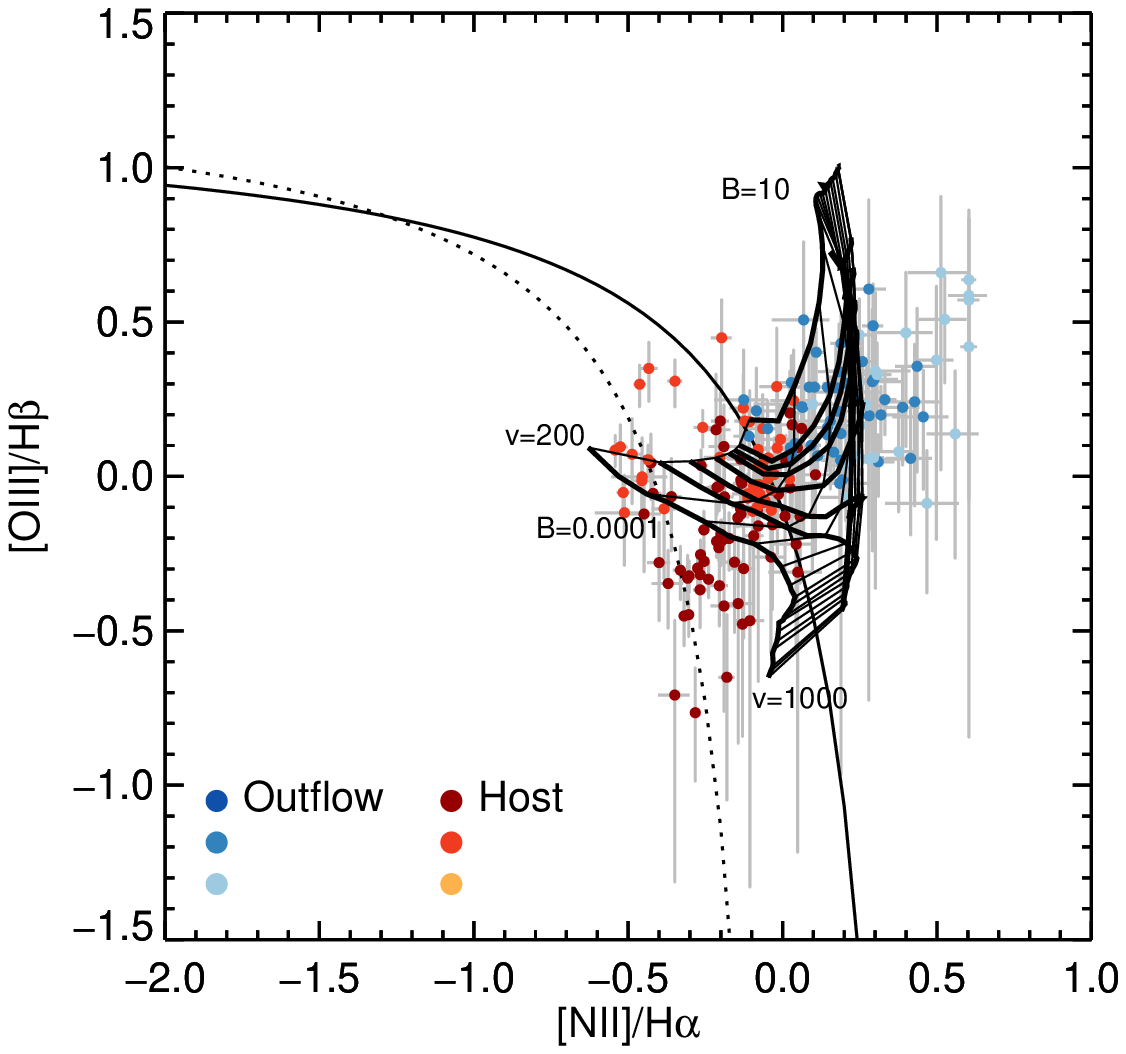}
		  \includegraphics[height=0.3\linewidth,trim=3.5cm 0.25cm 4.25cm 1cm,clip]{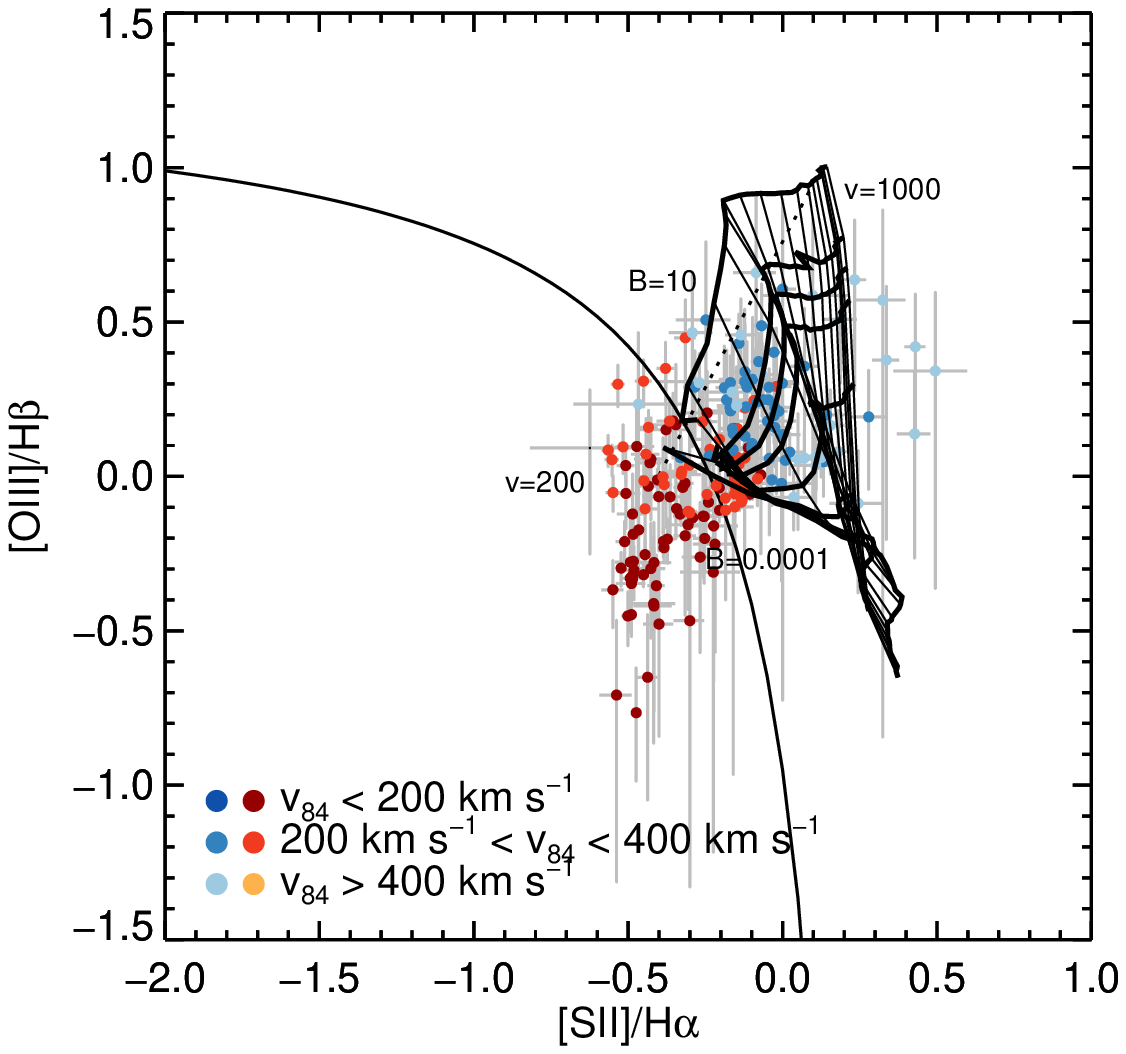}
          \includegraphics[height=0.3\linewidth,trim=3.5cm 0.25cm 4.25cm 1cm,clip]{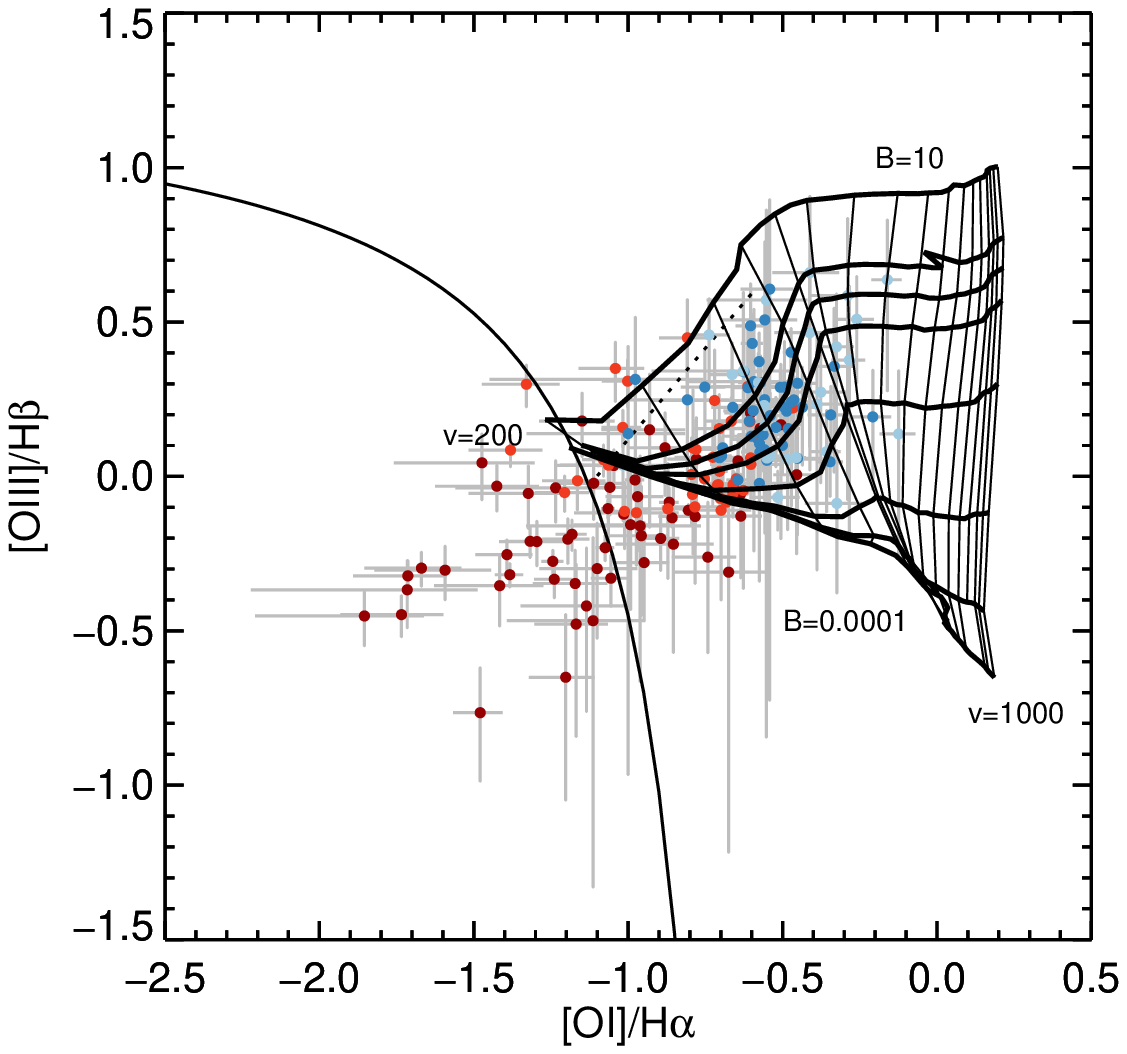}}
\caption{Comparisons between the line ratios observed in the host and outflow of F13218+0552 and the theoretical predictions from the shock models of Allen et al.\ (\citeyear{2008ApJS..178...20A}). The color-coding of the data points is based on the measured values of $v_{84}$. The thick lines show constant magnetic field parameter $B$ of 0.0001, 0.5, 1, 2, 3.23, 5, and 10 in units of $\mu$G cm$^{3/2}$. The thin lines show constant shock velocity $v_s$ from 200 to 1000 km s$^{-1}$ with contours separated by 100 km s$^{-1}$. The neutral hydrogen preshock density is frozen to $n_H$=1 cm$^{-3}$ and the metallicity is frozen to solar.\label{fig:f13218s}}
\end{figure*}

\begin{figure*}[htp!]
\gridline{\includegraphics[height=0.3\linewidth,trim=3.5cm 0.25cm 4.25cm 1cm,clip]{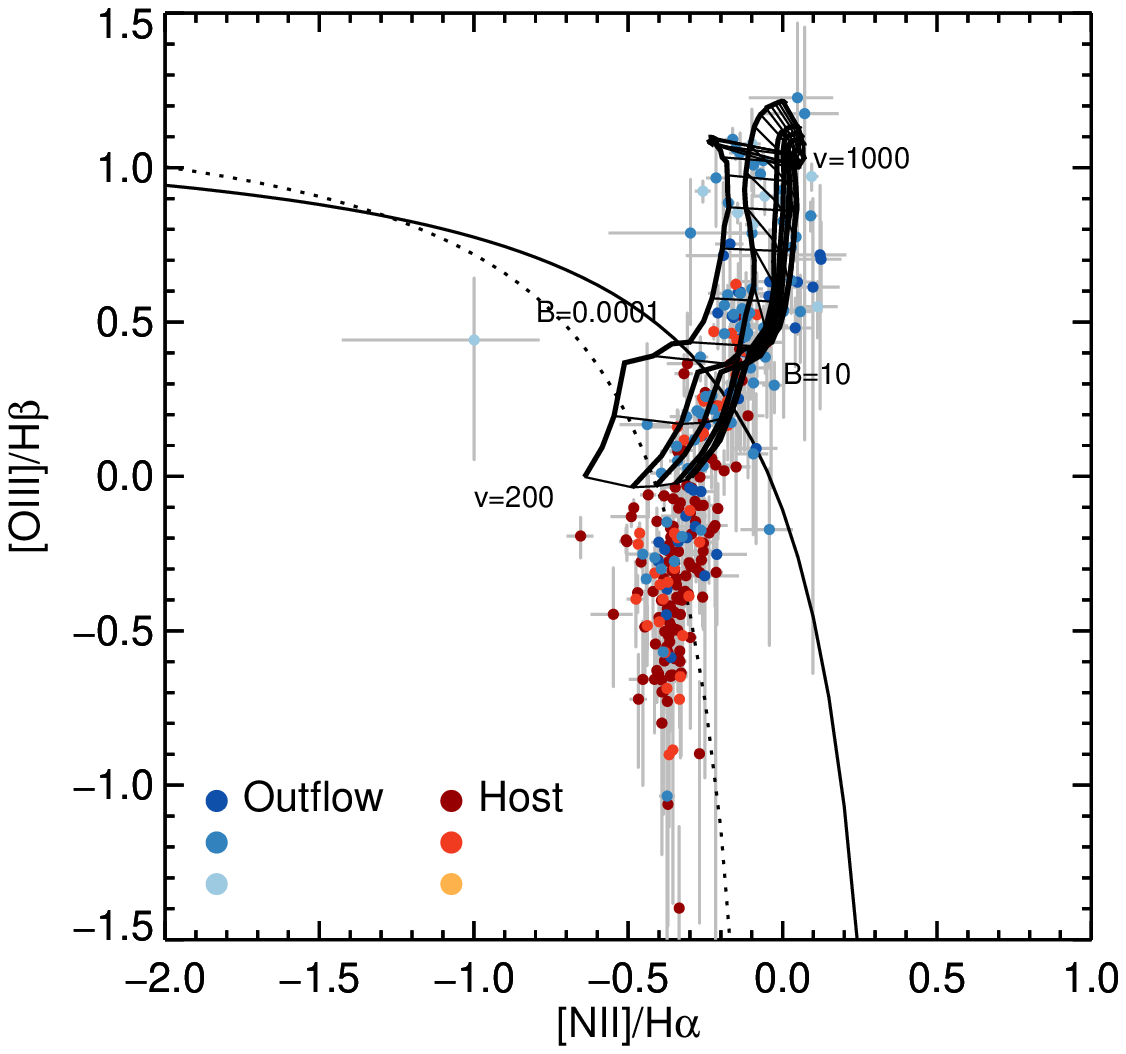}
		  \includegraphics[height=0.3\linewidth,trim=3.5cm 0.25cm 4.25cm 1cm,clip]{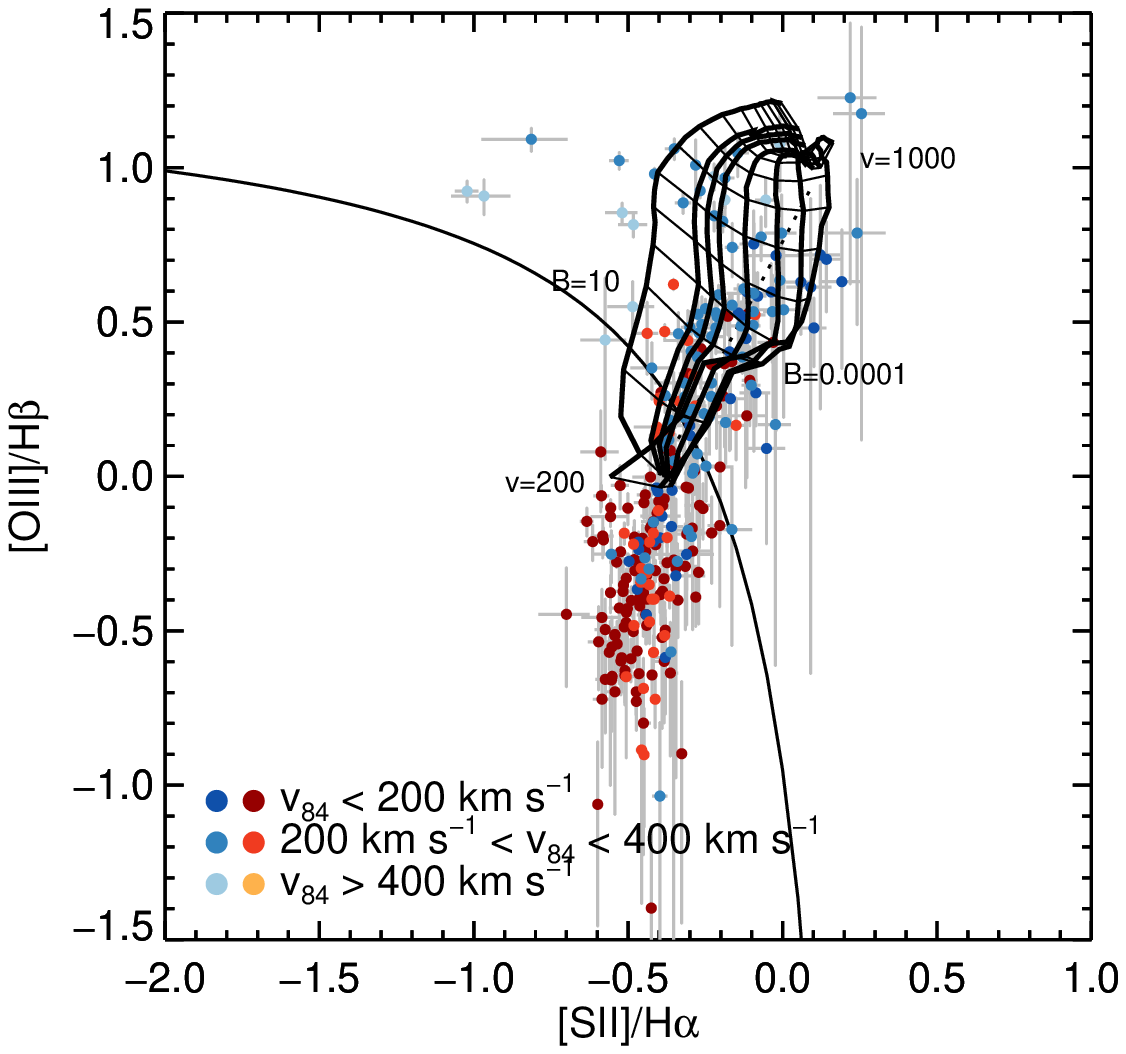}
          \includegraphics[height=0.3\linewidth,trim=3.5cm 0.25cm 4.25cm 1cm,clip]{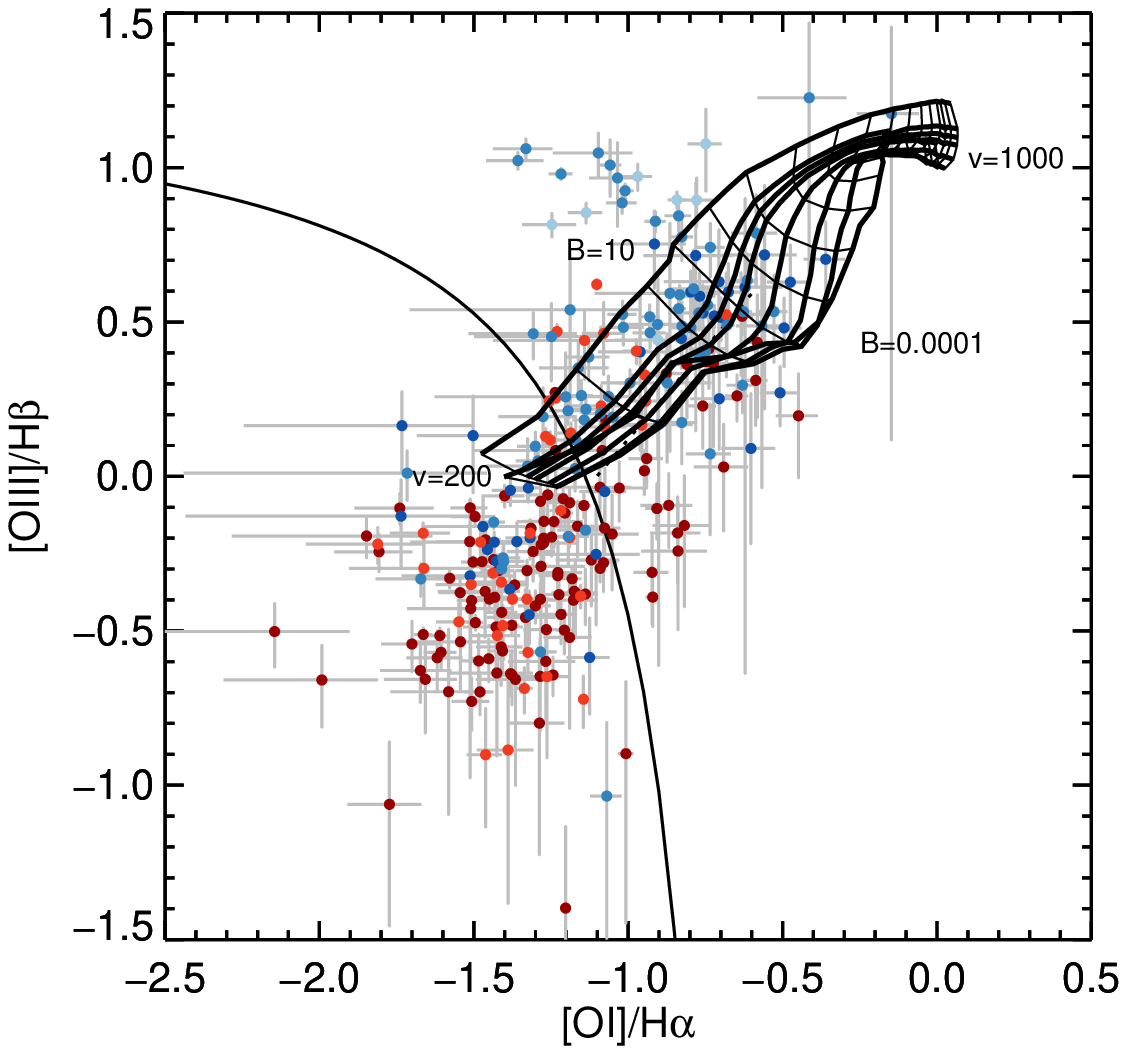}}
\caption{Same as Figure \ref{fig:f13218s} but for F13342+3932 and with a shock+precursor model.\label{fig:f13342sp}}
\end{figure*}

The results presented in \S 4 show that there is a broad diversity of excitation properties in the hosts and outflows of Type 1 quasars. Several qualitative statements were made about the role of AGN photoionization, hot star photoionization, and shocks in both components of each object. In the present section, we wish to quantify some of these statements by comparing the measured line ratios with the theoretical predictions from photoionization and shock models. For this exercise, we make use of ITERA (IDL Tool for Emission-line Ratio Analysis; Groves \& Allen \citeyear{2010NewA...15..614G}) to explore AGN photoionization (Groves et al.\ \citeyear{2004ApJS..153....9G}) and shock ionization (Allen et al.\ \citeyear{2008ApJS..178...20A}). We also compare our data with the theoretical starburst models of Dopita et al.\ \citeyear{2013ApJS..208...10D}. We choose to use these starburst models over those included in ITERA because they better reproduce the SDSS data of starburst galaxies presented in \S  4.

In Figures \ref{fig:density} and \ref{fig:denhist}, we compare the electron densities in the host and outflow based on the [S~II] $\lambda$6716 and $\lambda$6731 line ratios. A more detailed explanation of how these were calculated can be found in \citetalias{2017ApJ..850...40R}. From these figures, it is clear that there is a large range of densities in each of the components. Additionally, all but one of the objects (F13342+3932 is the only exception) display significantly different density distributions in the host galaxy and the outflow. This information, in conjunction with the theoretical models, will be used below to help us understand the different ionization states in these components.

\subsection{AGN Photoionization}\label{agn}

Since there is ample evidence that the gas in every object of our sample is exposed to the radiation field of the central AGN, we compare the predictions of AGN photoionization models with the line ratios of the host and outflowing gas components for all four objects in our sample.  We use the models of Groves et al.\ (\citeyear{2004ApJS..153....9G}) for this comparison. Based on the fiducial values selected in Groves et al.\ (\citeyear{2004ApJS..153....9G}), we present models that hold the neutral hydrogen density constant at $n_{H}$ = 1000 cm$^{-3}$ and the power law index is frozen to $\alpha=-1.4$. The metallicity $Z$ is varied from 0.25 to 4 times solar and the logarithm of the dimensionless ionization parameter $U$ (defined as the ratio of the number density of ionizing photons with energies above 13.6 eV and the electron density) is varied from $-$4 to 0. The results of these comparisons are shown in Figures \ref{fig:f05189agn} -- \ref{fig:pg1613agn}. 

The first important characteristic of these models is the extremely large area that they cover in the line ratio parameter space, especially in the diagram involving [\ion{N}{2}]/H$\alpha$. In addition, there are significant degeneracies between the line ratios and the parameters $U$ and $Z$. For example, the [\ion{O}{3}]/H$\beta$ ratio tends to increase with increasing ionization parameter, but this trend saturates and eventually turns over above log~$U$ $\sim$ $-$2. Similarly, the [\ion{N}{2}]/H$\alpha$, [\ion{S}{2}]/H$\alpha$ and [\ion{O}{1}]/H$\alpha$ line ratios correlate positively with the metallicity, but this correlation disappears above solar metallicities for  [\ion{S}{2}]/H$\alpha$ and [\ion{O}{1}]/H$\alpha$, but not for [\ion{N}{2}]/H$\alpha$.  One should therefore use caution when using these models to derive specific properties of the photoionized gas within a galaxy.

Figure \ref{fig:f05189agn} shows that most of the line ratios in the outflow of F05189$-$2524  can be reproduced with the AGN photoionization models of Groves et al.\ (\citeyear{2004ApJS..153....9G}), but these models tend to underpredict the [\ion{N}{2}]/H$\alpha$ and [\ion{O}{1}]/H$\alpha$ line ratios in the host galaxy, regardless of metallicity. Additionally, these models do not predict the points with high values of [\ion{O}{3}]/H$\beta$ and low values of the low-ionization line ratios. This can be explained if the clouds producing these lines are matter-bounded rather than ionization-bounded. In such a case, the truncated matter-bounded clouds produce strong high-ionization lines, but have very weak low-ionization lines (e.g., Binette et al.\ \citeyear{1996A&A...312..365B}). Notwithstanding this disagreement, these models indicate that the outflowing gas is characterized by a higher ionization parameter (log~$U$ = [$-$3, 0]) than the host (log~$U$ = [$-$3.5, $-$3]). This can be explained in two different ways: the gas in the outflow is more strongly exposed to the ionizing radiation from the central AGN than the gas in the host, or the density of the outflowing gas is lower than that of the host gas. Figures \ref{fig:density} and \ref{fig:denhist} seem to rule out this second possibility since the density of the outflowing gas in F05189$-$2524 is {\em higher} on average than the gas in the host galaxy. 

Figure \ref{fig:f13218agn} shows that the AGN photoionization models have difficulties explaining the broad range of values of the low-ionization line ratios in F13218+0552. The models tend to underpredict these line ratios in the outflow, and to overpredict [\ion{S}{2}]/H$\alpha$ and [\ion{O}{1}]/H$\alpha$ in the host. Assuming a larger metallicity in the outflow ($Z \ga 1$) compared with the host ($Z \la 0.5-1$) does not fully explain this discrepancy. Contrary to F05189$-$2524, the outflowing gas is characterized by an ionization parameter that is not significantly different from that of the host gas. 
The line ratios in the outflow of F13342+3932 are generally well reproduced by the AGN photoionization models (Figure \ref{fig:f13342agn}), except for a few spaxels with either large or very small [\ion{S}{2}]/H$\alpha$ and [\ion{O}{1}]/H$\alpha$ values. As mentioned in \S 4.3, shocks may also contribute to enhancing these line ratios (see also \S 5.3 below). The AGN models also have difficulties reproducing the many small [\ion{S}{2}]/H$\alpha$ and [\ion{O}{1}]/H$\alpha$ line ratios measured in the host of this galaxy. We return to this point below when we explore the starburst models (\S 5.2). Focusing only on the spaxels that are consistent with AGN photoionization, we note that some (but not all) of the outflowing gas is characterized by ionization parameters that are about +1 dex larger than the gas in the host. As in F05189$-$2524, this difference indicates that some of the outflowing gas is either exposed more directly to the ionizing radiation of the AGN than the gas in the host, or it is of lower density. Figures \ref{fig:density} and \ref{fig:denhist} do not show any significant difference in density between the outflow and host components in this object. The data thus favor the first scenario, as in F05189$-$2524. 

Figure \ref{fig:pg1613agn} confirms that the ionization of  both the host and outflow of PG1613+658 are dominated by AGN photoionization. The relatively large low-ionization line ratios of both components suggest metallicities that are approximately solar. As pointed out in \S 4.4, the outflowing gas shows, on average, systematically larger [\ion{O}{3}]/H$\beta$ line ratios and smaller low-ionization line ratios than that the gas in the host. This is interpreted once again as a difference in ionization parameter $U$ where the outflowing gas is either exposed more strongly to the ionizing radiation of the AGN than the gas in the host, or it is of lower density. As in F05189$-$2524, Figures \ref{fig:density} and \ref{fig:denhist} indicate that the density of the outflowing material in this object is {\em higher} on average than that of the host gas (Figs.\ \ref{fig:density} and \ref{fig:denhist}), so the first scenario is favored. The exact values of the ionization parameter for both components are difficult to derive from the AGN models since the line ratios lie in the region where the [\ion{O}{3}]/H$\beta$ line ratio becomes insensitive to the ionization parameter. We can only say that the line ratios in the outflow are consistent with log~$U$ $\ga$ $-$2, while those in the host indicate log~$U$ $\ga$ $-$3.

This apparent alignment between the outflowing gas and the ``ionization cones" of the narrow line region in F05189$-$2524, F13342+3932, and PG1613+658 has been seen in other nearby starburst galaxies, Seyfert galaxies, and quasars (e.g., Shopbell \& Bland-Hawthorn \citeyear{1998ApJ...493..129S}; Veilleux et al.\ \citeyear{2005ARA&A..43..769V}; Sharp \& Bland-Hawthorn \citeyear{2010ApJ...711..818S}; Kreimeyer \& Veilleux \citeyear{2013ApJ...772L..11K}; Fischer et al.\ \citeyear{2013ApJS..209....1F}; Revalski et al.\ \citeyear{2018ApJ...856...46R}, and references therein).

\subsection{Starburst Photoionization}\label{star}

We explore starburst photoionization models for objects where the line ratios of the outflowing component either overlap with the SDSS contours in Figures \ref{fig:f05189sdss}, \ref{fig:f13218sdss}, \ref{fig:f13342sdss}, and \ref{fig:pg1613sdss}, or are not reproduced by the AGN photoionization models in Figures \ref{fig:f05189agn} -- \ref{fig:pg1613agn}. The only objects that fit this description are F13218+0552 and F13342+3932. We compare the measured line ratios in these two objects with the starburst photoionization models of Dopita et al.\ (\citeyear{2013ApJS..208...10D}).  These models hold the electron density constant at $n_{e}$ = 10 cm$^{-3}$ and assume a Maxwell-Boltzmann distribution of electrons. The metallicity $Z$ is varied from 0.05 to 5 times solar and the logarithm of the ionization parameter $q = U c$ is varied from 6.5 to 8.5, where $q$ is in units of cm~s$^{-1}$.

The predicted line ratios from these models show good agreement with the SDSS data of starburst galaxies. Therefore, we can use these models in conjunction with the SDSS density contours in the three diagnostic line ratio diagrams to explore the possibility of photoionization by hot O and B type stars. However, as in the case of the AGN models, the starburst models have significant degeneracies affecting both $Z$ and $q$. These degeneracies are particularly pronounced in the diagrams involving [\ion{S}{2}]/H$\alpha$ and [\ion{O}{1}]/H$\alpha$. 

Figure \ref{fig:f13218hii} compares the line ratios of F13218+0552 presented in Figure \ref{fig:f13218sdss} with the starburst photoionization models  of Dopita et al.\ (\citeyear{2013ApJS..208...10D}). There is considerable overlap between these predictions and the data in the diagrams involving [\ion{S}{2}]/H$\alpha$ and [\ion{O}{1}]/H$\alpha$, but the models underpredict [\ion{N}{2}]/H$\alpha$. This comparison suggests that some of gas in the host is photoionized by hot stars rather than the AGN, and has a metallicity that is $\sim$ 1 $-$ 2 $\times$ solar. However, it is also clear that stellar photoionization is not a significant factor in the outflowing gas of this object. 

Figure \ref{fig:f13342hii} indicates that a significant fraction of the gas in the host galaxy of F13342+3932 is photoionized by hot OB stars from active star-forming regions, although the starburst models again underpredict the measured [\ion{N}{2}]/H$\alpha$ line ratios. Some of the outflowing material also appears to be photoionized by young stars. As mentioned in \S 4.3 and shown in Figure \ref{fig:f13342map}, the outflowing material that has these \ion{H}{2} region-like characteristics is located south of the nucleus, spatially coincident with \ion{H}{2} region-like gas in the host component. The outflowing material that is photoionized by OB stars is characterized by ionization parameters and metallicities that are similar to those of the host material at the same projected location. Figures \ref{fig:density} and \ref{fig:denhist} also show that the density of the outflowing material is similar to that of the host gas. The simplest explanation for the \ion{H}{2} region-like line ratios of this outflowing gas is therefore that it is being photoionized by hot stars in the host galaxy rather than by hot stars formed in-situ in the outflowing material itself (cf.\ Maiolino et al.\ \citeyear{2017Nature..544...202M} and Gallagher et al.\ \citeyear{2019MNRAS.tmp..556G}).

\subsection{Shocks}\label{shock+pre}

As pointed out in \S 3, the correlations between the line ratios and line widths in F13218+0552 and F13342+3932 suggest that shocks may play a role in these objects. The shock models of Allen et al.\ (\citeyear{2008ApJS..178...20A}) are used to quantify the role of shocks in these models. We consider shock models with and without a precursor component. The shock + precursor models takes into account the line emission from the gas that is photoionized by the EUV radiation emitted by the shocked gas, while the shock-only models do not. The shock + precursor models generally produce stronger [\ion{O}{3}]/H$\beta$ than the shock-only models. Both sets of models hold the neutral hydrogen preshock density constant at $n_{H}$=1 cm$^{-3}$ and assume a solar metallicity. The magnetic field parameter $B/n^{1/2}$ is varied from 0.0001 to 10 $\mu$G cm$^{3/2}$, while the shock velocity is varied from 200 to 1000 km s$^{-1}$, thus incorporating both slow and fast shocks.  Besides the high magnetic parameter lines in the [\ion{N}{2}]/H$\alpha$ diagram, these models are generally non-degenerate and therefore allow us to constrain both $B$ and $v_s$ in these galaxies, if shocks are indeed dominant. 

Figure \ref{fig:f13218s} shows the line ratios of F13218+0552 presented in Figure \ref{fig:f13218sdss}, overlaid with the shock-only models of Allen et al.\ \citeyear{2008ApJS..178...20A}. Here we use $v_{84}$ to color-code the data. We choose to use $v_{84}$ over the velocity dispersion $\sigma$ because it shows a broader range of values than $\sigma$, so it gives us more leverage to understand the velocity structure of the host and outflow. These models provide a good fit to most of the lines ratios in the outflowing gas, but not for the host material in general. The shock + precursor models of Allen et al.\ (not shown in Figure \ref{fig:f13218s}) provide a poorer fit to the data, generally overpredicting the [\ion{O}{3}]/H$\beta$ line ratios. This comparison suggests that shocks with velocities $\sim$  200 $-$ 500 km s$^{-1}$  may be significant in the outflowing gas of F13218+0552. 

The higher [\ion{O}{3}]/H$\beta$ line ratios in F13342+3932 favor the shock + precursor models over the shock-only models (see Figure \ref{fig:f13342sp}). Once again, the shock models are a good match to the lines ratios in the outflowing material but not for most of the host material, despite the correlations between line ratios and line  widths observed in the host galaxy component of this galaxy (Fig.\ \ref{fig:f13342rich}). The line ratios in the outflowing gas cover the entire range of shock velocities and magnetic parameters, suggesting a broad range of shock conditions in the outflow. An important caveat of this analysis is the fact that the AGN also contributes to the ionization of the outflowing gas, likely boosting the [\ion{O}{3}]/H$\beta$ line ratio that is used to assess the importance of the precursor in these shock models and constrain the shock velocities.

\section{Summary}\label{sec:conc}

In this paper, we have presented the emission-line properties of four Type 1 quasars, separating the outflowing gas from the quiescent gas in the host galaxies.  We used the classic diagnostic two-dimensional line ratio diagrams of \citetalias{1981PASP...93....5B} and \citetalias{1987ApJS..63...295V} and the kinematic information derived from the line profiles to assess the presence and importance of AGN photoionization, stellar photoionization, and shock ionization in each galaxy. This analysis was supplemented with comparisons of the data with the predictions of ionization models from the literature. The main results are the following:

\begin{itemize}

\item All of the objects in our sample have outflowing gas that is photoionized by the central quasar. The outflowing material in these objects is generally characterized by a higher ionization parameter than the host gas, an indication that this gas is more strongly irradiated by the central quasar. This apparent alignment between the outflowing gas and the ionization cones of the narrow line region has been seen in other nearby galaxies. 

\item Shocks may also contribute to the ionization of some of the outflowing material, particularly in F13218+0552 and F13342+3932, where we detect trends between the line ratios and line widths that are hard to explain otherwise. The AGN models have difficulties explaining the strong low-ionization lines in the outflowing material of these two objects. AGN photoionization is no doubt responsible for the large scatter in the correlations between line ratios and line widths and the elevated [\ion{O}{3}]/H$\beta$ in the outflowing material. However, the overlap between the line ratio predictions of the AGN and shock models do not allow us to precisely determine the relative contributions of AGN photoionization and shock ionization in the outflowing gas.

\item In the end, only one object, F13342+3932, in this small sample of four quasars harbors some outflowing gas that appears to be photoionized by hot young stars. All of the \ion{H}{2} region-like outflowing material is located south of the nucleus, near host material that is also photoionized by hot stars and characterized by very similar physical conditions (metallicity, ionization parameter). These data favor a scenario where the same hot OB stars that are photoionizing the material in the host are also responsible for the ionization of the outflowing material. Thus, in this case, there is no need for additional hot young stars in the outflow itself to reproduce the measured line ratios.

\end{itemize}

While it is hard to draw any general conclusion from a sample of only four objects, it is interesting to note that F13342+3932 has the lowest AGN fraction among the objects in the sample (0.69$^{+0.23}_{-0.24}$, i.e. the starburst in this object contributes $\sim$31\% of the bolometric luminosity of this system). Perhaps this simply reflects the fact that powerful starbursts are needed to dominate the ionization of the outflowing material over that of the central quasars. A more detailed discussion of possible trends between dominant ionization processes and host and outflow properties (e.g., black hole masses, outflow velocities and mass rates; Table 1) is not warranted here given the small sample size.

However, we feel that the analysis presented in this paper illustrates the potential of IFS studies in the future. Soon, large-scale surveys with IFS on 8-meter class telescopes will allow us to carry out two-dimensional spectroscopic analyses with a larger, more representative, sample of active galaxies. It will also be helpful to expand the parameter space by including galaxies covering a broader range of redshifts, to see how the relative influence of the AGN, starbursts, and shocks change relative to one another as function of look-back time.

\acknowledgments 
We would like to thank the anonymous referee for helpful comments that have improved this paper. S.V. was supported in part by NSF grant AST1009583 and NASA grant ADAP NNX16AF24G. S.V. acknowledges support from a Raymond and Beverley Sackler Foundation Distinguished Visitor fellowship, and thanks the host institute, the Institute of Astronomy, where some of this work was performed. S.V. also acknowledges support by the Science and Technology Facilities Council (STFC) and by the Kavli Institute for Cosmology, Cambridge. D.S.N.R. was supported in part by the J. Lester Crain Chair of Physics at Rhodes College and by a Distinguished Visitor grant from the Research School of Astronomy \& Astrophysics at Australian National University.

This work was based on observations obtained at the Gemini Observatory (program IDs GS-2011B-Q-64 and GN-2012A-Q-15), which is operated by the Association of Universities for Research in Astronomy, Inc., under a cooperative agreement with the NSF on behalf of the Gemini partnership: the National Science Foundation (United States), the National Research Council (Canada), CONICYT (Chile), Minist\'erio da Ci\^{e}ncia, Tecnologia e Inova\c{c}\~{a}o (Brazil) and Ministerio de Ciencia, Tecnolog\'ıa e Innovaci\'on Productiva (Argentina).

\end{document}